\documentclass[10pt]{article}
\usepackage{amsmath,amsfonts,amssymb,amsthm,bbm}
\usepackage{pdfsync}
\usepackage{graphicx}
\usepackage[latin1]{inputenc}
\usepackage{hyperref}
\usepackage[all]{xy}
\usepackage{xcolor}
\usepackage{colortbl}
\usepackage{stmaryrd}
\usepackage{rotating}
\usepackage{psfrag}
\usepackage{dsfont}
\usepackage{subcaption}
\usepackage{enumitem}
\usepackage{lipsum}  

\newcommand{\sixj}[2]{
\bgroup
\setlength{\arraycolsep}{0.05em} 
\left\lbrace \begin{array}{ccc}
#1 \\ 
#2
\end{array} \right\rbrace
\egroup
}

\newcommand{\njm}[2]{
\bgroup
\setlength{\arraycolsep}{0.1em} 
\left( \begin{array}{cccccc}
#1 \\ 
#2
\end{array} \right)
\egroup
}

\begin{document}

\title{\bf Infrared divergences in the EPRL-FK Spin Foam model}

\author{\Large{Pietro Don\`a\footnote{pxd81@psu.edu}}
\smallskip \\ 
\small{Institute for Gravitation and the Cosmos \& Physics Department,}\\
\small{Penn State, University Park, PA 16802, USA} \\
}
\date{\today}

%Report number: IGC-18/2-2

\maketitle

%----------------------------------------------------------------------------
\begin{abstract}
\noindent 
We provide an algorithm to estimate the divergence degree of the Lorentzian EPRL-FK spin foam amplitudes for arbitrary 2-complexes. We focus on the ``self-energy'' and ``vertex renormalization'' diagrams and find an upper bound estimate. We argue that our upper bound must be close to the actual value, and explain what numerical improvements are needed to verify this numerically. For the self-energy, this turns out to be significantly more divergent than the lower bound estimate present in the literature. We support the validity of our algorithm using 3-stranded  versions of the amplitudes (corresponding to a toy 3d model) for which our estimates are confirmed numerically. We also apply our methods to the simplified EPRLs model, finding an utterly convergent behavior, and to BF theory, independently recovering the divergent estimates present in the literature. 
\end{abstract}
%----------------------------------------------------------------------------

\tableofcontents

%----------------------------------------------------------------------------
\section{Introduction}
The spin foam formalism is an attempt to define the dynamics of loop quantum gravity in a background independent and Lorentz covariant way \cite{perez_spin-foam_2013,rovelli_covariant_2015}. It defines transition amplitudes for spin network states of the canonical theory in a form of a sum (or equivalently a refinement \cite{rovelli_quantum_2012}) over all the possible two-complexes having the chosen (projected) spin networks as boundary. This is equivalent to a sum over histories of quantum geometries providing in this way a regularised version of the quantum gravity path integral.
The state of the art is the model proposed by Engle, Pereira, Rovelli and Livine (EPRL) \cite{engle_lqg_2008,livine_new_2007,livine_solving_2008} and independently by Freidel and Krasnov (-FK) \cite{freidel_new_2008} and its extension to arbitrary spin network states \cite{kaminski_spin-foams_2010,ding_generalized_2011}. The model admits a quantum group deformation conjectured to describe the case of non-vanishing cosmological constant \cite{han_cosmological_2011,haggard_four-dimensional_2016} and notably, the large spin asymptotics of the 4-simplex vertex amplitude contains exponentials of the Regge action \cite{barrett_asymptotic_2009,barrett_lorentzian_2010}. The model is free of \textit{ultraviolet} divergences because there are no trans-Planckian degrees of freedom, however, there are potential large-volume \textit{infrared} divergences.\\

The presence of divergences may require some sort of renormalization procedure, and in general, their study and understanding is important in the definition of the continuum limit. This has been the subject of many studies and can be achieved in many ways: via refining of the 2-complex as proposed in \cite{dittrich_decorated_2016,dittrich_continuum_2017,bahr_investigation_2016}, or via a resummation, defined for instance using group field theory/random tensor models as proposed in \cite{bonzom_random_2012, benedetti_phase_2012, carrozza_renormalization_2014, geloun_functional_2016}.
The properties of these divergences have been studied in the context of the Ponzano-Regge model of 3d quantum gravity and discrete BF theory \cite{bonzom_bubble_2012}, group field theory \cite{baratin_melonic_2014} and EPRL model: with both Euclidean \cite{perini_self-energy_2009, krajewski_quantum_2010} and Lorentzian signature \cite{riello_self-energy_2013}. 

In particular \cite{riello_self-energy_2013} is, to our knowledge, the only analytic estimate of divergences in the Lorentzian EPRL model. It considers the ``self
energy'' (see Figure  \ref{fig:bubble4D}), finding a logarithmic divergence as a \emph{lower bound}. The computation is rather involved and relies on the techniques developed for the asymptotic analysis of the vertex amplitude of the model \cite{barrett_lorentzian_2010}. This approach requires an independent study of each geometrical sector: crucially, the logarithmic divergence is obtained by looking at the non-degenerate geometries, resulting in a lower bound estimate only. Our results suggest that this lower bound is close to 9 powers short. Moreover, even if in principle the same technique of \cite{riello_self-energy_2013} applies to \textit{any} spin foam diagram, doing it is a very challenging task. 
On the other hand, the various estimates provided in \cite{perini_self-energy_2009} for the Euclidean model of both the ``self-energy'' diagram and the ``vertex renormalization'' diagram, (see Figures \ref{fig:bubble4D} and \ref{fig:ball4D}) just rely on the scaling for large spins of SU(2) invariants, and they are easily applicable to \textit{any} spin foam diagram. Nevertheless, the extension of this technique to the Lorentzian model is not at all straightforward, due to the non-compactness of the Lorentz group.\\

In this work, we develop a simple algorithm to systematically determine the potential divergence of \textit{all} spin foam diagrams within the EPRL model. Instead of approaching it directly in its generality we proceed by increasing complexity a bit at a time: we will introduce our algorithm first for SU(2) BF theory, moving to a simplified version of the EPRL model and concluding with the full quantum gravity model. We review the three transition amplitudes and their relation in Section \ref{sec:gravityandBF}. In Section \ref{sec:diagrams} we introduce the four diagrams in analysis. Again, we opted to increase complexity gradually: before approaching the four stranded diagrams corresponding to a four dimensional triangulation (each four stranded edge is dual to a tetrahedron) we warm up with the analog three stranded diagrams corresponding to a three dimensional triangulation (each edge is dual to a triangle). Three dimensional spin foam diagrams are simpler than their four dimensional counterpart for the absence of edge intertwiners and the overall smaller number of internal faces. We will consider both three and four dimensional bubble and ball diagrams. In Sections \ref{sec:modelSU2}, \ref{sec:modelEPRLs}, \ref{sec:modelEPRL} we proceed with the study of the divergence of the diagrams one by one in order of complexity. We then conclude summarizing the algorithm and the results obtained. Let us for the impatient reader comment here the results. We estimate both the bubble and the ball amplitudes in the four dimensional EPRL model to be \textit{divergent} with the same power of the cutoff of the analog diagrams for SU(2) BF model. Furthermore, we also find convergence for all the diagrams in the simplified EPRL model and the three dimensional ones for the EPRL model.

\section{The EPRL model and its connection with BF theory}
\label{sec:gravityandBF}
We assume that the reader is familiar with the EPRL-FK\footnote{from now on we will call it just EPRL for notation convenience.} model, and refer to the original literature \cite{engle_lqg_2008,livine_new_2007,livine_solving_2008,freidel_new_2008} and existing reviews (e.g. \cite{perez_spin-foam_2013,rovelli_covariant_2015}) for motivations, details and its relation to Loop Quantum Gravity. In the following, we will use an unconventional notation for the partition function which was recently developed in \cite{speziale_boosting_2016}.

Given a closed 2-complex $\mathcal{C}$ the partition function is a state sum over $SU(2)$ spins $j_f$ and intertwiners $i_e$, associated respectively with faces $f$ and edges $e$: 
\begin{equation}
\label{partitionEPRL}
Z_{\mathcal{C}} = \sum_{j_f, i_e}  \prod_f A_f(j_f) \prod_e (2 i_e +1) \prod_v A_v \left(j_f, \  i_e\right) \ .
\end{equation}
We denoted with $A_f(j_f)$ the face weights: the requirement that the path integral at fixed boundary graph compose correctly under convolution fixes the face weight to be $A_f(j)\equiv 2 j +1$ \cite{bianchi_face_2010} but to compare to various other models present in the literature we will use a generalized face weight $A_f(j)\equiv \left(2j+1\right)^\mu$ (i.e. $\mu=1$ correspond to the choice made in the BF SU(2) model and the EPRL model, $\mu=2$ correspond to the BF $SO(4)$ model). To have more symmetric expressions we will also take the dimensions of the intertwiners on the edges to be $\left(2 i_e +1\right)\to \left(2 i_e +1 \right)^\mu$. The main goal of this paper is to find a systematic way to study the convergence of the multidimensional infinite sum $\sum_{j_f, i_e}$.

To each vertex $v$ of the two-complex a vertex amplitude is associated:
\begin{equation}
\label{vertexEPRL}
A_v \left(j_f, \, i_e\right) = \sum_{l_{fv}, k_{ev}} \left(\prod_{ev} \left(2k_{ev}+1\right) B_{n_{ev}}(j_{fv},l_{fv};i_{ev}, k_{ev})\right) \{3nj\}_v(l_{fv}, k_{ev}) \ ,
\end{equation}
it is defined as a superposition of $SU(2)$ invariants $\{3nj\}$\footnote{the specific invariant depend on the details of the vertex, if the vertex is dual to a 4-simplex the invariant is the ${15j}$ symbol.} weighted by one booster functions $B_{n_{ev}}$ per edge $ev$ touching the vertex $v$, with $n_{ev}$ the valency of the edge ${ev}$. The sums run over a set of auxiliary spins $l_{fv}$\footnote{that are effectively magnetic indices respect the group $SL(2,\mathbb{C})$} associated to each face $fv$ containing the vertex $v$, with $l_{fv}\geq j_{fv}$, and a set of auxiliary intertwiners $k_{ev}$ for each edge $ev$ connected to the vertex $v$.
Notice that the formulas for the partition function \eqref{partitionEPRL} and \eqref{vertexEPRL} are extendable to generalized spin foams with 2-complexes dual to arbitrary tesselations done with polyhedra being careful of using the appropriate dimension of the intertwiner space instead of $2 i_e +1$ and $2 k_{ev} + 1$ (i.e. for three valent %stranded? 
edges the intertwiner space associated to each edge is trivial and $i_e=k_{ev}=0$ on those edges; for five valent edges the intertwiner space associated to each edge is determined by two spins and the proper dimension to use is $\left(2 i_{e_1} +1\right)\left(2 i_{e_2} +1\right)$).

The booster functions encode all the details of the EPRL model, they are defined in the following way: 
\begin{equation}
\label{boosters}
B_n(j_a,l_a;i,k) = \frac{1}{4\pi} \sum_{p_a} \left(\begin{array}{c} j_a \\ p_a \end{array}\right)^{(i)} \left(\int_0^\infty \mathrm{d} r \sinh^2r \, 
\prod_{a=1}^n d^{(\gamma j_a,j_a)}_{j_a l_a p_a}(r) \right)
\left(\begin{array}{c} l_a \\ p_a \end{array}\right)^{(k)}\ ,
\end{equation}
where the boost matrix elements  $d^{(\rho,k)}(r)$ for $\gamma$-simple irreducible representation of $SL(2,\mathbb{C})$ in the principal series, $\gamma$ is the Immirzi parameter and the $(njm)$ symbols are reported in Appendix \ref{AppSU2}. We are using the notations used in \cite{speziale_boosting_2016}.
On one hand, the introduction of booster functions simplifies a lot the computation of spin foam transition amplitudes because it trades the problem of dealing with many high oscillatory integrals with the study a family of one dimensional integrals, which are easier to handle and manipulate. Analytical and numerical properties of these functions are work in progress \cite{speziale_boosting_2016, sarno_2-vertex_2018, citaPierre, citanoiEPRL}. On the other hand, the explicit evaluation of booster functions in spite of their rather simple form is still a very involved task: For $n=3$ we employ an expression for \eqref{boosters} in terms of finite sums of $\Gamma$ functions, for details see \cite{kerimov_clebsch-gordan_1978,speziale_boosting_2016}; for $n\geq 4$
a similar formula exists but features an integration over virtual labels\footnote{See Equation (41) of \cite{speziale_boosting_2016}.}, and in the end we found it less time consuming to numerically integrate directly the boost integrals. A C numerical code for the virtual irreps formula has been recently developed in \cite{citaGozzini}. The asymptotic behavior for large spins is still unknown: the properties we will need for our analysis will be inferred from numerical analysis.

As suggested in \cite{speziale_boosting_2016}, we introduce here a simplified version of the EPRL model, we will denote it EPRLs where \textit{s} stays for simplified. The reformulation of the EPRL amplitude as in \eqref{vertexEPRL} traded the major complexity of multiple integrals over the non-compact group $SL(2,\mathbb{C})$ with multiple infinite sums over the auxiliary spins $l$. We can for the moment put aside the proliferation of spin labels and fix all the new spins $l_{fv}$ to their minimal values $j_{fv}$:
\begin{equation}
\label{vertexEPRLs}
A_v \left(j_f, \,  i_e\right) = \sum_{k_{ev}} \left(\prod_{ev} \left(2k_{ev}+1\right) B_{n_{ev}}(j_{fv},j_{fv};i_{ev}, k_{ev})\right) \lbrace 3nj \rbrace _v (j_{fv}, k_{ev}) \ .
\end{equation}
We can also try to give a geometrical interpretation to this model. By removing the sums we fix the areas of the polyhedra on the edges the be fixed to the minimal ones, on the other hand, the shapes (associated to the intertwiners) are still allowed to be boosted from a vertex to the other. This is a dramatic simplification and it is not clear if this model can capture any feature of the full one, nevertheless it is a useful playground to study some properties in a simplified environment. There are some indications that the vertex amplitude of this model is dominated by Euclidean four dimensional geometries \cite{citanoiEPRL}.

Furthermore, notice that with the additional simplification $\left(2k_{ev}+1\right) B_{n_{ev}}(j_f,j_{f};i_e, k_{ev}) \to \delta_{i_e,k_{ev}} $ the vertex amplitude reduces to the one of the BF spin foam model:
\begin{equation}
\label{vertexBF}
A_v \left(j_f, \, i_e\right) = \{3nj\}_v(j_{fv}, i_{ev}) \ .
\end{equation}

In the following, we will study the divergences of these three models starting from the simpler one, BF model, for which the computation of the divergence of any diagram is also possible analytically, moving to the more complex EPRLs and finishing with the physically relevant EPRL.
\section{The diagrams}
\label{sec:diagrams}
In this Section, we will describe the four diagrams we will focus on in the rest of the paper. In spin foam models divergences turn out to be associated with bubbles in the triangulation. A bubble is a collection of faces in the cellular complex forming a closed 2-surface. Here we study the most elementary of such bubbles, and the potential divergences they give rise to, leaving the detailed characterization of all divergences of the whole theory to future works.

\begin{figure}[h!t]
    \centering
    \begin{subfigure}[t]{0.5\textwidth}
        \centering
        \includegraphics[scale=0.5]{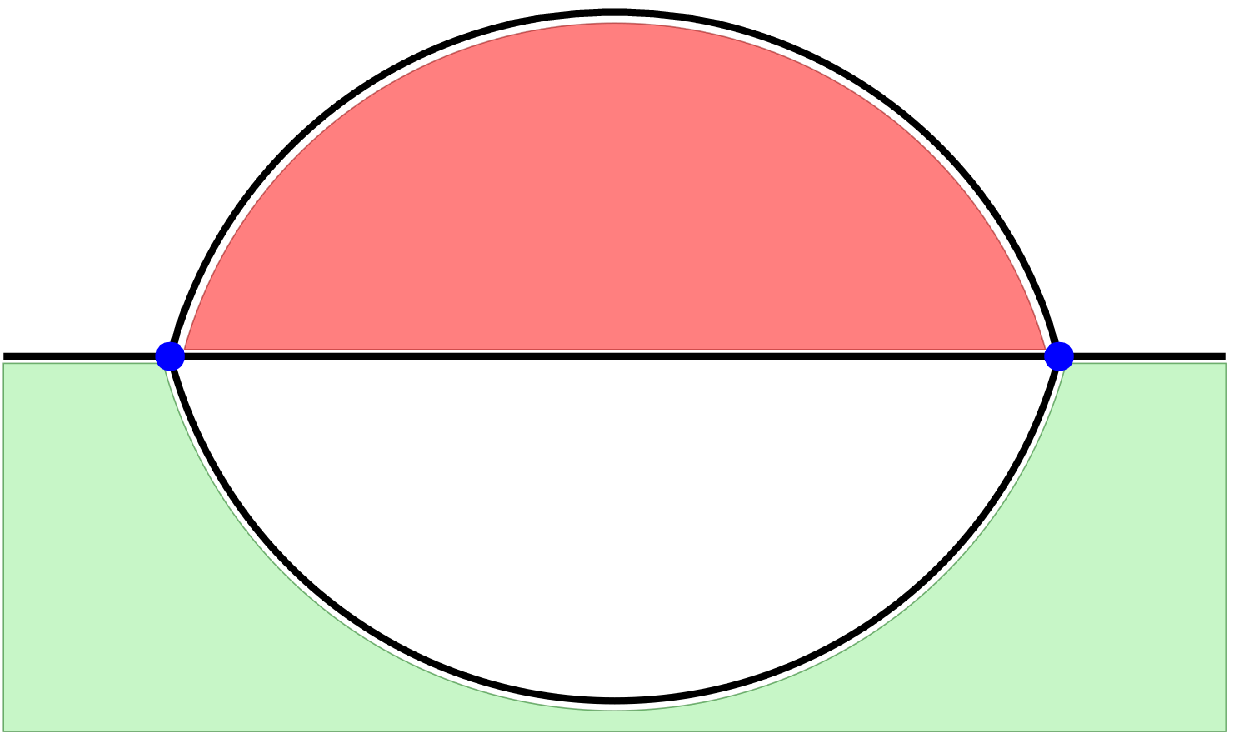}
        \caption{\label{fig:bubble3D}3D bubble diagram}
    \end{subfigure}%
    \begin{subfigure}[t]{0.5\textwidth}
        \centering
        \includegraphics[scale=0.5]{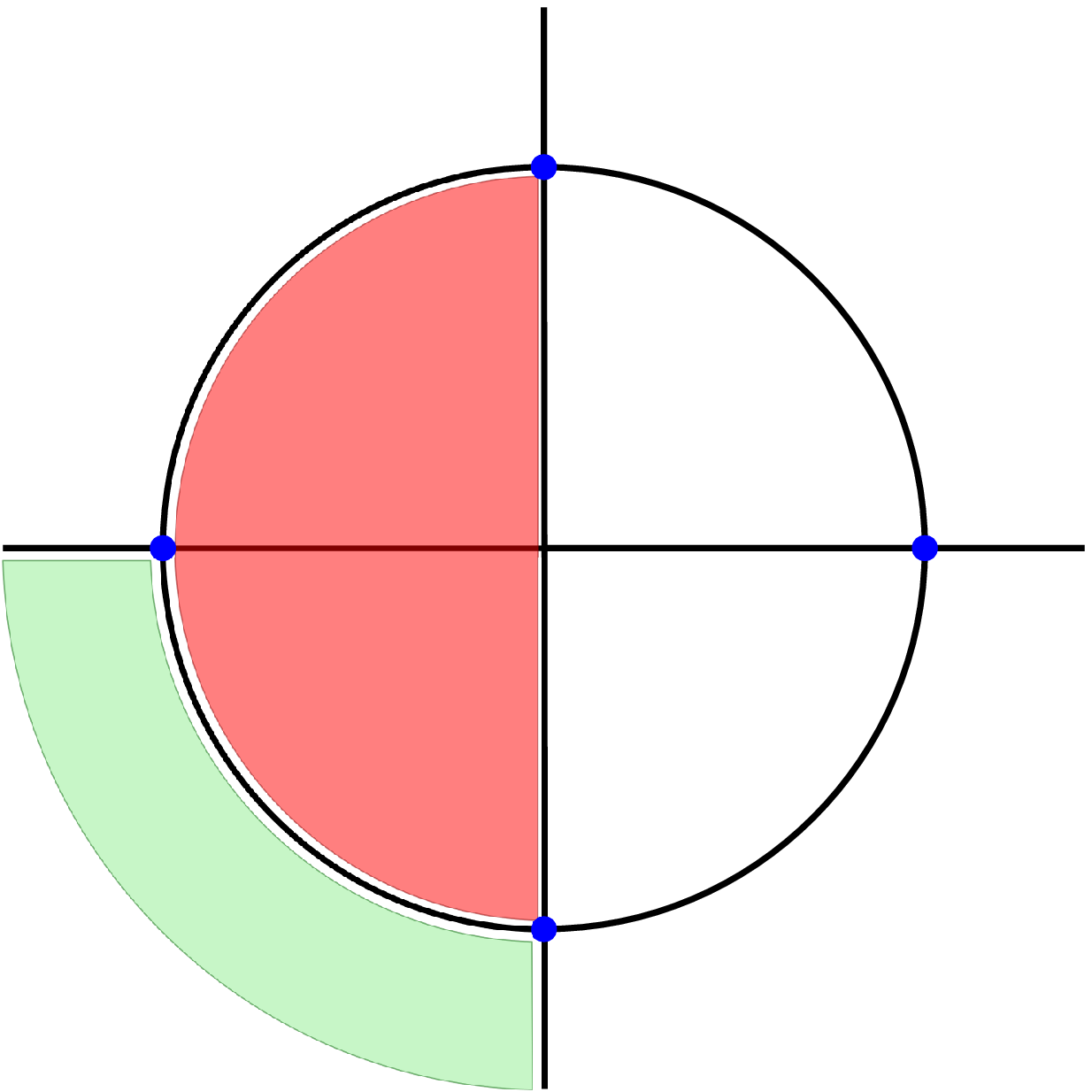}
        \caption{\label{fig:ball3D}3D ball diagram}
    \end{subfigure}
    \begin{subfigure}[t]{0.5\textwidth}
        \centering
        \includegraphics[scale=0.5]{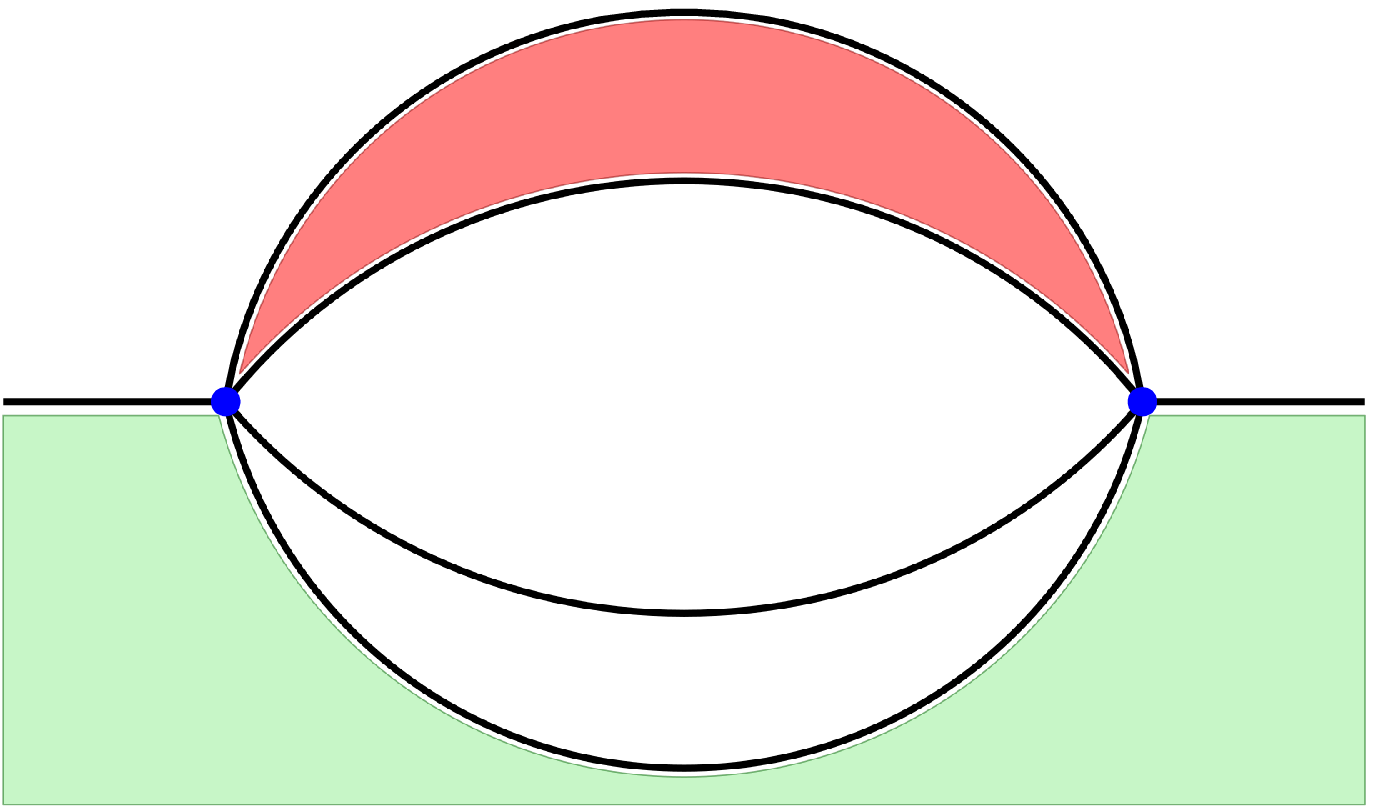}
        \caption{\label{fig:bubble4D}4D bubble diagram}
    \end{subfigure}%
    \begin{subfigure}[t]{0.5\textwidth}
        \centering
        \includegraphics[scale=0.5]{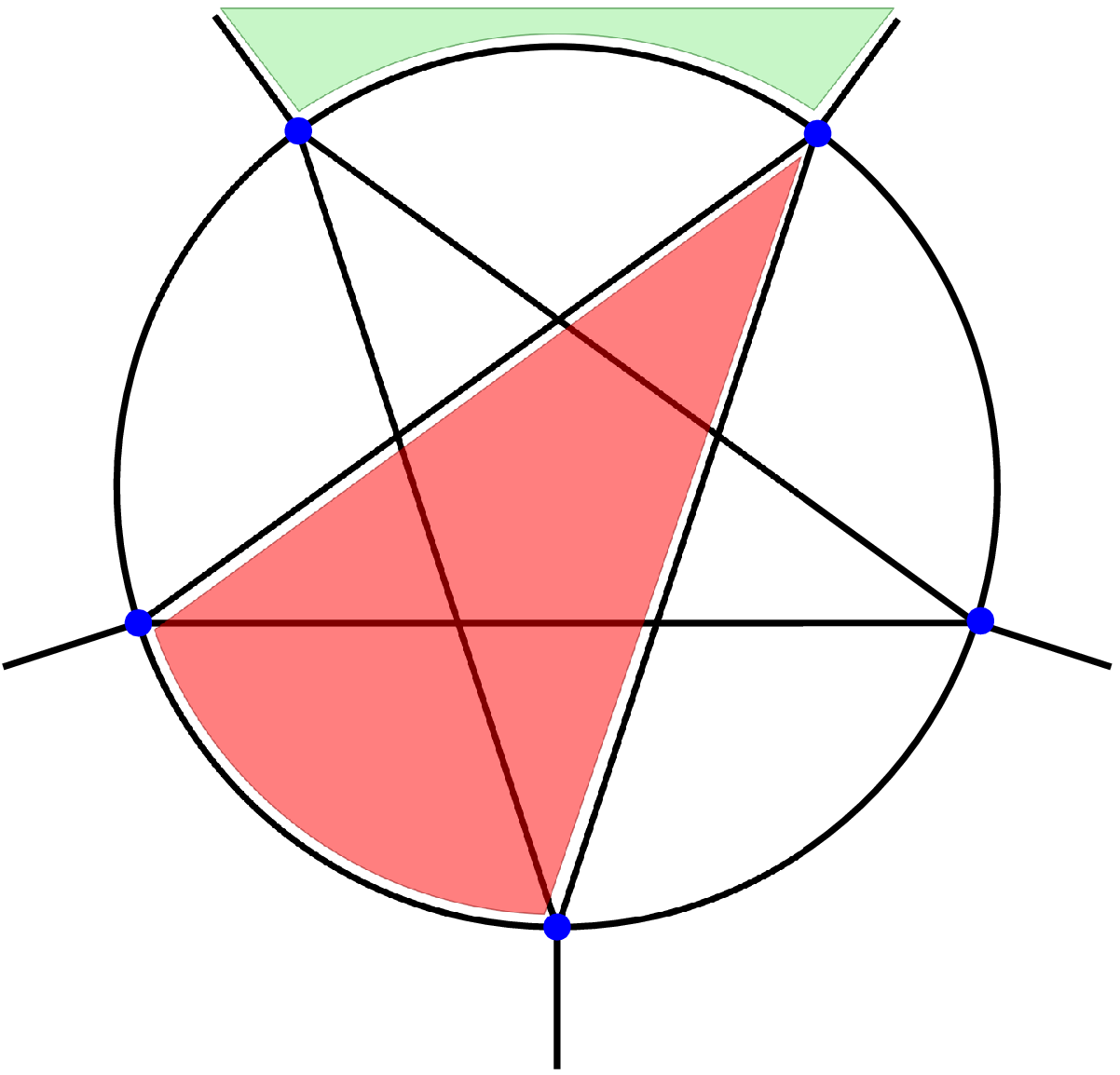}
        \caption{\label{fig:ball4D}4D ball diagram}
    \end{subfigure}
\caption{\label{fig:alldiagrams}We represent here the two-complex of the four diagrams we will study in the paper. The two diagrams on the top have three stranded edges. On the contrary, the diagrams on the bottom have four stranded edges and we will call them four dimensional, each edge is dual to a tetrahedron. We will refer to the diagrams on the left as bubble diagrams and to the diagrams on the right as ball diagrams. In each picture, we highlight in red an internal face and in green an external one.}
\end{figure}

We will focus on two classes of those diagrams represented in Figure \ref{fig:alldiagrams}: the bubble diagram (or to use the Feynman diagrams' language the self-energy), and the ball diagram (or vertex renormalization). The divergence of these two classes of diagrams can be viewed as the divergence on particularly simple
triangulations with boundaries or more in general as the divergence arising from a sub-triangulations of a larger triangulation.

Even if the physical implication of the three stranded diagrams on the top of \eqref{fig:alldiagrams} is not clear, we will look at them as a simpler prototype of the four stranded ones where is easier to test our algorithm and some of the assumptions we will make. We will refer to them as three dimensional because we can imagine the dual to the three stranded edge to be a triangle.

\paragraph{3D bubble diagram.} %%%%%%%%%%%%%%%%%%%%%%%%%%%%%%%%%%
The two-complex associated to the 3D bubble (Figure \ref{fig:bubble3D}) is composed by two vertices, three edges, three internal faces (one per couple of edges) and three external faces (one per edge).
The dual triangulation is formed by two tetrahedra joined by three triangles and its boundary is formed by two triangles joined by all their sides. Therefore, the boundary graph consists of two three valent nodes joined by all their links.

We will in the following use a general convention denoting with $k$s the boundary spins, $j$s the face spins, $t$s the boundary intertwiners and $i$s the edge intertwiners. In this specific case, the boundary graph is completely determined by the three spins of the boundary links $k_a$, $a = 1, \ldots , 3$. One spin is also associated to each internal face $j_{f}$, $f = 1, \ldots , 3$.

\paragraph{3D ball diagram.} %%%%%%%%%%%%%%%%%%%%%%%%%%%%%%%%%%
The two-complex associated to the 3D ball (Figure \ref{fig:ball3D}) is composed by four vertices, six edges, four internal faces (one per triple of vertices) and six external faces (one per internal edge). It can be interpreted as a tetrahedron expanded with a 1-4 Pachner move. The boundary of the dual triangulation is formed by four triangles joined to form a tetrahedron. Therefore, the boundary graph consists of four three-valent nodes joined in a complete graph. We associate a spin $k_a$, where $a = 1, \ldots , 6$, to each link of the boundary graph and a spin $j_{f}$ with $f = 1, \ldots , 4$ to each internal face.

\paragraph{4D bubble diagram.} %%%%%%%%%%%%%%%%%%%%%%%%%%%%%%%%%%
The two-complex associated to the 4D bubble (Figure \ref{fig:bubble4D}) is composed by two vertices, four edges, six internal faces (one per couple of edges) and four external faces (one per edge).
The dual triangulation is formed by two 4-simplices joined by four tetrahedra. The boundary of the dual triangulation is formed by two tetrahedra joined by all their four faces, therefore the boundary graph is formed by two four valent nodes joined by all the links.
Therefore, the boundary graph consists of two four valent node joined by all their links. We denote with $k_a$, where $a = 1, \ldots , 4$ the spins of the boundary graph links and $t_1$ and $t_2$ the intertwiners at the two nodes in the recoupling base $(k_1 ,k_2)$. We attach a spin $j_{f}$ with $f = 1, \ldots , 6$ to each face and an intertwiner $i_e$ with $e = 1, \ldots , 4$ to each edge. 

\paragraph{4D ball diagram.} %%%%%%%%%%%%%%%%%%%%%%%%%%%%%%%%%% 
Finally, the two-complex associated to the 4D ball (Figure \ref{fig:ball4D}) is composed by five vertices, ten edges, ten internal faces (one per triple of vertices) and ten external faces (one per internal edge). It can be interpreted as a 4-simplex expanded with a 1-5 Pachner move into five 4-simplices. Such graph corresponds to a triangulation of a 3-ball with five 4-simplices and its divergence can be associated to the vertex renormalization of a simplicial spinfoam model.
The boundary of the dual triangulation is formed by five tetrahedra joined in a 4-simplex. Therefore, the boundary graph consists of five four-valent nodes connected in a complete graph. 
We denote with $k_a$, where $a = 1, \ldots , 10$ the spins of the boundary graph links and $t_n$ with $n=1, \ldots , 5$ the intertwiners of the five nodes, we will not specify the base choice for the moment. 
We attach a spin $j_{f}$ with $f = 1, \ldots , 10$ to each face and an intertwiner $i_e$ with $e = 1, \ldots , 10$ to each edge.

\section{Divergences estimation in SU(2) BF spin foam model}
\label{sec:modelSU2}
We warm up by testing our techniques with the simplest of the three models we are going to look at: the SU(2) BF spin foam model. For this model is possible to compute any diagram analytically, we refer to Appendix \ref{AppA} for the analytic evaluation of the diagrams considered in this section.
The vertex amplitude \eqref{vertexBF} for three stranded edges spin foams is a $\{6j\}$ symbol while for four stranded edges spin foams is a $\{15j\}$ symbol.
\subsection{3D bubble diagram - self-energy}
\label{sec:bubble3D}
The transition amplitude for the 3D bubble diagram (Figure \ref{fig:bubble3D}) is:
\begin{equation}
\label{amplitudeSE3DBF}
W^{\mathrm{BF}\, \mathrm{3D}}_\mathrm{bubble} = \sum_{j_1, j_2, j_3} \prod_{f=1}^3 \left(2 j_f +1 \right)^\mu \sixj{k_1 & k_2 & k_3}{j_1 & j_2 & j_3}^2 \ .
\end{equation}
Not all the sums are unbounded, to isolate them is useful to make a change of variable: $\lambda_{1}=j_{1}$, $\lambda_{2}=j_{2}-j_{1}$, $\lambda_{3}=j_{3}-j_{1}$. Triangular inequalities implies that the sums over $\left|\lambda_{2}\right|=\left|j_{2}-j_{1}\right|\leq k_{3}$ and $\left|\lambda_{3}\right|=\left|j_{3}-j_{1}\right|\leq k_{2}$ are bounded. We can rewrite \eqref{amplitudeSE3DBF} in terms of these new variables and obtain
\begin{align}
\label{sum6j2}
W^{\mathrm{BF}\, \mathrm{3D}}_\mathrm{bubble} =& \sum_{\lambda_1,\,\lambda_2,\,\lambda_3}  \left(2 \lambda_1 +1 \right)^\mu \left(2 \lambda_1 + 2\lambda_2 +1 \right)^\mu \left(2 \lambda_1+2\lambda_3 +1 \right)^\mu \sixj{k_1 & k_2 & k_3}{\lambda_1 & \lambda_1 + \lambda_2 & \lambda_1 +\lambda_3}^2 \\
& \approx  \sum_{\lambda_1,\,\lambda_2,\,\lambda_3}  \left(2\lambda_1\right)^{3\mu} \sixj{k_1 & k_2 & k_3}{\lambda_1 & \lambda_1 & \lambda_1}^2 \approx \sum_{\lambda_1}  \left(\lambda_1 \right)^{3\mu} \sixj{k_1 & k_2 & k_3}{\lambda_1 & \lambda_1  & \lambda_1}^2 \ .
\end{align}
Our final goal is to study the convergence of the infinite sum over the face spins. With that scope in mind we can assume that $\lambda_1$ is arbitrarily large and drop any contribution small respect to $\lambda_1$. At this stage the summand does not depend anymore on the bounded variables $\lambda_2$ and $\lambda_3$, so we can perform the sum explicitly and then omit the multiplicative factor $8^\mu \left(2 k_2 +1\right)\left(2 k_3 +1\right)$ that is irrelevant for our purposes and cumbersome to keep track of. We use the symbol $\approx$ to indicate this equivalence. The asymptotic behavior of the $\{6j\}$ symbol with $3$ small spins and $3$ large spins is well known \cite{varshalovich_quantum_1988}:
\begin{equation}
\label{semiclassical6j}
\sixj{k_1 & k_2 & k_3}{\lambda_1 & \lambda_1  & \lambda_1} \propto \lambda_1^{-1/2} \ ,
\end{equation}
where we are ignoring an irrelevant multiplicative factor. If we introduce a cutoff $\Lambda$ to the sum over $\lambda_1$ and use the asymptotic expression \eqref{semiclassical6j} we obtain an estimate for the divergence of the amplitude:
\begin{equation}
W^{\mathrm{BF}\, \mathrm{3D}}_\mathrm{bubble}\left(\Lambda \right) \approx \sum_{\lambda_1}^{\Lambda}  \left(\lambda_1 \right)^{3\mu} \left( \lambda_1^{-1/2}\right)^2 \approx \Lambda^{3\mu} \ .
\end{equation}
For a trivial face amplitude $\mu=1$ we reproduce the divergence $ \Lambda^{3}$ we can compute analytically (see Appendix \ref{AppA} for more details).

\subsection{3D ball diagram - vertex renormalization}
\label{sec:ball3D}
By carefully placing the internal and external spins, the transition amplitude for the 3D ball diagram (Figure \ref{fig:ball3D}) is:
\begin{equation}
W^{\mathrm{BF}\, \mathrm{3D}}_\mathrm{ball} = \sum_{\substack{j_1, j_2,\\j_3, j_4}} \left(\prod_{f=1}^4 \left(2 j_f +1 \right)^\mu\right) 
\sixj{k_{1} & k_{2} & k_{3}}{j_{4} & j_{1} & j_{3}}
\sixj{k_{3} & k_{4} & k_{5}}{j_{2} & j_{4} & j_{1}} 
\sixj{k_{2} & k_{5} & k_{6}}{j_{2} & j_{3} & j_{4}}
\sixj{k_{1} & k_{4} & k_{6}}{j_{2} & j_{3} & j_{1}}  
\end{equation}
We follow closely the discussion of Section \ref{sec:bubble3D}, the first step is to identify and isolate the unbounded sums performing the following change of variables $\lambda_{1}=j_{1}$, $\lambda_{2}=j_{2}-j_{1}$, $\lambda_{3}=j_{3}-j_{1}$ and $\lambda_{4}=j_{4}-j_{1}$. Triangular inequalities implies that the sums over the new variables $\left|\lambda_{2}\right|=\left|j_{2}-j_{1}\right|\leq k_{4}$, $\left|\lambda_{3}\right|=\left|j_{3}-j_{1}\right|\leq k_{1}$ and $\left|\lambda_{4}\right|=\left|j_{4}-j_{1}\right|\leq k_{3}$ are bounded, tighter bounds are possible but they are not relevant for our analysis. In terms of this new variables we can rewrite the amplitude as:
\begin{align}
W^{\mathrm{BF}\, \mathrm{3D}}_\mathrm{ball}  = \sum_{\substack{\lambda_1, \lambda_2,\\\lambda_3, \lambda_4}} & \left(2 \lambda_1 + 1 \right)^\mu \left(2 \lambda_1 +2 \lambda_2+ 1 \right)^\mu \left(2 \lambda_1 + 2 \lambda_3+ 1 \right)^\mu \left(2 \lambda_1+2\lambda_4 + 1 \right)^\mu \\
&\sixj{k_{1} & k_{2} & k_{3}}{\lambda_{1}+\lambda_{4} & \lambda_{1} & \lambda_{1}+\lambda_{3}}
\sixj{k_{3} & k_{4} & k_{5}}{\lambda_{1}+\lambda_{2} & \lambda_{1}+\lambda_{4} & \lambda_{1}} \\
&\sixj{k_{2} & k_{5} & k_{6}}{\lambda_{1}+\lambda_{2} & \lambda_{1}+\lambda_{3} & \lambda_{1}+\lambda_{4}}
\sixj{k_{1} & k_{4} & k_{6}}{\lambda_{1}+\lambda_{2} & \lambda_{1}+\lambda_{3} & \lambda_{1}} \ .
\end{align}
Neglecting all the small contributions respect to $\lambda_1$, the variable of the only unbounded sum, and neglecting irrelevant multiplicative factors we obtain:
\begin{align}
\label{amplitudeVert3DBF2}
W^{\mathrm{BF}\, \mathrm{3D}}_\mathrm{ball}  \approx \sum_{\lambda_1} & \left(\lambda_1\right)^{4\mu} \sixj{k_{1} & k_{2} & k_{3}}{\lambda_{1}& \lambda_{1} & \lambda_{1}}
\sixj{k_{3} & k_{4} & k_{5}}{\lambda_{1} & \lambda_{1} & \lambda_{1}}
\sixj{k_{2} & k_{5} & k_{6}}{\lambda_{1}& \lambda_{1} & \lambda_{1}}
\sixj{k_{1} & k_{4} & k_{6}}{\lambda_{1} & \lambda_{1}& \lambda_{1}}  \ .
\end{align}
We put a cutoff on the sum over $\lambda_1$ and we approximate the $\{6j\}$ symbol with its large spin expression \eqref{semiclassical6j} to get the estimate:
\begin{equation}
W^{\mathrm{BF}\, \mathrm{3D}}_\mathrm{ball} \left(\Lambda\right) \approx \sum_{\lambda_1}^{\Lambda}  \left(\lambda_1 \right)^{4\mu} \left(\lambda_1^{-1/2}\right)^4 \approx \Lambda^{4\mu-1} \ .
\end{equation}
Setting a trivial face amplitude ($\mu=1$) our estimate agrees with the analytical computation $\Lambda^3$ (see Appendix \ref{AppA} for more details).

\subsection{4D bubble diagram - self-energy}
\label{sec:bubble4D}
The transition amplitude for the 4D bubble diagram (Figure \ref{fig:bubble4D}) is:
\begin{equation}
\label{amplitudeSE4DBF}
W^{\mathrm{BF}\, \mathrm{4D}}_\mathrm{bubble} = \sum_{j_f, i_e} \prod_{f=1}^6 \left(2 j_f +1 \right)^\mu\prod_{e=1}^4 \left(2 i_e +1 \right)^\mu \left\lbrace 15 j\right\rbrace_1 \left\lbrace 15 j\right\rbrace_2 \ ,
\end{equation}
where the specification of the $\{15j\}$ symbol depends on the choice of intertwiner base of each spin foam edge. Even if the full amplitude is independent of this choice, it is convenient to choose the intertwiner bases that lead to a reducible $\{15j\}$ symbols, as already noted in \cite{dona_asymptotic_2017}, to easily derive the scaling for large spins of the $\{15j\}$ symbol: 
\begin{equation}
\label{15jreduced}
\left\lbrace 15 j\right\rbrace_v
=\left\{ \begin{array}{ccc}
t_{v} & k_{1} & k_{2}\\
j_{1} & i_{2} & i_{1}
\end{array}\right\} \left\{ \begin{array}{ccc}
t_{v} & k_{3} & k_{4}\\
j_{2} & i_{3} & i_{4}
\end{array}\right\} \left\{ \begin{array}{ccc}
i_{1} & i_{2} & t_{v}\\
j_{6} & j_{5} & i_{3}\\
j_{3} & j_{4} & i_{4}
\end{array}\right\} \ .
\end{equation}
Each edge carries a boundary spin, three face spins and an intertwiner. Triangular inequalities constrain the intertwiner to assume values in an interval centered on a face spin, implying that the sums over these intertwiners are bounded. In analogy to the three dimensional case it is useful to perform a change of variables to make it manifest.
We define new variables for the spin faces $\lambda_f = j_f$ for $f=1,\ldots,6$ and for the intertwiners $\iota_1 = i_1 - j_1$, $\iota_2 = i_2 - j_1$, $\iota_3 = i_3 - j_2$, $\iota_4 = i_4 - j_2$. The sums over the intertwiners $\iota_e$ are in fact bounded: $\left|\iota_{1}\right|\leq k_{1}$, $\left|\iota_{2}\right|\leq k_{2}$, $\left|\iota_{3}\right|\leq k_{4}$ and $\left|\iota_{4}\right|\leq k_{3}$. The sums over the $\lambda_f$ are all unbounded contrary to the three dimensional case. The $\left\{ 15j\right\}$ symbol \eqref{15jreduced} can be rewritten in terms of this new variables and the large spins asymptotic can be found in the literature  \cite{varshalovich_quantum_1988,haggard_asymptotics_2010,yu_asymptotic_2011,yu_semiclassical_2011,bonzom_asymptotics_2012}:
\begin{flalign}
\label{semiclassical15j}
\{15 j \}_v &\approx \left\{ \begin{array}{ccc}
t_{v} & k_{1} & k_{2}\\
\lambda_{1} & \lambda_{1}+\iota_{2} & \lambda_{1}+\iota_{1}
\end{array}\right\} \left\{ \begin{array}{ccc}
t_{v} & k_{3} & k_{4}\\
\lambda_{2} & \lambda_{2}+\iota_{3} & \lambda_{2}+\iota_{4}
\end{array}\right\} \left\{ \begin{array}{ccc}
\lambda_{1}+\iota_{1} & \lambda_{1}+\iota_{2} & t_{v}\\
\lambda_{6} & \lambda_{5} & \lambda_{2}+\iota_{3}\\
\lambda_{3} & \lambda_{4} & \lambda_{2}+\iota_{4}
\end{array}\right\} \\
&\approx \frac{1}{\lambda_1}\frac{1}{\lambda_2}\frac{1}{\sqrt{V(\lambda_f)}} \ ,
\end{flalign}
where $V(\lambda_f)$ is the volume of a Euclidean tetrahedron having for sides $\approx \lambda_f$ with $f=1,\ldots,6$. We are ignoring the oscillatory behavior of the ${9j}$ symbol: since the summand is proportional to the square of this oscillation, disruptive interference between terms is not possible and we expect the leading order of the divergence to be unaffected.

Notice that this formula is not valid for values of the spins such that $V = 0$. In these cases, the semiclassical approximation used to derive the asymptotic formula for the ${9j}$ symbol in \eqref{semiclassical15j} needs to be modified \cite{yu_semiclassical_2011}. The set of spins for which this happens form a measure zero set in the bigger set of face spins, so we expect they will not affect the divergence. For this reason, we ignore those points completely in the following analysis.

We can rewrite the whole amplitude in the new variables and expand at the leading order in $\lambda_f$:
\begin{flalign}
W^{\mathrm{BF}\, \mathrm{4D}}_\mathrm{bubble} = & 
\sum_{j_f, i_e} \left(\prod_{f=1}^6 \left(2 \lambda_f +1 \right)^\mu\right)
\left(2 \iota_1 +2 \lambda_1 +1 \right)^\mu\left(2 \iota_{2} +2 \lambda_1 +1 \right)^\mu \left(2 \iota_3 +2 \lambda_2 +1 \right)^\mu\\
&
\hspace*{3.6cm}\left(2 \iota_{4} +2 \lambda_2 +1 \right)^\mu \left\lbrace 15 j\right\rbrace_1 \left\lbrace 15 j\right\rbrace_2 \approx\\
&
\sum_{\lambda_f} \left(\prod_{f=1}^6 \left( \lambda_f\right)^\mu\right) \left(\lambda_1\right)^{2\mu}\left(\lambda_2\right)^{2\mu} \left(\frac{1}{\lambda_1}\frac{1}{\lambda_2}\frac{1}{\sqrt{V(\lambda_f)}}\right)^2 \ .\\
\end{flalign}
To proceed with the estimate we will assume that the only kind of relevant divergence, if any, comes from the radial direction of the sum and will neglect any angular contribution. The divergence of this diagram can be computed analytically and has been extensively studied in the literature \cite{ben_geloun_radiative_2011}, we will use these results to test our assumption. We wanted to stress that this hypothesis is not new: all the other computation of divergences within the EPRL model in the literature also assume it \cite{riello_self-energy_2013,perini_self-energy_2009}.

Calling $\lambda$ this radial coordinate and introducing a factor $\lambda^5$ as measure volume element and a cutoff $\Lambda$:
\begin{equation}
W^{\mathrm{BF}\, \mathrm{4D}}_\mathrm{bubble} \approx \sum_{\lambda}^\Lambda \lambda^5 \lambda^{10\mu} \left(\lambda^{-7/2}\right)^2 \approx \Lambda^{10\mu-1} \ .
\end{equation}

For trivial face amplitude ($\mu=1$) we can compare our estimate with the analytical evaluation (see Appendix \ref{AppA} for more details). We find perfect agreement, this corroborates our hypothesis that the divergence gets contribution mainly from the radial direction of the sum. This assumption will be also used in the estimates of the divergences of amplitudes in the EPRL model where, unfortunately, any alternative computation or checks are not possible.
\subsection{4D ball diagram - vertex renormalization}
\label{sec:ball4D}
The transition amplitude for the 4D ball diagram (Figure \ref{fig:ball4D}) is:
\begin{flalign}
\label{amplitudevertex4DBF}
W^{\mathrm{BF}\, \mathrm{4D}}_\mathrm{ball}=\sum_{j_f, i_e} \prod_{f=1}^{10} \left(2 j_f +1 \right)^\mu\prod_{e=1}^{10} \left(2 i_e +1 \right)^\mu \prod_{v=1}^5 \left\lbrace 15 j\right\rbrace_v
\end{flalign}
We choose the intertwiner basis of the ten edges in order to get the following $ \left\lbrace 15 j\right\rbrace$ symbols: 
\begin{flalign}
\label{tutti15j}
\left\lbrace 15 j\right\rbrace_1
&=\left\{ \begin{array}{ccc}
t_{1} & k_{1} & k_{2}\\
j_{1} & i_{2} & i_{1}
\end{array}\right\} \left\{ \begin{array}{ccc}
t_{1} & k_{4} & k_{3}\\
j_{6} & i_{3} & i_{4}
\end{array}\right\} \left\{ \begin{array}{ccc}
i_{4} & i_{3} & t_{1}\\
j_{5} & j_{4} & i_{2}\\
j_{3} & j_{2} & i_{1}
\end{array}\right\} \ ,\\
\left\lbrace 15 j\right\rbrace_2
&=\left\{ \begin{array}{ccc}
t_{2} & k_{1} & k_{5}\\
j_{1} & i_{5} & i_{1}
\end{array}\right\} \left\{ \begin{array}{ccc}
i_{1} & i_{6} & i_{7}\\
j_{9} & j_{3} & j_{2}
\end{array}\right\} \left\{ \begin{array}{ccc}
k_{7} & k_{6} & t_{2}\\
i_{7} & i_{6} & i_{1}\\
j_{8} & j_{7} & i_{5}
\end{array}\right\} \ ,\\
\left\lbrace 15 j\right\rbrace_3
&=\left\{ \begin{array}{ccc}
t_{3} & k_{5} & k_{2}\\
j_{1} & i_{2} & i_{5}
\end{array}\right\} \left\{ \begin{array}{ccc}
i_{8} & i_{9} & i_{2}\\
j_{5} & j_{4} & j_{10}
\end{array}\right\} \left\{ \begin{array}{ccc}
k_{9} & k_{8} & t_{3}\\
i_{9} & i_{8} & i_{2}\\
j_{8} & j_{7} & i_{5}
\end{array}\right\} \ ,\\
\left\lbrace 15 j\right\rbrace_4
&=\left\{ \begin{array}{ccc}
t_{4} & k_{3} & k_{10}\\
j_{6} & i_{10} & i_{3}
\end{array}\right\} \left\{ \begin{array}{ccc}
t_{4} & k_{8} & k_{6}\\
j_{7} & i_{6} & i_{8}
\end{array}\right\} \left\{ \begin{array}{ccc}
i_{3} & i_{10} & t_{4}\\
j_{5} & j_{10} & i_{8}\\
j_{2} & j_{9} & i_{6}
\end{array}\right\}  \ ,\\
\left\lbrace 15 j\right\rbrace_5
&=\left\{ \begin{array}{ccc}
t_{5} & k_{9} & k_{7}\\
j_{8} & i_{7} & i_{9}
\end{array}\right\} \left\{ \begin{array}{ccc}
t_{5} & k_{10} & k_{4}\\
j_{6} & i_{4} & i_{10}
\end{array}\right\} \left\{ \begin{array}{ccc}
i_{4} & i_{10} & t_{5}\\
j_{5} & j_{10} & i_{9}\\
j_{3} & j_{9} & i_{7}
\end{array}\right\} \ .
\end{flalign}
Analougusly to the analysis performed in the previous section we define new variables for the spin faces $\lambda_f = j_f$ for $f=1,\ldots,10$ and for the edge intertwiners:
\begin{equation*}
\bgroup
\setlength{\arraycolsep}{1.5em} 
\begin{array}{ccccc}
\iota_1 = i_1 -j_1 & \iota_2 = i_2 -j_1 & \iota_3 = i_3 -j_6 & \iota_4 = i_4 -j_6 & \iota_5 = i_5 -j_1 \\
\iota_6 = i_6 -j_7 & \iota_7 = i_7 -j_8 & \iota_8 = i_8 -j_7 & \iota_9 = i_9 -j_8 & \iota_{10} = i_{10} -j_6  
\end{array} 
\egroup
\end{equation*}
%
%$\iota_e = i_e - j_{f_e}$ where $j_{f_e}$ is the face spin in the edge $e$ that couple with the external spin $k_{l_e}$ in the intertwiner base we choose for that edge. 
In terms of these new variables all sums over $\iota_e$ are manifestly bounded,% $\left|\iota_{e}\right|\leq k_{l_e}$,
 while the sums over $\lambda_f$ are all unbounded. Even if the invariants in \eqref{tutti15j} do not have the small spins in the same places their large spin scaling, omitting again the oscillations, are similar: 
\begin{equation}
\begin{array}{lll}
\displaystyle
\left\lbrace 15 j\right\rbrace_1  \approx \frac{1}{\lambda_{1}}\frac{1}{\lambda_{6}}\frac{1}{\sqrt{V(\lambda_{f_{v1}})}}
& \hspace{2cm}
&\displaystyle\left\lbrace 15 j\right\rbrace_2  \approx \frac{1}{\lambda_{1}}\frac{1}{\sqrt{\lambda_{7}}}\frac{1}{\sqrt{\lambda_{8}}}\frac{1}{\sqrt{V(\lambda_{f_{v2}})}}\\ 
\displaystyle\left\lbrace 15 j\right\rbrace_3 \approx \frac{1}{\lambda_{1}}\frac{1}{\sqrt{\lambda_{7}}}\frac{1}{\sqrt{\lambda_{8}}}\frac{1}{\sqrt{V(\lambda_{f_{v3}})}}
&
&\displaystyle\left\lbrace 15 j\right\rbrace_4  \approx \frac{1}{\lambda_{6}}\frac{1}{\lambda_{7}}\frac{1}{\sqrt{V(\lambda_{f_{v4}})}}\\
\displaystyle\left\lbrace 15 j\right\rbrace_5  \approx \frac{1}{\lambda_{6}}\frac{1}{\lambda_{8}}\frac{1}{\sqrt{V(\lambda_{f_{v5}})}}
\end{array} 
\end{equation}
where $\lambda_{f_{vn}}$ are all the face spins entering in the $n$-th vertex, i.e. for the 4-th vertex $f_{v4} = 2, 5, 6, 7, 9, 10$.

We assume also in this case that there is no angular contribution to the divergence and we change to radial coordinates. Imposing a cutoff $\Lambda$ to the radial summation 
\begin{equation}
W^{\mathrm{BF}\, \mathrm{4D}}_\mathrm{bubble} \approx \sum_{\lambda}^\Lambda \lambda^9 \lambda^{20\mu} \left(\lambda^{-7/2}\right)^5 \approx \Lambda^{20\mu-15/2}
\end{equation}

If we set a trivial face amplitude $\mu=1$ we do not reproduce the divergence obtained with analytical methods (see Appendix \ref{AppA} for more details). We stress that all our estimate are upper bounds since we are neglecting any oscillations. Even if we are overestimating the divergence, neglecting the interference between the terms of the sum, we still get a result very close to the analytic evaluation.

\section{Divergences estimation in the simplified EPRL model}
\label{sec:modelEPRLs}
Before trying to estimate divergences in the full EPRL model, it is useful to test our technique on the simpler EPRLs model we introduced at the end of Section \ref{sec:gravityandBF}. The vertex amplitude of the EPRLs model \eqref{vertexEPRLs} differs from the SU(2) BF one in the introduction of the booster functions and in the extra summations over a set of auxiliary ``boosted'' intertwiners per vertex. While the latter requires minimal modification in the logic described in the previous Sections, how to deal with the booster functions will be the main novelty of this Section.

\begin{figure}[h!t]
\centering
    \includegraphics[width=0.495 \textwidth]{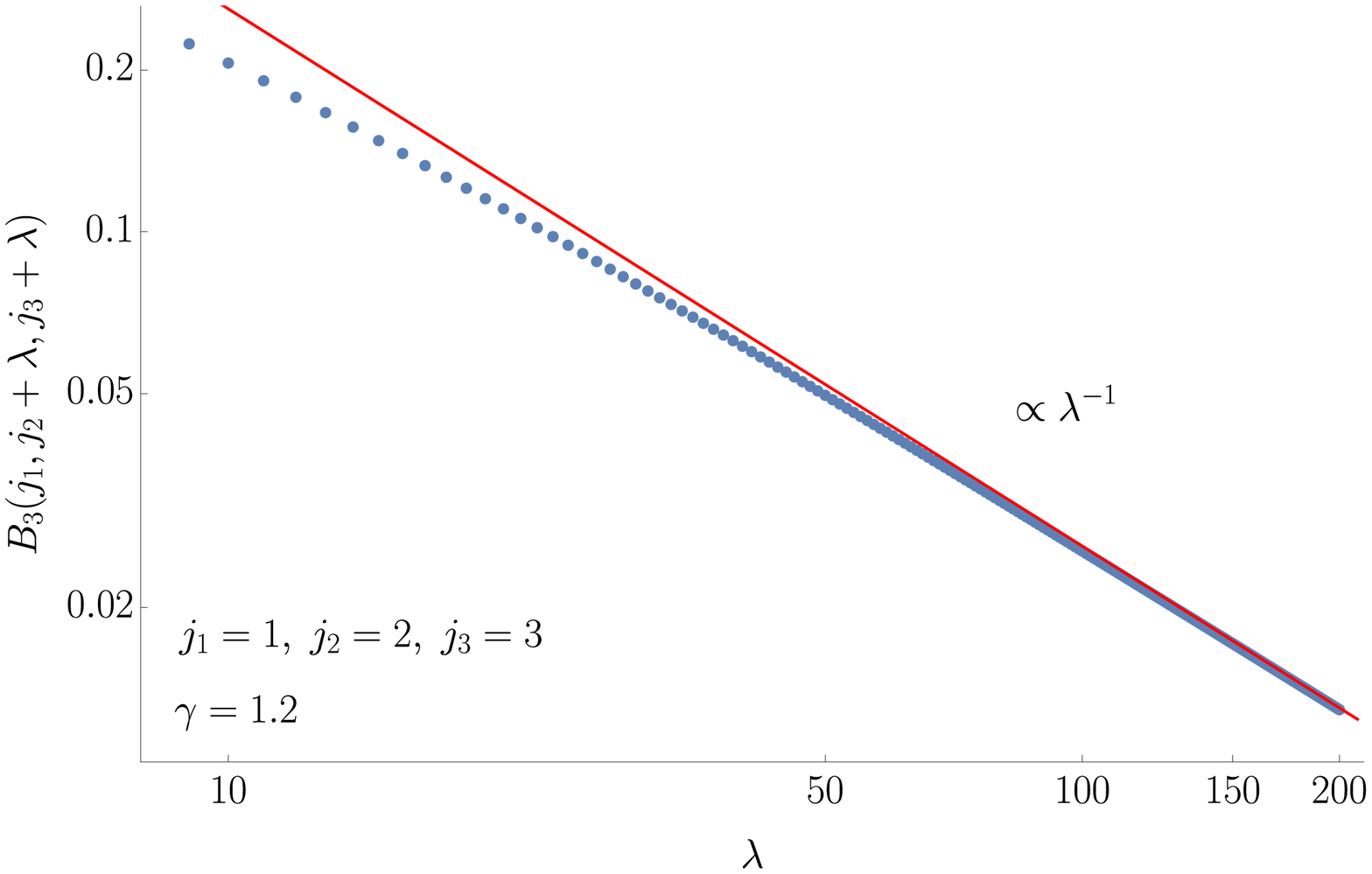}
    \includegraphics[width=0.495 \textwidth]{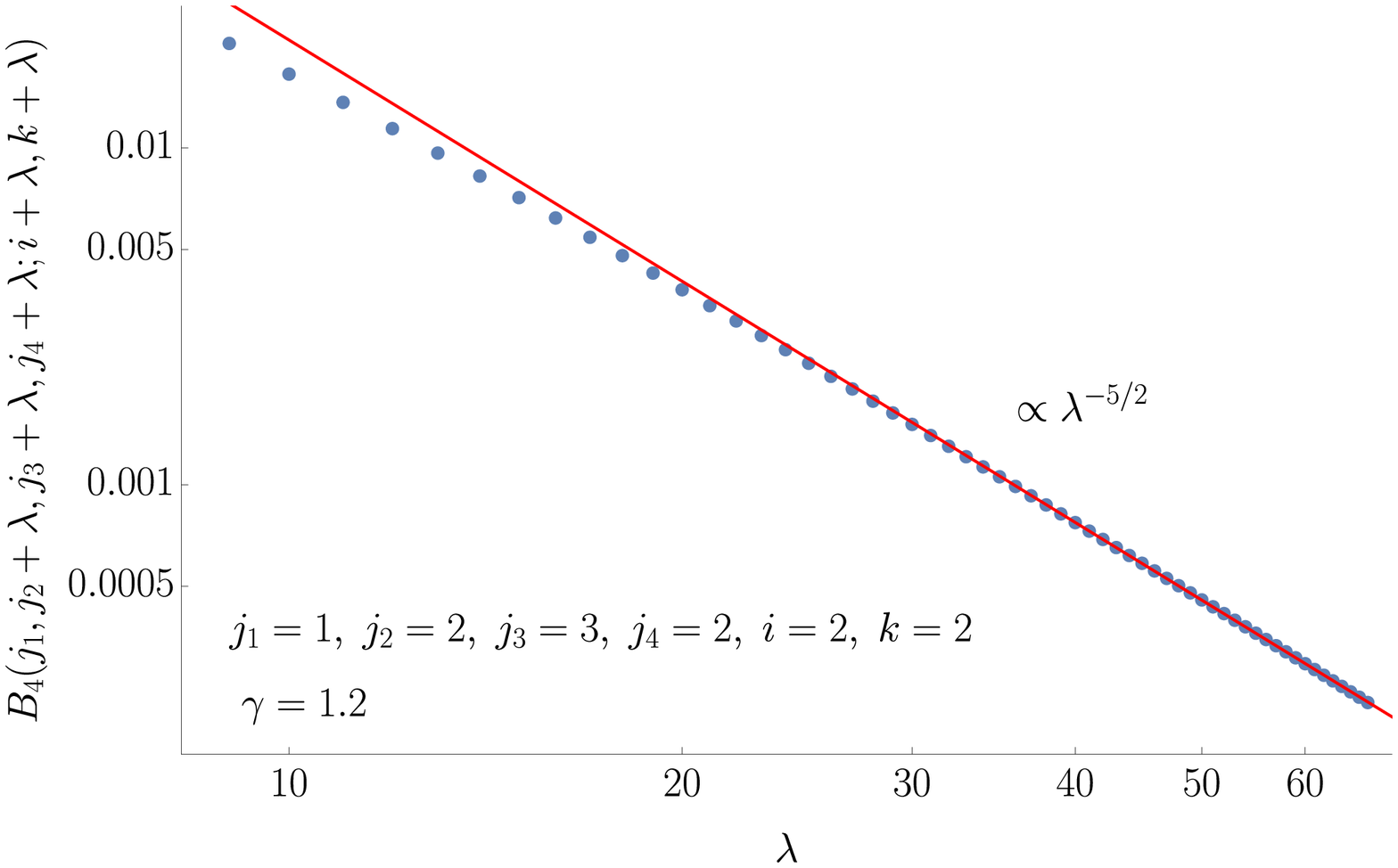}
\caption{\label{fig:Bnscaling} {\small{\emph Numerical scaling of booster functions.} Left panel: {\emph Non-isotropic scaling of the booster function $B_3\left(j_ 1,j_ 2+\lambda,j_ 3+\lambda \right)$ compared with the best fit $f(\lambda) = 2.6 \lambda^{-1}$. We rescaled the booster function by its $\lambda=0$ value.} Right panel: {\emph Non-isotropic scaling of the booster function $B_4\left(j_1,j_2+\lambda,j_3+\lambda , j_4 + \lambda ; i + \lambda , k + \lambda \right)$ compared with the best fit $f(\lambda) = 5.012\ \lambda^{-2.38}$. We rescaled the booster function by its $\lambda=0$ value.} The difference in the range is due to additional resources needed to compute the $B_4$ respect to the $B_3$. We also expect, comparing with the behavior of the $B_3$s, that the proper asymptotic region for the $B_4$ boosters functions is reached for larger spins of the one plotted. To give an idea to the reader while we were able to compute all the points in the left panel on a normal laptop, the plot on the right required 64 cores in a cluster working for approximately 80 hours of walltime.}}
\end{figure}

The main ingredient of the recipe we will describe in the following is the large spins scaling of both the $B_3$ and $B_4$ booster functions, where a spin is kept small and the others become large uniformly. The analytic study of the booster functions is very difficult and it is still work in progress \cite{citaPierre}. This forces us to employ numerical methods to extract the scaling we are looking for. A similar property is already been investigated in \cite{sarno_2-vertex_2018} and we independently confirm it here. We infer from our numerics the following scaling for the booster functions (refer to Figure \ref{fig:Bnscaling} and Appendix \ref{AppB} for more details):
\begin{flalign}
\label{semiclassicalBoosterSIMPL}
&B_3\left(k_1 , j_2 + \lambda , j_3 + \lambda \right) \approx \lambda^{-1} \ ,\\
\label{semiclassicalBoosterSIMPL2}
&B_4\left(k_1 , j_2 + \lambda, j_3 + \lambda, j_4 + \lambda; i+\lambda, i'+\lambda\right) \approx \lambda^{-\frac{5}{2}} \ ,
\end{flalign}
with $\lambda\gg k_1,\ j_2,\ j_3,\ j_4$ and $i$ or $i'$. To keep the expressions compact, we employed, and we will employ in the rest of the paper, a short-hand notation for the booster functions: 
\begin{flalign}
\label{shorthand1}
B_3(j_1,j_2,j_3) &\equiv B_3(j_1,j_2,j_3;j_1,j_2,j_3) \ ,\\
B_4(j_1,j_2,j_3,j_4;i,i') &\equiv B_4(j_1,j_2,j_3,j_4;j_1,j_2,j_3,j_4;i,i') \ .
\end{flalign}

\subsection{3D bubble diagram - self-energy}
The transition amplitude associated to this diagram in the EPRLs model is the following:
\begin{equation}
\label{amplitudeSE3DEPRLs}
W^{\mathrm{EPRLs}\, \mathrm{3D}}_\mathrm{bubble}  = \sum_{j_1,j_2,j_3} \prod_{f=1}^3 \left(2 j_f +1 \right)^\mu \left(\sixj{k_1 & k_2 & k_3}{j_1 & j_2 & j_3} B_3(k_1,j_2,j_3) B_3(j_1,k_2,j_3) B_3(j_1,j_2,k_3)\right)^2 \ .
\end{equation}
We can estimate the divergence of this diagram following the strategy used in Section \ref{sec:bubble3D}. We proceed by performing the same change of variable to isolate the unbounded summations and we drop all the irrelevant multiplicative terms:
\begin{align}
\label{sum6j2b}
W^{\mathrm{EPRLs}\, \mathrm{3D}}_\mathrm{bubble}  & \approx \sum_{\lambda_1}  \left(\lambda_1 \right)^{3\mu} \left(\sixj{k_1 & k_2 & k_3}{\lambda_1 & \lambda_1  & \lambda_1}B_3(k_1,\lambda_1,\lambda_1) B_3(\lambda_1,k_2,\lambda_1) B_3(\lambda_1,\lambda_1,k_3)\right)^2 \ .
\end{align}
We introduce a cutoff $\Lambda$ in the unbounded sum over $\lambda_1$ and we approximate the summand with its asymptotic behavior obtained combining the large spin scaling of the $\{ 6j \}$ symbol \eqref{semiclassical6j} and of the booster functions \eqref{semiclassicalBoosterSIMPL}:
\begin{equation}
W^{\mathrm{EPRLs}\, \mathrm{3D}}_\mathrm{bubble}\left(\Lambda \right) \approx \sum_{\lambda_1}^{\Lambda}  \left(\lambda_1 \right)^{3\mu} \left(\lambda_1^{-1/2}\, \lambda_1^{-3} \right)^2 \approx \Lambda^{3\mu-6} \ .
\end{equation}

Notice that for the standard choice of face weight $\mu=1$ the amplitude is convergent, where for the SU(2) BF model it was cubically divergent.

\begin{figure}[h!t]
\centering
    \includegraphics[width=0.5 \textwidth]{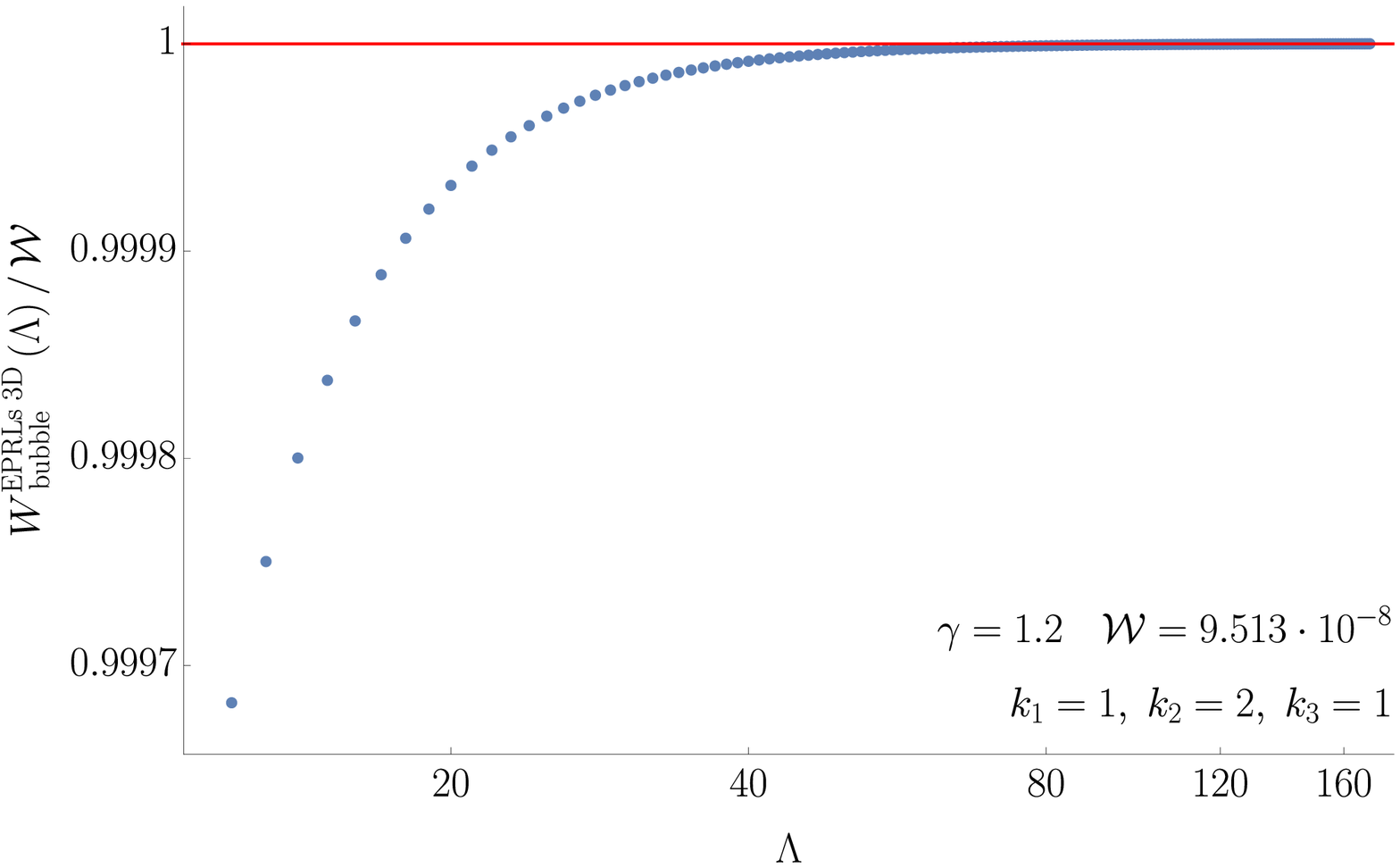}\includegraphics[width=0.5 \textwidth]{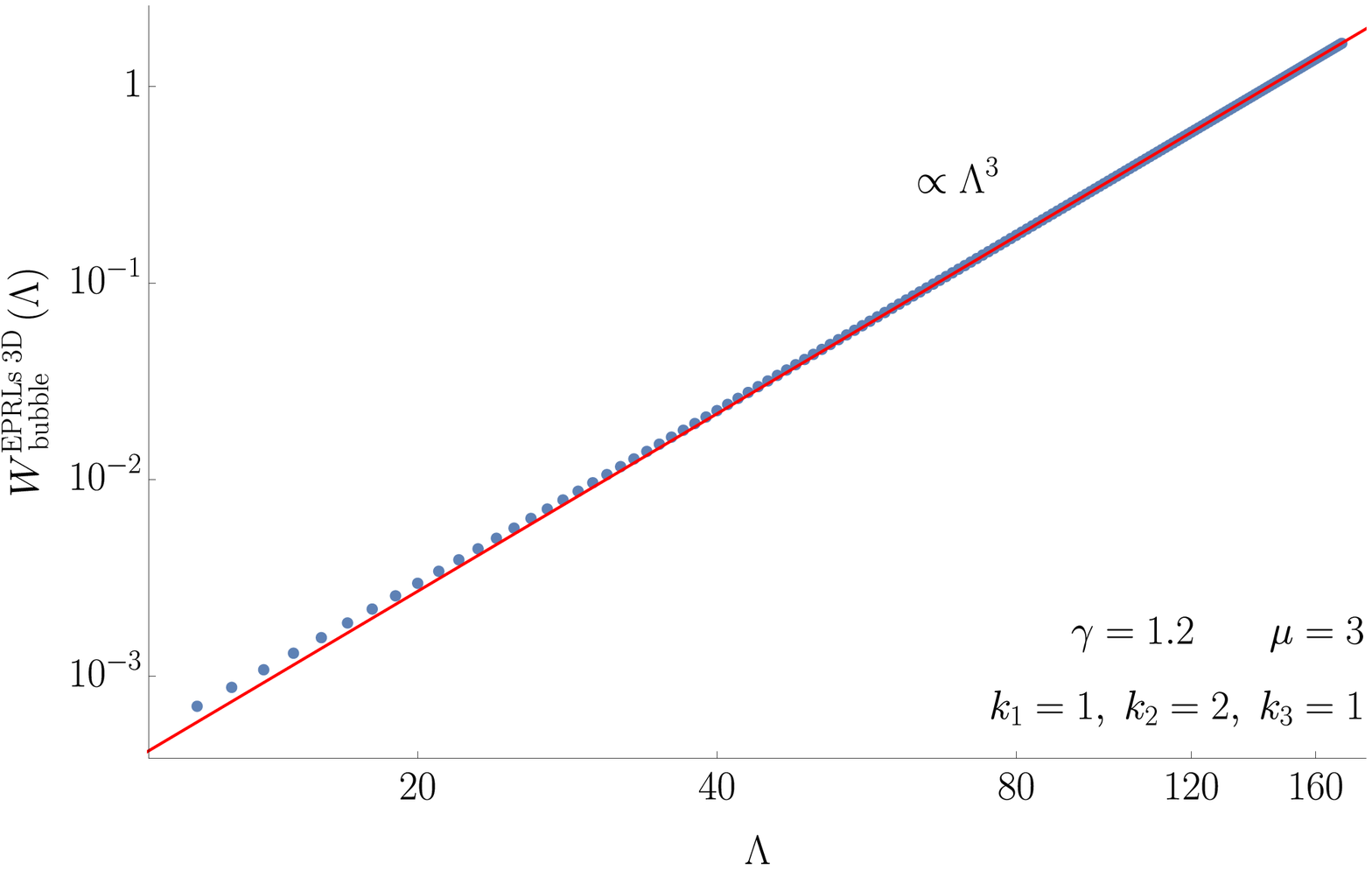}
\caption{\label{fig:SE3DEPRLs_data} {\small \emph{Numerical evaluation of the transition amplitude \eqref{amplitudeSE3DEPRLs} as a function of the cutoff in logarithmic scale. We choose the external spins to be $k_1 = 1$, $k_2 = 2$, $k_3 = 3$, Immirzi parameter $\gamma = 1.2$.} Left panel: \emph{for face weight $\mu = 1$ the amplitude is convergent to the best fit $\mathcal{W}=9.513 \cdot 10^{-8}$ in red. The plot is rescaled to allow a clearer reading.}
Right panel: \emph{for face weight $\mu = 3$, the amplitude diverge cubically. We plot for comparison the best fit function $3.385 \cdot 10^{-7}\, \Lambda^3$ in red.}}}
\end{figure}

We do not have in this case an analytical computation to compare to, but the system is simple enough to allow us to evaluate numerically the amplitude \eqref{amplitudeSE3DEPRLs} as a function of the cutoff $\Lambda$. We show the numerical result in Figure \ref{fig:SE3DEPRLs_data}, we see a remarkable agreement with our estimate. To have a better comparison we artificially make the amplitude divergent by setting $\mu=3$.

One can wonder where and if there is any dependence in the Immirzi parameter. Our analysis is not sensitive to it since it mainly focuses on the power of the cutoff. It will for sure play a role in the multiplicative factor that we ignored.

\subsection{3D ball diagram - vertex renormalization}
\label{sec:ball3DEPRLs}
\begin{figure}[h!t]
\centering
    \includegraphics[width=0.5 \textwidth]{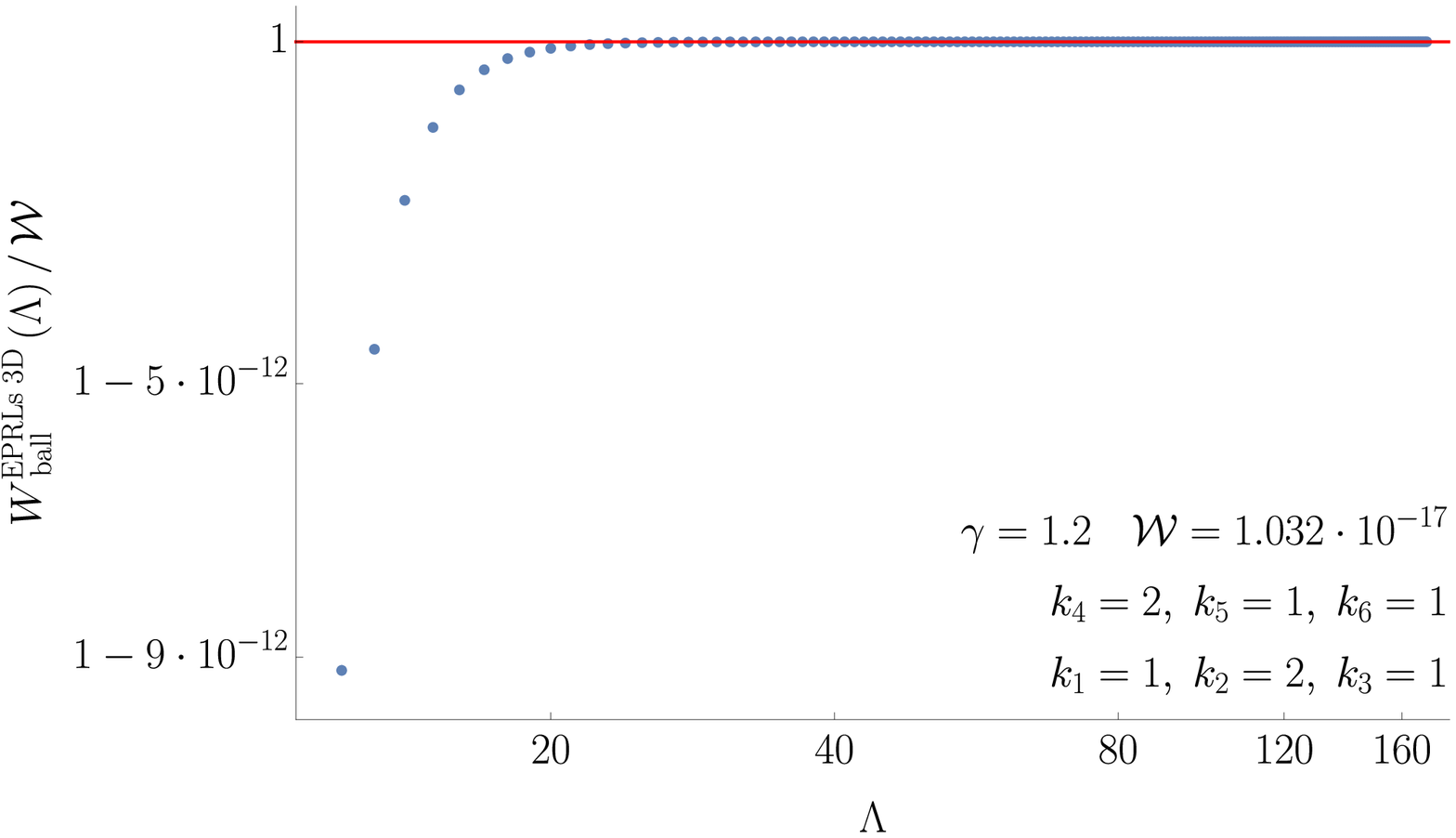}\includegraphics[width=0.5 \textwidth]{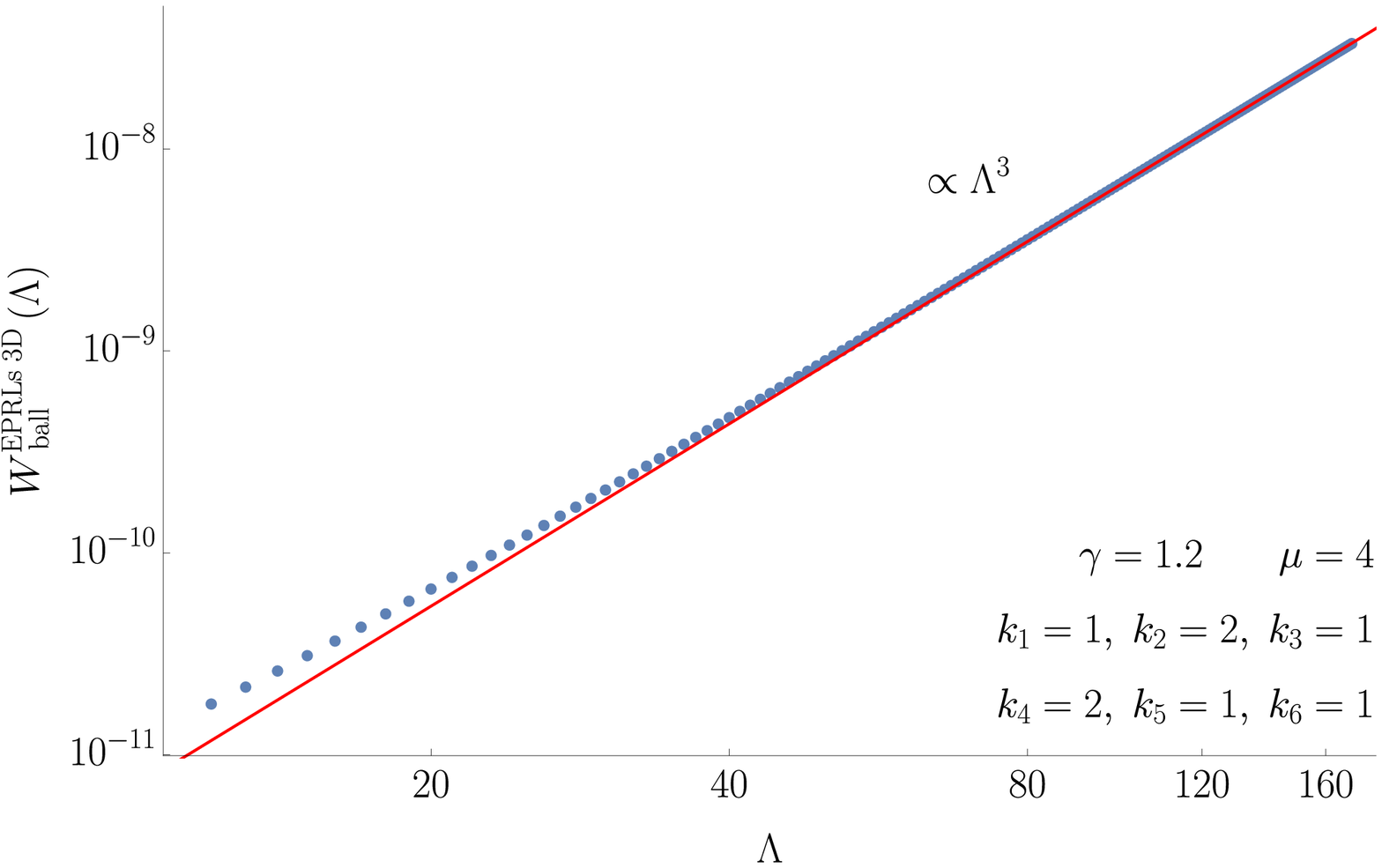}
\caption{\label{fig:vertex3DEPRLs_data} {\small \emph{Numerical evaluation of the transition amplitude \eqref{amplitudevert3DEPRLs} as a function of the cutoff in logarithmic scale. We choose the external spins to be $k_1 = 1$, $k_2 = 2$, $k_3 = 1$, $k_4 = 2$, $k_5 = 1$, $k_6 = 1$ and the Immirzi parameter is set to $\gamma = 1.2$.} Left panel: \emph{for face weight $\mu = 1$ the amplitude is convergent to the best fit $\mathcal{W}=1.032 \cdot 10^{-17}$ in red. The plot is rescaled to allow a clearer reading.}
Right panel: \emph{for face weight $\mu = 4$ the amplitude diverge cubically. We plot for comparison the best fit function $6.811 \cdot 10^{-15}\, \Lambda^3$ in red.}}}
\end{figure}

The transition amplitude associated to the 3D ball diagram in the EPRLs model is the following:
\begin{equation}
\label{amplitudevert3DEPRLs}
W^{\mathrm{EPRLs}\, \mathrm{3D}}_\mathrm{ball} = \sum_{\substack{j_1, j_2,\\j_3, j_4}} \left(\prod_{f=1}^4 \left(2 j_f +1 \right)^\mu\right) \prod_{v=1}^4 A_v
\end{equation}
where 
\begin{flalign*}
A_1=&\sixj{k_{1} & k_{2} & k_{3}}{j_{4} & j_{1} & j_{3}} B_3(k_1,j_1,j_3)B_3(j_4,k_2,j_3)B_3(j_4,j_1,k_3) \ , \\
A_2=&\sixj{k_{3} & k_{4} & k_{5}}{j_{2} & j_{4} & j_{1}} B_3(k_3,j_4,j_1)B_3(j_2,k_4,j_1)B_3(j_2,j_4,k_5) \ , \\
A_3=&\sixj{k_{2} & k_{5} & k_{6}}{j_{2} & j_{3} & j_{4}} B_3(k_2,j_3,j_4)B_3(j_2,k_5,j_4)B_3(j_2,j_3,k_6) \ , \\
A_4=&\sixj{k_{1} & k_{4} & k_{6}}{j_{2} & j_{3} & j_{1}} B_3(k_1,j_3,j_1)B_3(j_2,k_4,j_1)B_3(j_2,j_3,k_6) \ .
\end{flalign*}
We proceed by performing the same change of variable of Section \ref{sec:ball3D} to isolate the unbounded summations and we drop all the irrelevant multiplicative terms: 
\begin{flalign*}
\label{amplitudevert3DEPRLs2}
A_1\approx &
\sixj{k_{1} & k_{2} & k_{3}}{\lambda_{1}& \lambda_{1} & \lambda_{1}} B_3(k_1,\lambda_1,\lambda_1)B_3(\lambda_1, k_2,\lambda_1)B_3(\lambda_1,\lambda_1, k_3) \ , \\
A_2\approx&\sixj{k_{3} & k_{4} & k_{5}}{\lambda_{1} & \lambda_{1} & \lambda_{1}} B_3(k_3,\lambda_1,\lambda_1)B_3(\lambda_1, k_4,\lambda_1)B_3(\lambda_1, \lambda_1, k_5) \ , \\
A_3\approx&\sixj{k_{2} & k_{5} & k_{6}}{\lambda_{1}& \lambda_{1} & \lambda_{1}} B_3(k_2,\lambda_1, \lambda_1)B_3(\lambda_1, k_5 \lambda_1)B_3(\lambda_1, \lambda_1, k_6)\ ,  \\
A_4\approx&\sixj{k_{1} & k_{4} & k_{6}}{\lambda_{1} & \lambda_{1}& \lambda_{1}} B_3(k_1,\lambda_1, \lambda_1)B_3(\lambda_1, k_4,\lambda_1)B_3(\lambda_1, \lambda_1, k_6) \ .
\end{flalign*}
As we did in the previous section we introduce a cutoff $\Lambda$ in the unbounded sum over $\lambda_1$ and we approximate the summand with its asymptotic behavior obtained combining the large spin scaling of the $\left\lbrace 6j \right\rbrace$ symbol \eqref{semiclassical6j} and of the booster functions \eqref{semiclassicalBoosterSIMPL}:
\begin{equation}
W^{\mathrm{EPRLs}\, \mathrm{3D}}_\mathrm{ball}\left(\Lambda\right) \approx \sum_{\lambda_1}^{\Lambda}  \left(\lambda_1 \right)^{4\mu}  \left( \lambda_1^{-1/2}\, \lambda_1^{-3} \right)^4 \approx \Lambda^{4\mu-13}
\end{equation}
The amplitude is convergent for the standard choice of face weight $\mu=1$ while is cubically divergent for $\mu=4$. The amplitude \eqref{amplitudevert3DEPRLs} is also simple enough to allow us to evaluate it exactly as a function of the cutoff $\Lambda$. The results are shown in Figure \ref{fig:vertex3DEPRLs_data} and we see an excellent agreement with our estimate for both $\mu=1$ and $\mu=4$.

\subsection{4D bubble diagram - self-energy}
The transition amplitude associated to the 4D bubble diagram (Figure \ref{fig:ball4D}) in the EPRLs model is:
\begin{equation}
\label{amplitudeSE4DEPRLs}
W^{\mathrm{EPRLs}\, \mathrm{4D}}_\mathrm{bubble} = \sum_{j_f, i_e}\prod_{f=1}^6 \left(2 j_f +1 \right)^\mu\prod_{e=1}^4 \left(2 i_e +1 \right)^\mu A_1 \cdot A_2
\end{equation}
where 
\begin{flalign*}
A_v =\sum_{i_e^{(v)}} \left(\prod_{e=1}^4 (2 i_e^{(v)} + 1) \right)\left\lbrace 15 j\right\rbrace_v & B_4(k_1, j_1,j_3,j_6; i_1, i_1^{(v)})B_4(k_2, j_1,j_4,j_5; i_2, i_2^{(v)})\\
&B_4(k_4, j_2,j_5,j_6; i_3, i_3^{(v)})B_4(k_3, j_2,j_3,j_4; i_4, i_4^{(v)}) \ .
\end{flalign*}
where the $\{15j\}_v$ symbols are the one defined in \eqref{15jreduced} with the substitution $i_e \to i_e^{(v)}$. Once again, we perform the same change of variable of Section \ref{sec:bubble4D} to isolate the unbounded summations. The main difference is that we have to deal with two additional summations over two sets of intertwiners $i_e^{(1)}$ and $i_e^{(2)}$, with that purpose we define some $\iota_e^{(v)}$ such that 
\begin{equation*}
\bgroup
\setlength{\arraycolsep}{1.5em} 
\begin{array}{ccccc}
& \iota_1^{(v)} = i_1^{(v)} - j_1 \ , & \iota^{(v)}_2 = i^{(v)}_2 - j_1 \ , & \iota^{(v)}_3 = i^{(v)}_3 - j_2 \ , & \iota^{(v)}_4 = i^{(v)}_4 - j_2 \ ,\\ 
\end{array}
\egroup
\end{equation*}
for each vertex $v$. The booster functions are nonvanishing only if the auxiliary intertwiners satisfy the same triangular inequality as the normal ones. As a direct consequence the summations over both $\iota^{(v)}$ variables are bounded by a boundary spin. Expanding at the leading order in  $\lambda_f$ and dropping all the irrelevant multiplicative terms, the vertex amplitudes read:
\begin{flalign}
\label{vamplEPRLs4D}
A_v \approx \lambda_1^2 \lambda_2^2 \{ 15 j\} \left(\lambda_f\right) \ 
& B_4(k_1, \lambda_1,\lambda_3,\lambda_6; \lambda_1, \lambda_1) B_4(k_2, \lambda_1,\lambda_4,\lambda_5; \lambda_1, \lambda_1)\\
& B_4(k_4, \lambda_2,\lambda_5,\lambda_6; \lambda_2, \lambda_2) B_4(k_3, \lambda_2,\lambda_3,\lambda_4; \lambda_2, \lambda_2) \ , \nonumber
\end{flalign}
where  $\left\lbrace 15 j\right\rbrace \left(\lambda_f\right)$ is the same \eqref{semiclassical15j} where all the $\iota_e^{(v)}$ variables have been ignored since they are small respect to the $\lambda_f$.
We introduce a radial coordinate $\lambda$ in the $\lambda_f$ sum and we assume that there is no contribution to the divergence coming from the angular summation. In terms of the radial coordinate the vertex amplitudes $A_1$ and $A_2$ become:
\begin{flalign*}
A_v \approx \lambda^4 \left\lbrace 15 j\right\rbrace \left(\lambda\right) \
& B_4(k_1, \lambda,\lambda,\lambda; \lambda, \lambda) B_4(k_2, \lambda,\lambda,\lambda; \lambda, \lambda) \\
& B_4(k_4, \lambda,\lambda,\lambda; \lambda, \lambda)  B_4(k_3,\lambda,\lambda,\lambda; \lambda, \lambda) \ ,
\end{flalign*}
We substitute to the $\left\lbrace 15j \right\rbrace$ symbol and to the boosters functions their asymptotic expressions \eqref{semiclassical15j} and \eqref{semiclassicalBoosterSIMPL2}.
\begin{flalign*}
A_v \approx \lambda^4 \lambda^{-\frac{7}{2}} \left(\lambda^{-\frac{5}{2}}\right)^4
\end{flalign*}
We introduce a factor $\lambda^5$ as volume element and we put a cutoff $\Lambda$, the amplitude \eqref{amplitudeSE4DEPRLs} reads:
\begin{flalign}
W^{\mathrm{EPRLs}\, \mathrm{4D}}_\mathrm{bubble}\left(\Lambda\right) \approx \sum_{\lambda}^\Lambda \lambda^5 \lambda^{10\mu} \left(\lambda^4 \lambda^{-\frac{7}{2}} \left(\lambda^{-\frac{5}{2}}\right)^4 \right)^2 \approx \Lambda^{10\mu-13}
\end{flalign}
We notice that for trivial face weight $\mu=1$ the amplitude result convergent. At present time there are no analytical or numerical checks to verify this estimate. We are not aware of any code or technique able to compute the booster functions and the sum over the six faces fast enough to be able to compute \eqref{amplitudeSE4DEPRLs} exactly in a reasonable amount of computational time. A lot of work is being done in this direction at the moment \cite{citaFrancois, citanoiEPRL}: we believe we will be able to evaluate numerically this amplitude in a not so distant future.

\subsection{4D ball diagram - vertex renormalization}
The transition amplitude for the 4D ball diagram (Figure \ref{fig:ball4D}) in the EPRLs model is: 
\begin{flalign}
\label{amplitudevert4DEPRLs}
W^{\mathrm{BF}\, \mathrm{4D}}_\mathrm{ball}=\sum_{j_f, i_e} \prod_{f=1}^{10} \left(2 j_f +1 \right)^\mu\prod_{e=1}^{10} \left(2 i_e +1 \right)^\mu \prod_{v=1}^5 A_v \ ,
\end{flalign}
where we used the same intertwiner basis of Section \eqref{sec:ball4D}. To not distract the reader we will focus exclusively on just the first vertex amplitude, we treat the others in an analogous way, in the end, they will contribute in the same way to the divergence, and we write them explicitly in Appendix \ref{AppC1}:
\begin{flalign}
\label{vertexamplitudeBallEPRLS}
A_1 =\sum_{i^{(1)}_{ev}} \left(\prod_{ev} (2 i_{ev}^{(1)} + 1) \right) \left\lbrace 15 j\right\rbrace_1 \  &B_4(k_1, j_{1} ,j_{3},j_{2}; i_{1}, i^{(1)}_{1})B_4(k_2, j_{1} ,j_{5},j_{4}; i_{2}, i^{(1)}_{2})\\[-1.5em]
&B_4(k_3, j_{6} ,j_{2},j_{4}; i_{3}, i^{(1)}_{3})B_4(k_4, j_{6} ,j_{3},j_{5}; i_{4}, i^{(1)}_{4})\ . \nonumber
\end{flalign}
We denoted with $\left\lbrace 15 j\right\rbrace_v$ the same symbols defined in \eqref{tutti15j} with the substitution $i_e \to i_e^{(v)}$. The summation over the auxiliary intertwiners $i_{ev}^{(v)}$, a set of four per vertex $(v)$, is carried over the edges connected to the vertex $v$ (i.e. in the 1-st vertex $e1=1,2,3,4$).
To manifestly identify the bounded sums and unbounded sums we make the same change of variables on $j_f$ and $i_e$ of Section \ref{sec:ball4D} and in addition
\begin{equation*}
\bgroup
\setlength{\arraycolsep}{1.5em} 
\begin{array}{ccccc}
A_1:& \iota^{(1)}_1 = i^{(1)}_1 -j_1 \ , & \iota^{(1)}_2 = i^{(1)}_2 -j_1 \ , & \iota^{(1)}_3 = i^{(1)}_3 -j_6 \ , & \iota^{(1)}_4 = i^{(1)}_4 -j_6 \ . \\
\end{array} 
\egroup
\end{equation*}
In terms of these new variables all sums over $\iota_e$ and $\iota^{(v)}_e$ are bounded, while the sums over $\lambda_f$ are all unbounded. Expanding at the leading order in  $\lambda_f$, the vertex amplitudes $A_1$, $\ldots$, $A_5$ are recasted in the following form:
\begin{flalign*}
A_1 \approx \lambda_1^2\lambda_6^2 \left\lbrace 15 j\right\rbrace_1 \ &
B_4(k_1, \lambda_{1} ,\lambda_{3},\lambda_{2}; \lambda_{1}, \lambda_{1}) B_4(k_2, \lambda_{1} ,\lambda_{5},\lambda_{4}; \lambda_{1}, \lambda_{1})\\
& B_4(k_3, \lambda_{6} ,\lambda_{2},\lambda_{4}; \lambda_6, \lambda_6)B_4(k_4, \lambda_{6} ,\lambda_{3},\lambda_{5}; \lambda_6, \lambda_6)\ .
\end{flalign*}
We introduce a radial coordinate $\lambda$ in the $\lambda_f$ sum and assume that there is no contribution to the divergence coming from the angular summation in the $\lambda_f$ space. If we substitute to the $\left\lbrace 15j\right\rbrace$ symbol and to the boosters functions their asymptotic expressions \eqref{semiclassical15j} and \eqref{semiclassicalBoosterSIMPL2} each vertex amplitude gives the same contribution. Introducing a factor $\lambda^5$ as volume element and a cutoff $\Lambda$ in the radial sum, the amplitude \eqref{amplitudevert4DEPRLs} reads:

\begin{flalign}
W^{\mathrm{EPRLs}\, \mathrm{4D}}_\mathrm{ball}\left(\Lambda\right) \approx \sum_{\lambda}^\Lambda \lambda^9 \lambda^{10\mu} \lambda^{10\mu}  \left(\lambda^4 \lambda^{-\frac{7}{2}} \left(\lambda^{-\frac{5}{2}}\right)^4 \right)^5 \approx \Lambda^{20\mu-\frac{75}{2}} \ .
\end{flalign}
For trivial face weight $\mu=1$ the amplitude is convergent. Similarly to the 4D bubble, we hope to be able to numerically check this result soon.

\section{Divergences estimation in the full EPRL model}
\label{sec:modelEPRL}
Finally, in this section, we will compute the divergence of the transition amplitudes of the four diagrams in Figure \ref{fig:alldiagrams} in the full EPRL model. The additional complication in the EPRL vertex amplitude compared to the EPRLs vertex amplitude is the presence of additional sums over the auxiliary spins $l_{fv}$, one per face including the vertex $v$ in consideration. The way we will deal with these additional sums will be explained in details in the various examples. 

From now on we will write the $l_{fv}$ variables in the vertex amplitude \eqref{vertexEPRL}, taking values from $j_{fv}$ to infinity, as $l_{fv}= j_f + \Delta l_{fv}$ where $\Delta l_{fv}$ takes values from $0$ to infinity.

\begin{figure}[h!t]
\centering
    \includegraphics[width=0.495 \textwidth]{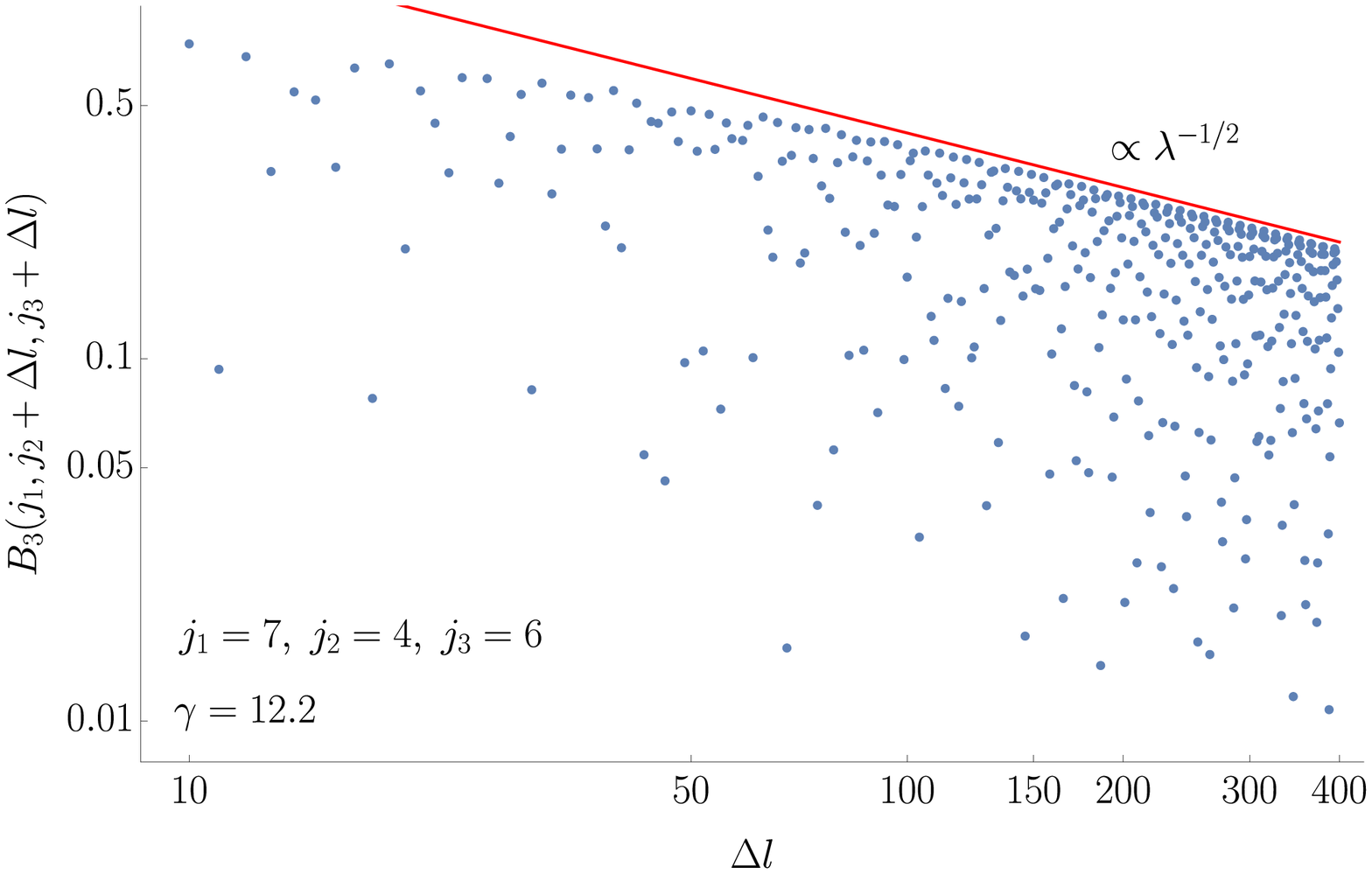}
    \includegraphics[width=0.495 \textwidth]{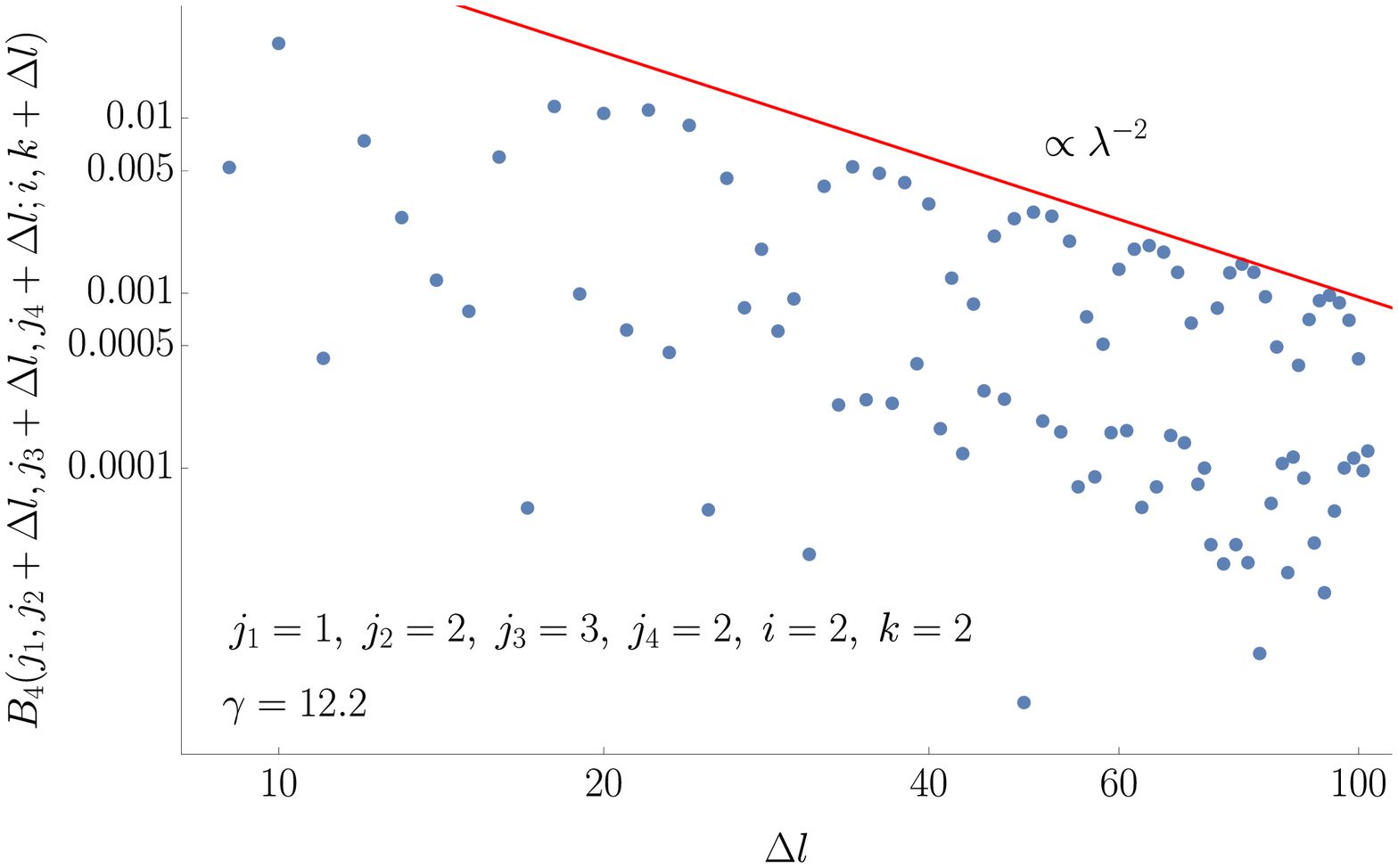}
\caption{\label{fig:BnLscaling} {\small{\emph Numerical scaling of booster as a function of the magnetic spins $l$s.} Left panel: {\emph Non-isotropic scaling of the booster function $B_3\left(j_ 1,j_ 2+\Delta l,j_ 3+\Delta l\right)$ in the auxiliary spins $\Delta l$ compared with the curve $f(\Delta l) = 4.2 \Delta l^{-1/2} $. We rescaled the booster function by its $\Delta l=0$ value.} Right panel: {\emph Non-isotropic scaling of the booster function $B_4\left(j_1,j_2+\Delta l,j_3+\Delta l, j_4 +\Delta l; i , k +\Delta l\right)$ compared with the curve $f(\Delta l) = 9.5 \Delta l^{-2}$. We rescaled the booster function by its $\lambda=0$ value. We would prefer to accumulate more point to have a more definite estimate since by comparing with the plot on the left the asymptotic region is reached at larger spins, unfortunately our software needs to be improved to treat boosters with spins larger than 100 with sufficient precision. Luckily the analysis we are going to perform is not very sensitive to the value of this coefficient.} 
}}
\end{figure}

In the following, we will need the large $\Delta l$s scaling of both the $B_3$ and $B_4$ booster functions, we will infer it from a numerical analysis. This particular kind of scaling has not been explored before, we summarize our findings here and in Figure \ref{fig:BnLscaling}:
\begin{flalign}
\label{semiclassicalBooster1}
&B_3\left(k_1 , j_2 + \Delta l , j_3 + \Delta l \right) \approx \left(\Delta l\right)^{-\frac{1}{2}} \ ,\\
\label{semiclassicalBooster2}
&B_4\left(k_1 , j_2 + \Delta l, j_3 + \Delta l, j_4 + \Delta l; i , i'+\Delta l\right) \approx \left( \Delta l\right)^{-2} \ ,
\end{flalign}
for $\Delta l\gg k_1,\ j_2,\ j_3,\ j_4$ and $i$ or $i'$. To keep the expressions compact, we employed, and we will employ in the rest of this paper, a short-hand notation for the booster functions: 
\begin{flalign}
\label{shorthand2}
B_3(j_1+ \Delta l_1,j_2 + \Delta l_2, j_3 + \Delta l_3) &\equiv B_3(j_1,j_2,j_3;j_1+\Delta l_1, j_2+ \Delta l_2, j_3 + \Delta l_3) \ ,\\
B_4(j_1+ \Delta l_1,j_2 + \Delta l_2, j_3 + \Delta l_3, j_4+ \Delta l_4; i, i') &\equiv B_4(j_1,j_2,j_3,j_4;j_1+\Delta l_1, j_2 + \Delta l_2, j_3 + \Delta l_3, j_4+ \Delta l_4; i, i') \ .
\end{flalign}
We will refer to this short hand notation only if any $\Delta l$ is written explicitely, to not make confusion with the one introduced in the previous section. However, notice that when all the $\Delta l$ variables vanishes \eqref{shorthand2} reduces to \eqref{shorthand1}.

\noindent Combining the scalings we obtained in \eqref{semiclassicalBoosterSIMPL} and \eqref{semiclassicalBooster1} we infer that for $\lambda \gg k$ and $\Delta l \gg k$ 
\begin{flalign}
\label{semiclassicalBooster}
&B_3\left(k,\lambda+\Delta l,\lambda+\Delta l\right) \approx \left(\lambda\right)^{-\frac{1}{2}} \left(\lambda + \Delta l\right)^{-\frac{1}{2}} \\
&B_4\left(k,\lambda+\Delta l,\lambda+\Delta l, \lambda +\Delta l; \lambda +\Delta l, \lambda +\Delta l\right) \approx \left(\lambda\right)^{-\frac{1}{2}} \left(\lambda + \Delta l\right)^{-2}
\end{flalign}

Notice the oscillatory behavior of the booster functions in Figure \ref{fig:BnLscaling}. In our estimates for the scaling of the booster \eqref{semiclassicalBooster} these oscillations are neglected, corresponding to the scaling of the maximum of the oscillations. The consequence is that the estimates we will do have to be interpreted as an upper bound on the degree of divergence of the diagram. 
 In fact, for the amplitude of any diagram we can write the following inequalities:
\begin{flalign*}
 W_{diagram} \leq& \left|\sum_{j_f, i_e} \prod_f (2j_f+1)^\mu \prod_e (2i_e +1)^\mu \prod_v A_v \left(j_f,  i_e\right) \right| \leq \\
 &\sum_{j_f, i_e}  \prod_f (2j_f+1)^\mu \prod_e (2i_e +1)^\mu \left|\prod_v A_v \left(j_f, i_e\right) \right|\leq\\
  & \sum_{j_f, i_e}  \prod_f (2j_f+1)^\mu \prod_e (2i_e +1)^\mu \prod_v A^{scal}_v \left(j_f, i_e\right)
\end{flalign*}
where $A^{scal}_v$ is the quantity we estimated using \eqref{semiclassicalBooster}.

\subsection{3D bubble diagram - self-energy}
\label{sec:bubble3DEPRL}
The amplitude associated to this diagram in the EPRL model is the following:
\begin{equation}
\label{amplitudeSE3DEPRL}
W^{\mathrm{EPRL}\, \mathrm{3D}}_\mathrm{bubble} = \sum_{j_1,j_2,j_3} \prod_{f=1}^3 \left(2 j_f +1 \right)^\mu A_1 A_2,
\end{equation}
where
\begin{flalign}
\label{amplitudeA1A2}
A_v=\sum_{\Delta l_1,\Delta l_2,\Delta l_3}&\sixj{k_1 & k_2 & k_3}{j_1 + \Delta l_1 & j_2 + \Delta l_2 & j_3+ \Delta l_3} B_3(k_1,j_2+\Delta l_2,j_3 + \Delta l_3)\\
&\ B_3(j_1+\Delta l_1,k_2,j_3+\Delta l_3) B_3( j_1+\Delta l_1,j_2+\Delta l_2,k_3) \ .
\end{flalign}
We proceed by changing variables like we did for the other models $\lambda_{1}=j_{1}$, $\lambda_{2}=j_{2}-j_{1}$, $\lambda_{3}=j_{3}-j_{1}$ and analogously we also define the variables $\delta_{1}=\Delta l_{1}$, $\delta_{2}=\Delta l_{2}-\Delta l_{1}$, $\delta_{3}=\Delta l_{3}-\Delta l_{1}$. Triangular inequalities imply that the sums over $\left|\lambda_{2}\right|=\left|j_{2}-j_{1}\right|\leq k_{3}$ and $\left|\lambda_{3}\right|=\left|j_{3}-j_{1}\right|\leq k_{2}$ are bounded as expected, analogously the sums over $\delta_2$ and $\delta_3$ are also bounded. In fact:
\begin{flalign}
\left|\delta_{2}\right| &= \left| \Delta l_{2}-\Delta l_{1}\right| =\left| \Delta l_{2}-j_2-\Delta l_{1}+j1 +j_2 - j_1\right|\leq \left| \Delta l_{2}-j_2-\Delta l_{1}+j1\right| + \left|j_2 - j_1\right| \leq  2k_{3} \ ,\\
\left|\delta_{3}\right| &= \left| \Delta l_{3}-\Delta l_{1}\right| =\left| \Delta l_{3}-j_3-\Delta l_{1}+j1 +j_3 - j_1\right|\leq \left| \Delta l_{3}-j_3-\Delta l_{1}+j1\right| + \left|j_3 - j_1\right| \leq  2k_{2}\ .
\end{flalign}
We can eliminate the variable $j_f$ and $\Delta l_f$ from \eqref{amplitudeSE3DEPRL} in favor of $\lambda_f$ and $\delta_f$. We expand the summand at the first order in $\lambda_1$ and $\delta_1$ and drop all the subleading terms and multiplicative factors\footnote{Remember that we are only interested in the divergent part of the amplitude so we can choose the lower bound of the sums in $\lambda_1$ and $\delta_1$ arbitrarily large.} to obtain:
\begin{flalign*}
A_v\approx \sum_{\delta_1}&\sixj{k_1 & k_2 & k_3}{\lambda_1 + \delta_1 & \lambda_1 + \delta_1 & \lambda_1 + \delta_1} B_3(k_1,\lambda_1 + \delta_1 ,\lambda_1 + \delta_1) B_3(\lambda_1+\delta_1,k_2,\lambda_1+\delta_1) B_3(\lambda_1+\delta_1,\lambda_1+\delta_1,k_3)
\end{flalign*}
If we replace the booster functions and the $\{ 6j \}$ symbol with their large spin scaling \eqref{semiclassicalBooster} and \eqref{semiclassical6j} the vertex amplitude reduces to
\begin{equation}
A_v\approx \sum_{\delta_1} \left(\lambda_1 + \delta_1 \right)^{-\frac{1}{2}}  \left(\lambda_1\right)^{-\frac{3}{2}} \left(\lambda_1 + \delta_1\right)^{-\frac{3}{2}}
\end{equation}

The summation over $\delta_1$, from a lower bound big enough to justify the asymptotic expansion, is convergent and, at leading order in $\lambda_1$, it does not depend on the choice of the lower bound and it gives a contribution $\lambda_1^{-\frac{5}{2}}$. 

Moreover, notice that the result of the summation over $\delta_1$ does not depend on the details of the scaling \eqref{semiclassicalBooster} as long as it is convergent and the scaling of the booster functions in $\lambda_1$ and $\delta_1$ is power law. In particular we will obtain the exact same result if
\begin{equation}
B_3\left(k,\lambda+\delta,\lambda+\delta\right) \approx \left(\lambda\right)^{-\alpha} \left(\lambda + \delta\right)^{-\beta} \text{ with } \beta> \frac{1}{4}\text{ and }\alpha + \beta = -1 \ ,
\end{equation}
where the requirement $\alpha + \beta = -1 $ is necessary to be compatible with the scaling in the simplified model \eqref{semiclassicalBoosterSIMPL}. The effect in the scaling in $\lambda$ of the summation over $\delta$ is to add one power per unbounded sum over the auxiliary spins $l_{fv}$ per vertex. This step is the key to dealing with these summations that are typical of the EPRL model and were the major obstacle in all the previous attempts to similar computations.

Finally introducing a cutoff $\Lambda$ in the sum over $\lambda_1$ the transition amplitude \eqref{amplitudeSE3DEPRL} reduces to
\begin{flalign}
W^{\mathrm{EPRL}\, \mathrm{3D}}_\mathrm{bubble}\left(\Lambda\right) \approx \sum_{\lambda_1} \lambda_1^{3\mu} \left(\left(\lambda_1\right)^{-\frac{3}{2}}\left(\lambda_1\right)^{-1} \right)^2 \approx \Lambda^{3\mu - 4}
\end{flalign}

\begin{figure}[h!t]
\centering
    \includegraphics[width=0.5 \textwidth]{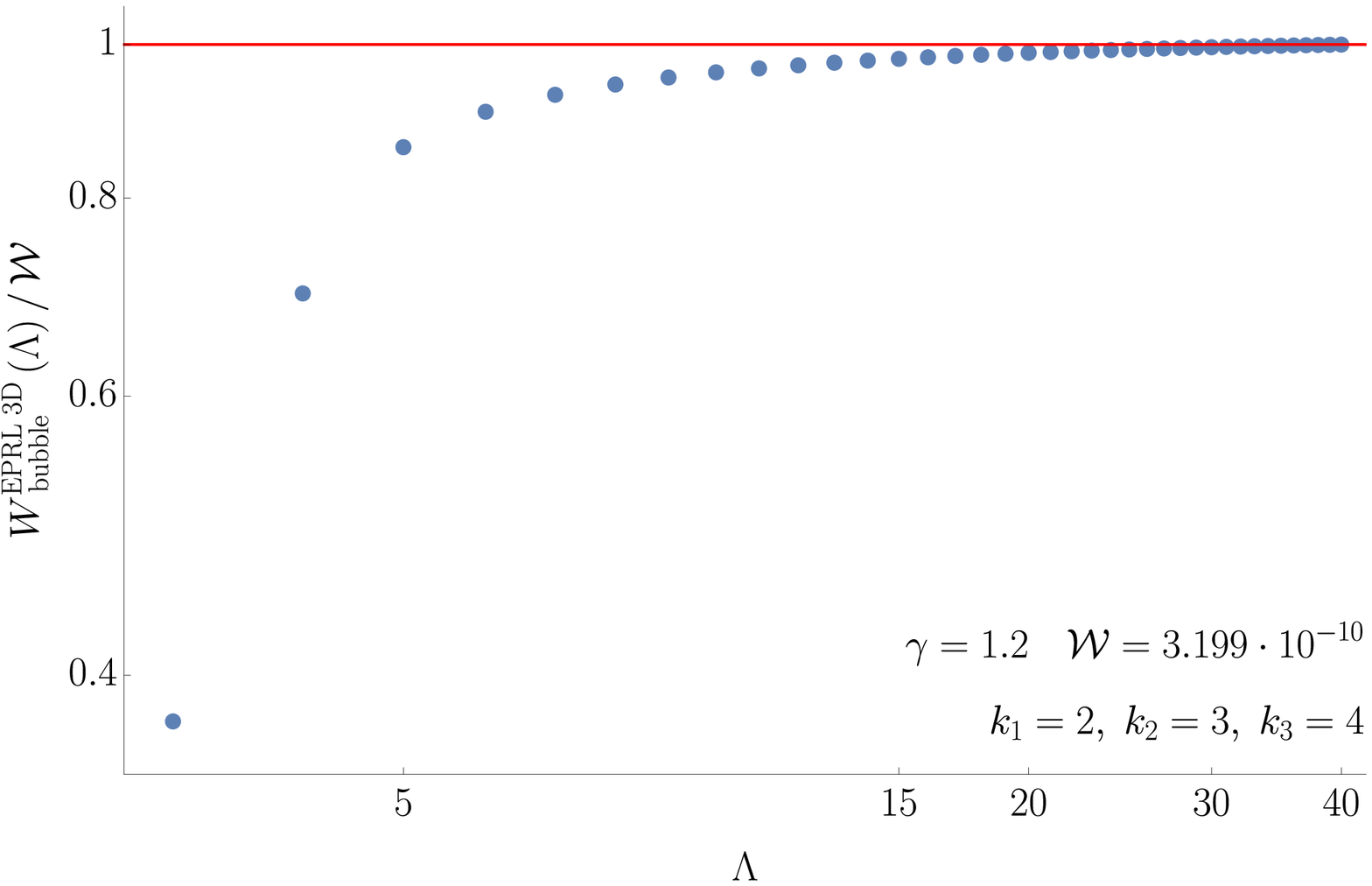}\includegraphics[width=0.5 \textwidth]{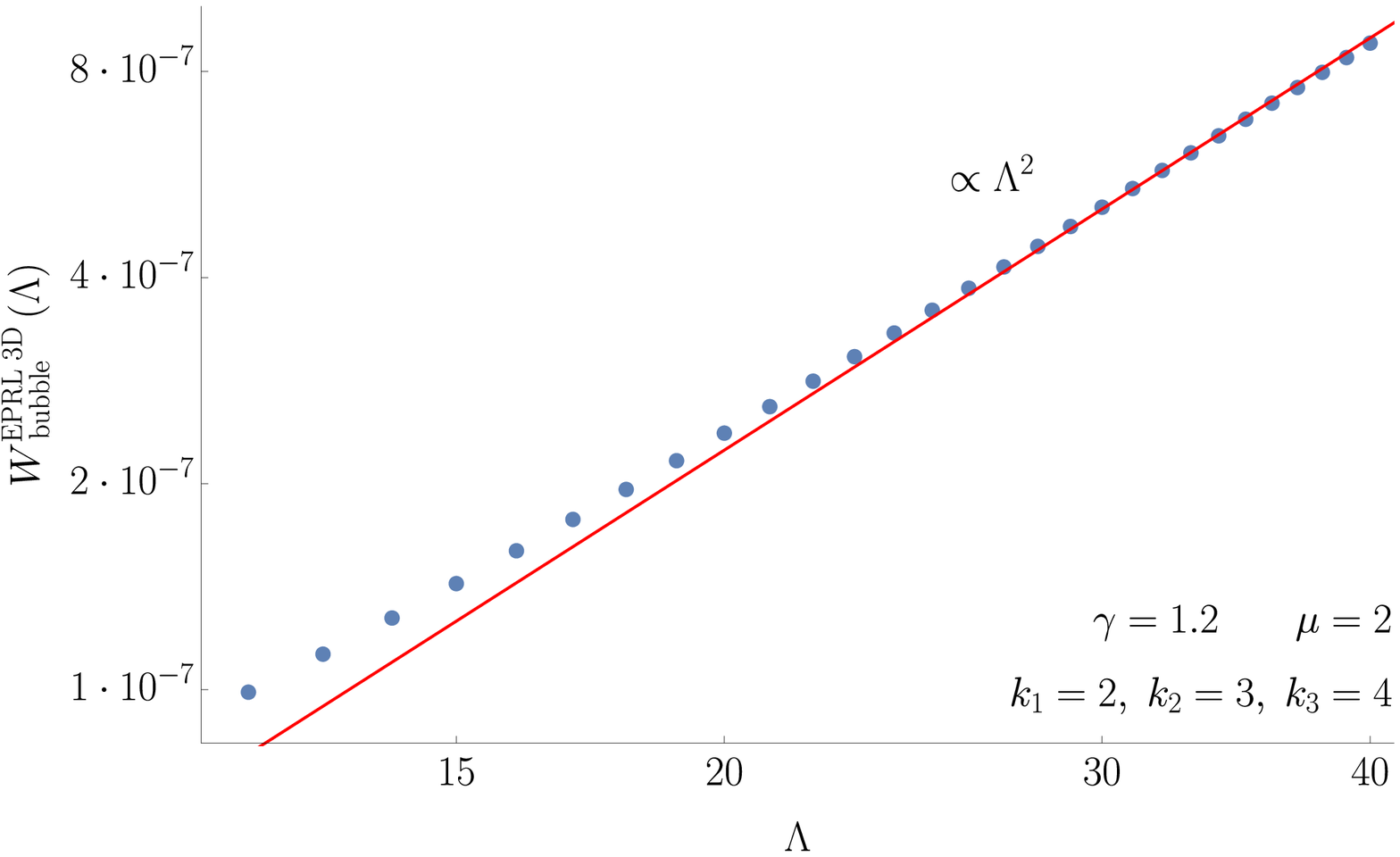}
\caption{\label{fig:SE3DEPRL_data} {\small \emph{Numerical evaluation of the transition amplitude \eqref{amplitudeSE3DEPRL} as a function of the cutoff in logarithmic scale. The external spins are $k_1 = 2$, $k_2 = 3$, $k_3 = 4$ and the Immirzi parameter is set to $\gamma = 1.2$.} Left panel: \emph{for face weight $\mu = 1$ the amplitude is convergent to the best fit $\mathcal{W}=3.199 \cdot 10^{-10}$. The plot is rescaled to allow a clearer reading.}
Right panel: \emph{for face weight $\mu = 2$ the amplitude diverge quadratically. We plot for comparison the best fit function $5.60 \cdot 10^{-10}\, \Lambda^2$ in red.}}}
\end{figure}

Independent analytical confirmations for this estimate are not available but, similarly to what we did for the EPRLs model, we are able to evaluate the amplitude\eqref{amplitudeSE3DEPRL} numerically almost exactly. ``Almost'' because we need to truncate the sums over $\Delta l_f$ in the vertex amplitudes at a certain value. These sums are convergent so we arbitrarily decided to truncate them at $\Delta l_f\approx 50$, checking a posteriori that adding one additional term change the value of the sum by a relative factor of order $10^-9$ (for more details about the numerical errors see Appendix \ref{AppB}). Our estimate is extremely accurate as reported in Figure \ref{fig:SE3DEPRL_data}: for a face weight $\mu = 1$ the amplitude is, in fact, convergent, while for a face weight $\mu=2$ diverge quadratically.

\subsection{3D ball diagram - vertex renormalization}
The amplitude associated to this diagram in the EPRL model is the following:
\begin{equation}
\label{amplitudevert3DEPRL}
W^{\mathrm{EPRL}\, \mathrm{3D}}_\mathrm{ball} = \sum_{\substack{j_1, j_2,\\j_3, j_4}} \left(\prod_{f=1}^4 \left(2 j_f +1 \right)^\mu\right) A_1 A_2 A_3 A_4 \ ,
\end{equation}%
To not distract the reader we will focus exclusively on just the first vertex amplitude, we treat the others in an analogous way, in the end they will contribute in the same way to the divergence, and we write them explicitly in Appendix \ref{AppC2}:
\begin{flalign}
\label{vertexamplitude3DBallEPRL}
A_1= \sum_{\Delta l_1^{(1)}, \Delta l_3^{(1)}, \Delta l_4^{(1)}}&\sixj{k_{1} & k_{2} & k_{3}}{j_4+\Delta l_{4}^{(1)} & j_1+\Delta l_{1}^{(1)} & j_3+\Delta l_{3}^{(1)}} B_3(k_1,j_1 + \Delta l_1^{(1)},j_3+ \Delta l_3^{(1)})\\
\nonumber &\quad B_3(j_4+ \Delta l_4^{(1)},k_2,j_3+ \Delta l_3^{(1)})B_3(j_4+ \Delta l_4^{(1)},j_1+ \Delta l_1^{(1)},k_3) \ .
\end{flalign}
Notice the triple sum over the auxiliary spins $\Delta l_f^{(v)}$ at each vertex $v$. We perform a change of variable similar to the one in Section \ref{sec:ball3DEPRLs}: we introduce a a new variable for the face spins $\lambda_{1}=j_{1}$, $\lambda_{2}=j_{2}-j_{1}$, $\lambda_{3}=j_{3}-j_{1}$ and $\lambda_{4}=j_{4}-j_{1}$ and analogously a set of $\delta$s for each vertex, for the first vertex:
\begin{equation*}
\bgroup
\setlength{\arraycolsep}{2em} 
\begin{array}{llll}
A_1: & \delta_{1}^{(1)}=\Delta l_{1}^{(1)} & \delta_{3}^{(1)}=\Delta l_{3}^{(1)}-\Delta l_{1}^{(1)} & \delta_{4}^{(1)}=\Delta l_{4}^{(1)}-\Delta l_{1}^{(1)}
\end{array} 
\egroup
\end{equation*}
The sums over $\lambda_2$, $\lambda_3$, $\lambda_4$ are bounded as lengthly discussed in the previous sections. Triangular inequalities force the sums over $\delta_3^{(1)}$, $\delta_4^{(1)}$, and analogously a couple of $\Delta l^{(v)}$ for the other vertices, to be bounded. Each vertex then has only one unbounded sum. We expand at leading order in the unbounded variables and we drop the irrelevant multiplicative factors to obtain (we drop the ${}^{(1)}$ to improve readability):   
\begin{flalign*}
A_1\approx & \sum_{\delta_1} \sixj{k_{1} & k_{2} & k_{3}}{\lambda_1+\delta_1 & \lambda_1+\delta_1 & \lambda_1+\delta_1} B_3(k_1,\lambda_1+\delta_1,\lambda_1+\delta_1) B_3(\lambda_1+\delta_1, k_2, \lambda_1+\delta_1)B_3(\lambda_1+\delta_1, \lambda_1+\delta_1, k_3) \ .
\end{flalign*}
If we use the large spin scaling for both the booster functions \eqref{semiclassicalBooster} and the $\left\lbrace 6j \right\rbrace$ symbol \eqref{semiclassical6j} all the vertex amplitudes give the same contribution:
\begin{flalign*}
A_v\approx &\left(\sum_{\delta_1}\left(\lambda_1 + \delta_1 \right)^{-\frac{1}{2}}  \left(\lambda_1\right)^{-\frac{3}{2}} \left(\lambda_1 + \delta_1\right)^{-\frac{3}{2}} \right) \ .
\end{flalign*}
The sum over $\delta_1$ is convergent and, at the leading order in $\lambda_1$, it contributes with a factor $\lambda_1^{-\frac{5}{2}}$ to the main sum over the face spins. Introducing a cutoff $\Lambda$ in the sum over $\lambda_1$ we are left with
\begin{flalign}
W^{\mathrm{EPRL}\, \mathrm{3D}}_\mathrm{ball}\left(\Lambda\right) \approx \sum_{\lambda_1}^\Lambda \lambda_1^{4\mu} \left(\lambda_1^{-\frac{5}{2}}\right)^4 \approx \Lambda^{4\mu-9}
\end{flalign}

\begin{figure}[h!t]
\centering
    \includegraphics[width=0.5 \textwidth]{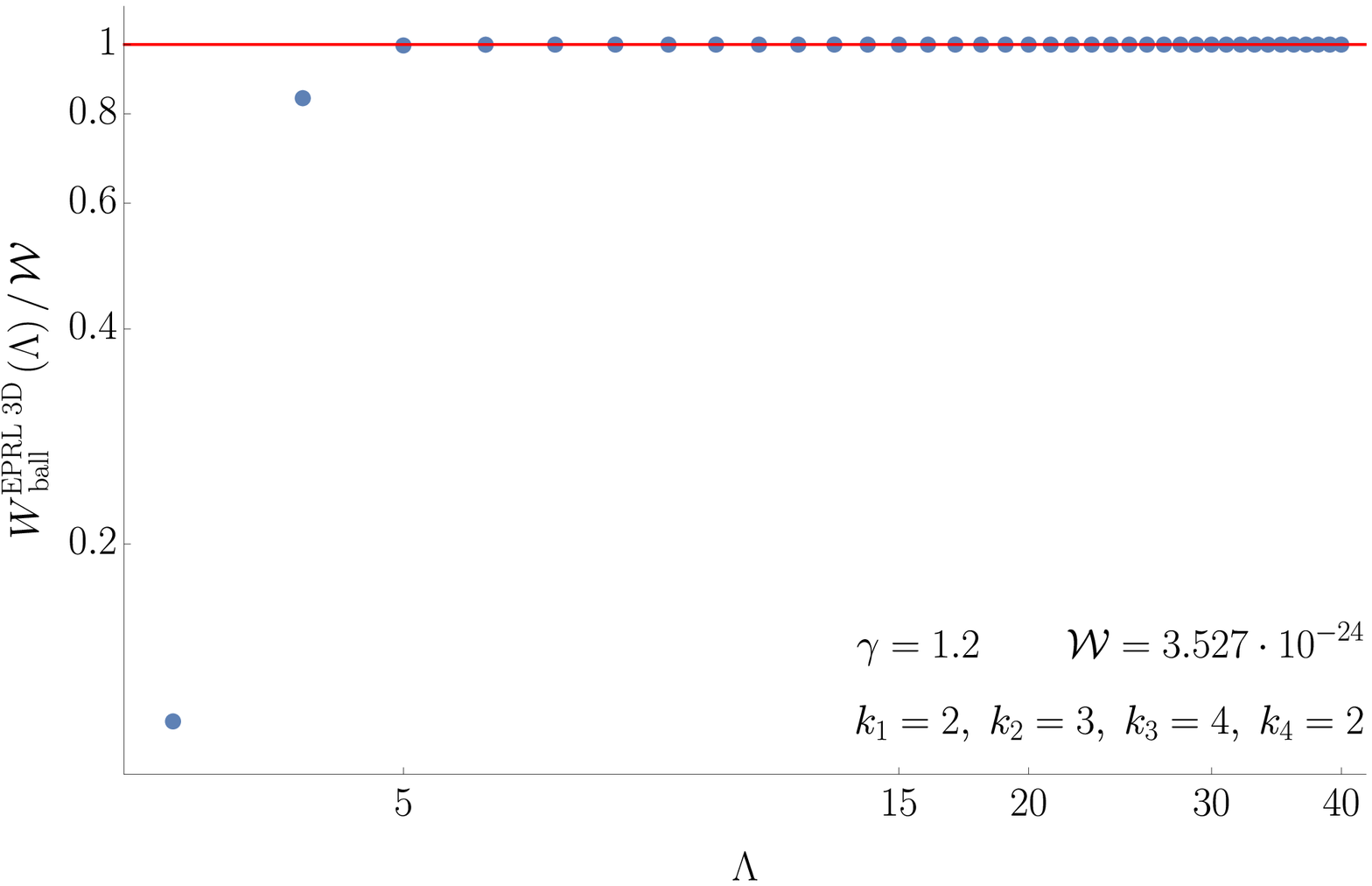}\includegraphics[width=0.5 \textwidth]{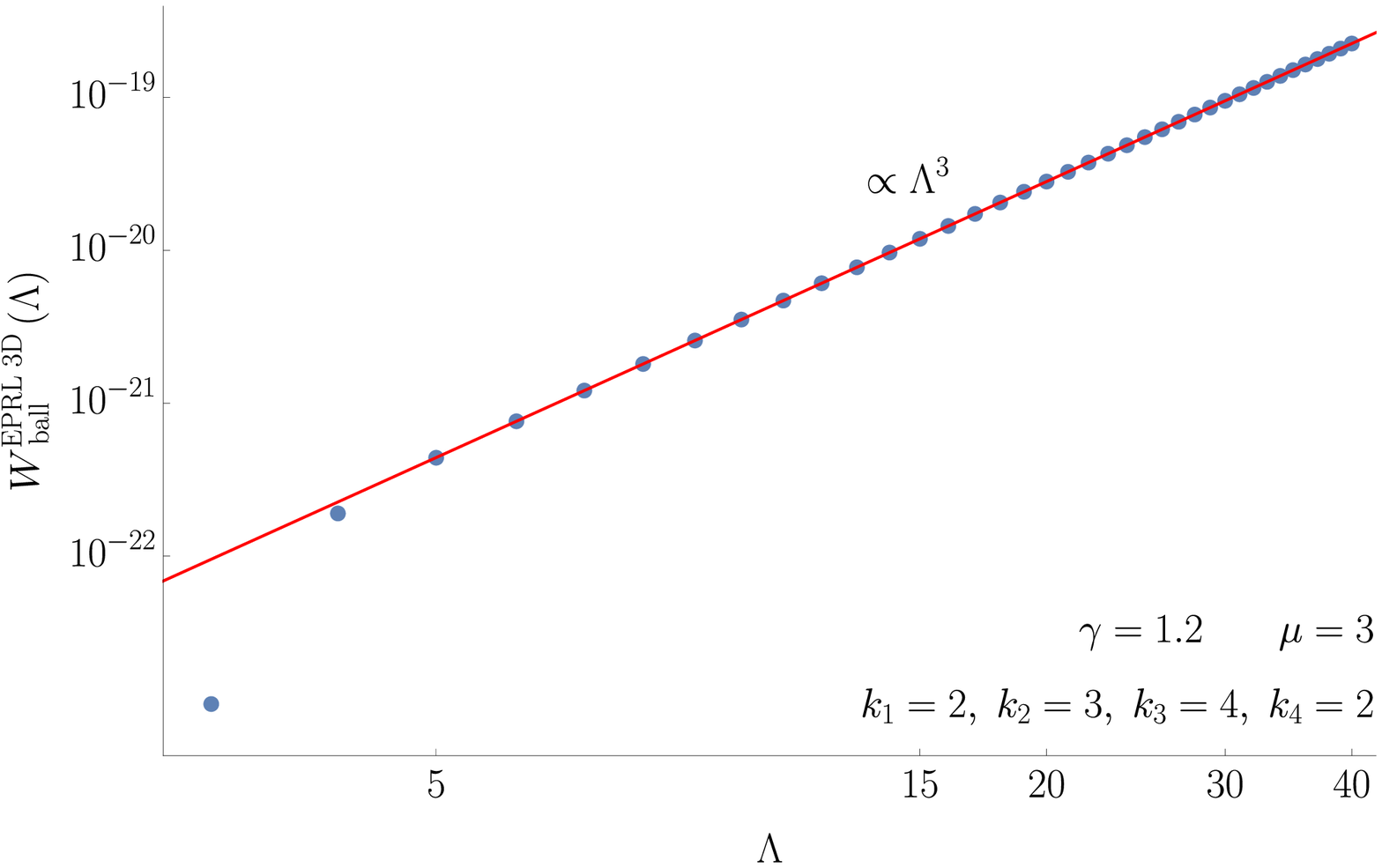}
\caption{\label{fig:vert3DEPRL_data} {\small \emph{Numerical evaluation of the transition amplitude \eqref{amplitudevert3DEPRL} as a function of the cutoff in logarithmic scale. We choose the external spins to be $k_1 = 2$, $k_2 = 3$, $k_3 = 4$, $k_4 = 2$ and Immirzi parameter $\gamma = 1.2$.} Left panel: \emph{for face weight $\mu = 1$ the amplitude is convergent to the best fit $\mathcal{W}=3.527 \cdot 10^{-24}$. The plot is rescaled to allow a clearer reading.}
Right panel: \emph{for face weight $\mu = 3$ the amplitude diverge cubically. We plot for comparison the best fit function $4.52 \cdot 10^{-24}\, \Lambda^3$ in red.}}}
\end{figure}

Independent analytical estimates of the divergence of this diagram, to our knowledge, do not exist but, similarly to what we did for the EPRLs model, we are able to evaluate \eqref{amplitudevert3DEPRL} numerically. With a truncation of the sum over $\Delta l$s our estimate is very accurate as reported in Figure \ref{fig:vert3DEPRL_data}: for a face weight $\mu = 1$ the amplitude is, in fact, convergent, while for a face weight $\mu=3$ diverge cubically.

\subsection{4D bubble diagram - self-energy}
The transition amplitude for the 4D bubble diagram (Figure \ref{fig:bubble4D}) in the EPRL model is 
\begin{equation}
\label{amplitudeSE4DEPRL}
W^{\mathrm{EPRL}\, \mathrm{4D}}_\mathrm{bubble} = \sum_{j_f, i_e}\prod_{f=1}^6 \left(2 j_f +1 \right)^\mu
\prod_{e=1}^4 \left(2 i_e +1 \right)^\mu  A_1 A_2 \ ,
\end{equation}
where the vertex amplitudes are
\begin{flalign*}
A_v=\sum_{\Delta l_f^{(v)}, i_e^{(v)}} \left(\prod_{e=1}^4 \left(2 i_e^{(v)} +1 \right)\right) \left\lbrace 15 j\right\rbrace_v \ &B_4(k_1, j_1 + \Delta l_1^{(v)},j_3 + \Delta l_3^{(v)},j_6 + \Delta l_6^{(v)}; i_1, i_1^{(v)})\\[-1.5em]
&B_4(k_2, j_1 + \Delta l_1^{(v)},j_4 + \Delta l_4^{(v)},j_5 + \Delta l_5^{(v)}; i_2, i_2^{(v)}) \\
&B_4(k_4, j_2  + \Delta l_2^{(v)},j_5 + \Delta l_5^{(v)},j_6  + \Delta l_6^{(v)}; i_3, i_3^{(v)})\\
&B_4(k_3, j_2  + \Delta l_2^{(v)},j_3 + \Delta l_3^{(v)},j_4 + \Delta l_4^{(v)}; i_4, i_4^{(v)})\ ,
\end{flalign*}
with the $\{15j\}_v$ symbols defined in \eqref{15jreduced} with the substitution $i_e \to i_e^{(v)}$ and  $j_f \to j_f + \Delta l_f^{(v)}$. We perform a change of variable on the face spins $j_f$, edge intertwiners $i_e$, auxiliary spins $\Delta l_f^{(v)}$ and auxiliary intertwiners $i_e^{(v)}$ to identify and isolate the independent bounded sums. For the spins we take $\lambda_f = j_f$ and $\delta_f^{(v)} = \Delta l_f^{(v)}$ while for the intertwiners:
\begin{equation*}
\bgroup
\setlength{\arraycolsep}{0.6em} 
\begin{array}{llll}
\iota_1 = i_1 - j_1 \ , & \iota_2 = i_2 - j_1 \ , & \iota_3 = i_3 - j_2 \ , & \iota_4 = i_4 - j_2 \ , \\
\iota^{(v)}_1 = i^{(v)}_1 - j_1 - \Delta l_1^{(v)} \ , & \iota^{(v)}_2 = i^{(v)}_2 - j_1 - \Delta l_1^{(v)} \ , & \iota^{(v)}_3 = i^{(v)}_3 - j_2 - \Delta l_2^{(v)} \ , & \iota^{(v)}_4 = i^{(v)}_4 - j_2 - \Delta l_2^{(v)} \ .\\
\end{array}
\egroup
\end{equation*}
Using all the triangular inequalities encoded in the booster functions it is possible to show that all the sums over the intertwiners $\iota_e$, $\iota_e'$, $\iota_e''$ are bounded by the boundary spins. Performing this change of variable, expanding first order in $\lambda_f$, $\delta_f$ and $\delta_f'$ and dropping irrelevant multiplicative factors the vertex amplitudes reduce to:
\begin{flalign*}
A_v \approx \sum_{\delta^{(v)}_f} \left(\lambda_1 + \delta^{(v)}_1 \right)^2 \left(\lambda_2 + \delta^{(v)}_2 \right)^2 \left\lbrace 15 j\right\rbrace_v
 & B_4(k_1, \lambda_1 + \delta^{(v)}_1,\lambda_3 + \delta^{(v)}_3,\lambda_6 + \delta^{(v)}_6; \lambda_1, \lambda_1 + \delta^{(v)}_1)\\[-1.5em]
 & B_4(k_2, \lambda_1 + \delta^{(v)}_1,\lambda_4 + \delta^{(v)}_4,\lambda_5 + \delta^{(v)}_5; \lambda_1, \lambda_1 + \delta^{(v)}_1) \\
 & B_4(k_4, \lambda_2  + \delta^{(v)}_2,\lambda_5 + \delta^{(v)}_5,\lambda_6  + \delta^{(v)}_6; \lambda_2 , \lambda_2 + \delta^{(v)}_2)\\
 &B_4(k_3, \lambda_2  + \delta^{(v)}_2,\lambda_3 + \delta^{(v)}_3,\lambda_4 + \delta^{(v)}_4; \lambda_2 , \lambda_2 + \delta^{(v)}_2) \ ,
\end{flalign*}
The sums over the auxiliary spins $\delta^{(v)}_f$ are now six dimensional. To estimate their behavior we will assume that there are no angular contribution to the divergence, then all the face spins and auxiliary spins in the radial direction scale uniformly:
\begin{equation}
\lambda_f \propto \lambda \ , \qquad \delta^{(v)}_f \propto \delta^{(v)} \ .
\end{equation}
In doing so we can rewrite the vertex amplitudes as a sum over the radial direction by taking into account the proper measure element:
\begin{flalign*}
A_v \approx \sum_{\delta} \left(\delta^{(v)}\right)^5 \left(\lambda + \delta^{(v)} \right)^4 \left\lbrace 15 j\right\rbrace_v
& B_4(k_1, \lambda + \delta^{(v)},\lambda + \delta^{(v)},\lambda + \delta^{(v)}; \lambda, \lambda + \delta^{(v)})\\[-1em]
& B_4(k_2, \lambda + \delta^{(v)},\lambda + \delta^{(v)},\lambda + \delta^{(v)}; \lambda, \lambda + \delta^{(v)}) \\
& B_4(k_4, \lambda  + \delta^{(v)},\lambda + \delta^{(v)},\lambda  + \delta^{(v)}; \lambda , \lambda + \delta^{(v)})\\
& B_4(k_3, \lambda  + \delta^{(v)},\lambda + \delta^{(v)},\lambda + \delta^{(v)}; \lambda , \lambda + \delta^{(v)}) \ ,
\end{flalign*}
We can substitute to the $\{15j\}_v$ symbol and to the booster functions their asymptotic expansions \eqref{semiclassical15j} and \eqref{semiclassicalBooster2}. The two vertex amplitudes gives than the same contribution at leading order in $\lambda$:
\begin{equation}
A_v \approx \sum_{\delta^{(v)}} \left(\delta^{(v)}\right)^5 \left(\lambda + \delta^{(v)} \right)^4 \left( \lambda + \delta^{(v)} \right)^{-\frac{7}{2}}
\left( \left(\lambda\right)^{-\frac{1}{2}} \left(\lambda + \delta^{(v)} \right)^{-2} \right)^4 \approx \lambda^{-\frac{7}{2}} \ .
\end{equation}
We introduce a factor $\lambda^5$ as volume element and we put a cutoff $\Lambda$, the amplitude \eqref{amplitudeSE4DEPRL} reads:
\begin{equation}
W^{\mathrm{EPRL}\, \mathrm{4D}}_\mathrm{bubble}\left(\Lambda \right) \approx \sum_{\lambda}^\Lambda \lambda^5 \left(\lambda\right)^{10\mu} \left(\lambda^{-\frac{7}{2}}\right)^2 \approx \left(\Lambda\right)^{10\mu -4} \ .
\end{equation}
For trivial face weight $\mu=1$ the amplitude is divergent with the same power of the cutoff as the SU(2) BF model. This estimate is compatible with the only alternative computation in the literature \cite{riello_self-energy_2013} since as the authors points out they are providing a lower bound of the divergence. To be honest we need to stress that our result is just an upper bound to the divergence, but in all the cases where we were able to perform independent computations (analytical or numerical) our estimate was extremely accurate. 

\subsection{4D ball diagram - vertex renormalization}
The transition amplitude for 4D ball diagram (Figure \ref{fig:ball4D}) in the EPRL model is: 
\begin{flalign}
\label{amplitudevert4DEPRL}
W^{\mathrm{BF}\, \mathrm{4D}}_\mathrm{ball}=\sum_{j_f, i_e} \prod_{f=1}^{10} \left(2 j_f +1 \right)^\mu\prod_{e=1}^{10} \left(2 i_e +1 \right)^\mu \prod_{v=1}^5 A_v \ ,
\end{flalign}
where we used the same intertwiner basis of Section \eqref{sec:ball4D}. To not distract the reader we will focus exclusively on just the first vertex amplitude, we treat the others in an analogous way, in the end they will contribute at the same way to the divergence, and we write them explicitely in Appendix \ref{AppC2}:
\begin{flalign}
\label{vertexamplitudeBallEPRL}
A_1 =\sum_{\Delta l^{(1)}_{fv}, i^{(1)}_{ev}} \left(\prod_{ev} (2 i_{ev}^{(1)} + 1) \right) \left\lbrace 15 j\right\rbrace_1 \ &  B_4(k_1, j_{1} +\Delta l^{(1)}_{1} ,j_{3} +\Delta l^{(1)}_{3},j_{2} +\Delta l^{(1)}_{2}; i_{1}, i^{(1)}_{1})\\[-1.5em]
&B_4(k_2, j_{1} +\Delta l^{(1)}_{1},j_{5}+\Delta l^{(1)}_{5},j_{4}+\Delta l^{(1)}_{4}; i_{2}, i^{(1)}_{2})\\
&B_4(k_3, j_{6} +\Delta l^{(1)}_{6},j_{2}+\Delta l^{(1)}_{2},j_{4}+\Delta l^{(1)}_{4}; i_{3}, i^{(1)}_{3})\\
&B_4(k_4, j_{6} +\Delta l^{(1)}_{6} ,j_{3} +\Delta l^{(1)}_{3},j_{5}+\Delta l^{(1)}_{5}; i_{4}, i^{(1)}_{4})\ .
\end{flalign}
We denoted with $\left\lbrace 15 j\right\rbrace_v$ the same symbols defined in \eqref{tutti15j} with the substitution $i_e \to i_e^{(v)}$ and  $j_f \to j_f + \Delta l_f^{(v)}$. The summation over the auxiliary intertwiners $i_{ev}^{(v)}$, a set of four per vertex $(v)$, is carried over the edges that are attached to the vertex $v$ (i.e. $v=1$ implies $ev=1,2,3,4$). The summation over the auxiliary spins $\Delta l^{(v)}_{fv}$, a set of six per vertex $(v)$, is carried over the faces that contain the vertex $v$ (i.e. $v=1$ implies $fv=1,2,3,4,5,6$).
To make the bounded sums and unbounded sums manifest we make the same change of variables on $j_f$ and $i_e$ of Section \ref{sec:ball4D} and in addition $\Delta l^{(v)}_{fv} = \delta^{(v)}_{fv}$,
\begin{equation*}
\begin{array}{ccccc}
A_1:& \iota^{(1)}_1 = i^{(1)}_1 -j_1 - \Delta l^{(1)}_{1} \ , & \iota^{(1)}_2 = i^{(1)}_2 -j_1 - \Delta l^{(1)}_{1} \ ,& \iota^{(1)}_3 = i^{(1)}_3 -j_6 - \Delta l^{(1)}_{6} \ ,& \iota^{(1)}_4 = i^{(1)}_4 -j_6 - \Delta l^{(1)}_{6} \ .\\
\end{array}
\end{equation*}
In terms of these new variables all sums over $\iota_e$ and $\iota^{(v)}_e$ are bounded, while the sums over $\lambda_f$ and $\delta^{(v)}_{fv}$ are all unbounded.
Expanding the vertex amplitudes at the first order in $\lambda_f$, $\delta^{(v)}_{fv}$ and dropping irrelevant multiplicative factors the vertex amplitudes reduce to: \begin{flalign*}
A_1 \approx \sum_{\delta^{(1)}_{fv}}\left(\lambda_1 + \delta_1^{(1)}\right)^2\left(\lambda_6 + \delta^{(1)}_6\right)^2 \left\lbrace 15 j\right\rbrace_1 \ 
& B_4(k_1, \lambda_{1}  + \delta^{(1)}_1,\lambda_{3} + \delta^{(1)}_3,\lambda_{2} + \delta^{(1)}_2; \lambda_1, \lambda_{1} + \delta^{(1)}_1)\\[-1.5em]
& B_4(k_2, \lambda_{1} + \delta^{(1)}_1,\lambda_{5}+ \delta^{(1)}_5,\lambda_{4}+ \delta^{(1)}_4; \lambda_1, \lambda_{1}+ \delta^{(1)}_1)\\
& B_4(k_3, \lambda_{6} + \delta^{(1)}_6,\lambda_{2}+ \delta^{(1)}_2,\lambda_{4}+ \delta^{(1)}_4; \lambda_6, \lambda_6+ \delta^{(1)}_6)\\
& B_4(k_4, \lambda_{6} + \delta^{(1)}_6,\lambda_{3}+ \delta^{(1)}_3,\lambda_{5}+ \delta^{(1)}_5; \lambda_6, \lambda_6+ \delta^{(1)}_6)\ .
\end{flalign*}
The sums over the auxiliary spins $\delta_f^{(v)}$ are now six dimensional. To estimate their behavior we will assume that there are no angular contribution to the divergence, than all the face spins and the auxiliary spins in the radial direction scale uniformly:
\begin{equation}
\lambda_f \propto \lambda \ , \qquad \delta^{(v)}_{fv} \propto \delta^{(v)} \ .
\end{equation}
In doing so we can rewrite the vertex amplitudes as a sum over the radial direction by taking into account the proper measure element. We subtitute to the $\left\lbrace 15j\right\rbrace$ symbol and to the boosters functions their asymptotic expressions \eqref{semiclassical15j} and \eqref{semiclassicalBoosterSIMPL2}. Each vertex amplitude gives the same contribution at leading order in $\lambda$
\begin{equation}
A_v \approx \sum_{\delta^{(v)}} \left(\delta^{(v)}\right)^5 \left(\lambda + \delta^{(v)} \right)^4 \left( \lambda + \delta^{(v)} \right)^{-\frac{7}{2}}
\left( \left(\lambda\right)^{-\frac{1}{2}} \left(\lambda + \delta^{(v)} \right)^{-2} \right)^4 \approx \lambda^{-\frac{7}{2}} \ .
\end{equation}
We introduce a factor $\lambda^5$ as volume element and we put a cutoff $\Lambda$, the amplitude \eqref{amplitudevert4DEPRL} reads:
\begin{flalign}
W^{\mathrm{EPRL}\, \mathrm{4D}}_\mathrm{ball}\left(\Lambda\right) \approx \sum_{\lambda}^\Lambda \lambda^9 \lambda^{10\mu} \lambda^{10\mu}  \left(  \lambda^{-\frac{7}{2}} \right)^5 \approx \Lambda^{20\mu-\frac{17}{2}} \ .
\end{flalign}
For trivial face weight $\mu=1$ the amplitude is divergent with the same power of the cutoff as the SU(2) BF model. To our knowledge this is the first estimate in the literature of this divergence.

%--------------------------------------------------
\section{Conclusions}
%--------------------------------------------------
In this paper, we estimated the large volume divergence of the bubble and ball diagrams in three and four dimensions in the EPRL model at fixed boundary states. This is formally done with the artificial insertion of a uniform cut-off $\Lambda$ on all the spins associated with the faces of the spin foam diagrams.
As a collateral product, we were able to estimate the divergence of the same diagrams in the EPRLs model and in the SU(2) BF model. Two assumptions are made in the computation: 
\begin{enumerate}
\item the main contribution to the divergence comes from the uniform scaling of all the spins;
\item there is no interference between various terms of the sum.
\end{enumerate}
The first assumption is the one we have the least control over, nevertheless, we can test this hypothesis in the SU(2) BF model, where analytical computations are possible, and it seems to be verified. We also note that the same supposition is also made in similar works in the literature like \cite{perini_self-energy_2009} and \cite{riello_self-energy_2013}.
The second assumption can be freely relaxed if we interpret our estimate as an upper bound on the divergence of the diagram as we already discussed at the end of Section \ref{sec:modelEPRL}.
The first assumption is crucial for the success of the algorithm. This hypothesis has an enlightening analog in the study of convergence at infinity of multi-dimensional integrals. There we can perform a radial coordinate change and immediately see that, \textit{if the angular integration is regular}, the only source of divergence is the radial asymptotic behavior of the integrand. If this is not the case, the divergence will be in general of higher order. Similarly in our case, if the assumption 1 is not verified we should expect a divergence of higher order then the one we estimate. Nevertheless, in the simpler models we considered, like the BF SU(2) models, this hypothesis can be explicitly checked and happens to be satisfied.\\

\noindent Using some examples, we proposed a general algorithm to estimate the divergence of any spin foam transition amplitude. We summarize it in the following:\\
First, we determine the \textit{scaling of each vertex amplitude} \eqref{vertexEPRL} in a uniform face amplitude rescaling:
\begin{enumerate}
\item Find the unbounded sums over the auxiliary spins and intertwiners at that vertex using edge triangular inequalities.
\item Combine the scaling of the SU(2) invariant at the vertex with the scaling of the booster functions attached to the vertex and the dimensions of the auxiliary intertwiners.
\item The so obtained scaling is raised by one power for each unbounded sum found in point 1.
\end{enumerate}

\noindent Then we determine the \textit{scaling of the whole amplitude} \eqref{partitionEPRL}:
\begin{enumerate}[resume]
\item Find the unbounded sums over the face spins and intertwiners using again edge triangular inequalities.
\item Combine the scaling of each vertex amplitude with the face amplitude and the dimension of the intertwiners on the edges. 
\item The divergence of the diagram as a function of a cutoff is the scaling just obtained raised by a power for each unbounded sum found in point 4.
\end{enumerate}

\noindent This being said, the estimate of the divergences of the four diagrams in the various models we considered are summarized in the following table:
\begin{center}
\bgroup
\def\arraystretch{1.5}%  1 is the default, change whatever you need
\begin{tabular}{c|c|c|c|c}
 & bubble 3D & ball 3D & bubble 4D & ball 4D \\ 
\hline 
 BF & \cellcolor{red!25}$\Lambda^{3\mu}$ & \cellcolor{red!25}$\Lambda^{4\mu-1}$ & \cellcolor{red!25}$\Lambda^{10\mu-1}$ & \cellcolor{red!25}$\Lambda^{20\mu-15/2}$ \\ 
\hline 
EPRLs & \cellcolor{green!25}$\Lambda^{3\mu-6}$ & \cellcolor{green!25}$\Lambda^{4\mu-13}$ & \cellcolor{green!25}$\Lambda^{10\mu-13}$ & \cellcolor{green!25}$\Lambda^{20\mu-75/2}$ \\
\hline 
EPRL & \cellcolor{green!25}$\Lambda^{3\mu-4}$ & \cellcolor{green!25}$\Lambda^{4\mu-9}$ &  \cellcolor{red!25}$\Lambda^{10\mu-1}$ & \cellcolor{red!25}$\Lambda^{20\mu-15/2}$  \\ 
\end{tabular} 
\egroup
\end{center}
To facilitate the reading of the table we highlighted in green the diagrams that for the standard choice of face amplitude $\mu=1$ have a convergent amplitude and in red the divergent one. 
All the transition amplitude we considered diverge in the SU(2) BF model. The degree of divergence we compute is in excellent agreement with the analytical evaluation of the diagram. Moreover, all the considered transition amplitude in the EPRLs model are convergent. Even if analytical evaluations are not possible for the three dimensional diagrams we were able to evaluate the amplitude numerically without any approximations, finding perfect agreement with our estimate and growing confidence on the validity of our work hypothesis. We believe that, with the development of more performant numerical methods to treat the booster functions, we will be able in the future to evaluate also the amplitudes of the four dimensional diagrams.
The transition amplitudes of the three dimensional diagrams in the EPRL model are convergent. We are able to evaluate the sum almost exactly (some truncations are needed but the numerics is not very sensible on them) showing that our estimates are very accurate. 
The amplitudes of both the four dimensional diagrams in the EPRL model are divergent. Our result, even if not directly comparable with the computation done in \cite{riello_self-energy_2013} because of the different techniques, it is still compatible since they provide effectively a lower bound for the divergence (logarithmic in the cutoff) while we provide an upper bound. 
For the simpler 3-stranded amplitudes we found a precise numerical confirmation. This suggests that the 4-stranded divergences are also close to the upper bound we estimate. A possible source for a value close to but not exactly at the bound comes from the fact that the oscillations present in the $B_4$ functions could give rise to destructive interference. The ongoing work on improving the understanding of booster asymptotics and numerical codes should allow us to settle this question in the near future.

We should also comment that a non-vanishing cosmological constant can be incorporated in the theory with a conjectured quantum group deformation studied in \cite{han_cosmological_2011,haggard_four-dimensional_2016}. The divergences we studied are likely to be effectively regulated in this formulation in terms of the quantum group. This is consistent with the fact that q-deformed amplitudes are suppressed for large spins, correspondingly to the fact that the presence of a cosmological constant sets a maximal distance.

%--------------------------------------------------------------------------------------------------
\section*{Acknowledgments}
This work was supported in part by the NSF grant PHY-1505411, the Eberly research funds of Penn State. The computations were carried out at the Institute for CyberScience at Penn State. I am greatly indebted to Simone Speziale for constant encouragement throughout the production period of this work. I wish to thank Hal Haggard for countless hours of discussion on $SU(2)$ invariants. I also want to thank Giorgio Sarno and Fran\c{c}ois Collet for discussion on the numerics of this work and for sharing part of their code for the evaluation of the booster functions. 
%--------------------------------------------------------------------------------------------------

%----------------------------------------------------------------------------
\newpage
\appendix
%----------------------------------------------------------------------------

%----------------------------------------------------------------------------
\section{SU(2) Symbols and Boosters}\label{AppSU2}
%----------------------------------------------------------------------------
In the following, we implicitly assume that the Clebsch-Gordan triangular inequalities are satisfied, else the evaluations vanish.
We use the definition for the Wigner's $(3jm)$ symbol reported in \cite{varshalovich_quantum_1988} with the following orthogonality properties
\begin{equation*}
\sum_{m_1,m_2} \njm{ j_1& j_2& j_3}{m_1 & m_2 & m_3} \njm{ j_1& j_2& l_3}{m_1 & m_2 & n_3}  = \frac{\delta_{j_{3}l_{3}}\delta_{m_3n_3}}{2 j_3 + 1} \ ,
\end{equation*}
implying they are normalized to one. We define the $(4jm)$ symbol as the contraction of two $(3jm)$ symbol via an intertwiner $i$
\begin{equation*}
\njm{ j_1& j_2& j_3 & j_4}{m_1 & m_2 & m_3 & m_4}^{(i)}\equiv \sum_{m_i} (-1)^{i-m_i}\njm{ j_1& j_2& i}{m_1 & m_2 & m_i} \njm{ i& j_3& j_4}{-m_i & m_3 & m_4}\ ,
\end{equation*}
respecting the following orthogonality relations
\begin{equation*}
\sum_{m_1,m_2,m_3} \njm{ j_1& j_2& j_3 & j_4}{m_1 & m_2 & m_3 & m_4}^{(i_1)}
\njm{ j_1& j_2& j_3 & l_4}{m_1 & m_2 & m_3 & n_4}^{(i_2)} = \frac{\delta_{i_1 i_2}}{2 i_1 + 1} \frac{\delta_{j_{4}l_{4}}\delta_{m_4n_4}}{2 j_4 + 1} \ ,
\end{equation*}
normalized to $ \frac{\delta_{i_1 i_2}}{2 i_1 + 1}$. In Section \ref{sec:gravityandBF} we used a short-hand notation for the general $(njm)$ symbol:
\begin{flalign}
\left(\begin{array}{c} j_a \\ p_a \end{array}\right)^{(i)} =& \quad \njm{ j_1& j_2& \cdots & j_n}{m_1 & m_2 & \cdots & m_n}^{(i_1,\ i_2,\ \cdots, i_{n-3})} = \nonumber \\
&\sum_{m_{i_s}} 
(-1)^{\sum_{s=1}^{n-3}\left(i_s-m_{i_s}\right)}\njm{ j_1& j_2& i_1}{m_1 & m_2 & m_{i_1}}  \njm{ i_1& j_3& i_2}{-m_{i_1} & m_3 & m_{i_2}}  \cdots \njm{ i_{n-3}& j_{n-1}& j_n}{-m_{i_{n-3}} & m_{n-1} & m_n} \ .\nonumber
\end{flalign}

The explicit $z$-boost matrix elements can be found in the literature in its general form \cite{ruhl_lorentz_1970,naimark_linear_1964,rashid_boost_1979,speziale_boosting_2016}, here we just report the explicit form of $z$-boost matrix elements for simple irreducible representation:
\begin{align*}
d^{(\gamma j,j)}_{jlp}(r) =&  
(-1)^{\frac{j-l}{2}} \frac{\Gamma\left( j + i \gamma j +1\right)}{\left|\Gamma\left(  j + i \gamma j +1\right)\right|} \frac{\Gamma\left( l - i \gamma j +1\right)}{\left|\Gamma\left(  l - i \gamma j +1\right)\right|} \frac{\sqrt{2j+1}\sqrt{2l+1}}{(j+l+1)!}  
\left[(2j)!(l+j)!(l-j)!\frac{(l+p)!(l-p)!}{(j+p)!(j-p)!}\right]^{1/2} \\
&\ \times e^{-(j-i\gamma j +p+1)r}
\sum_{s} \frac{(-1)^{s} \, e^{- 2 s r} }{s!(l-j-s)!} \, {}_2F_1[l+1-i\gamma j,j+p+1+s,j+l+2,1-e^{-2r}] \ .
\end{align*}
We refer to \cite{speziale_boosting_2016} for a more in depth definition of the booster functions.
%----------------------------------------------------------------------------
\section{Divergences of SU(2) BF}\label{AppA}
%----------------------------------------------------------------------------
For SU(2) BF spin foams is possible to compute the divergence of the various diagrams analytically by using the representation of the Dirac delta over the group in terms of characters
\begin{equation*}
\delta\left( U \right) = \sum_j (2j+1) \chi^j \left(U\right) \ .
\end{equation*}
The Dirac delta computed at the identity is divergent if we place a cutoff $\Lambda$ in the sum over the SU(2) irreducible representations
we can see that the delta diverge cubically in it.
\begin{equation}
\delta_\Lambda\left( \mathds{1} \right) = \sum_j^\Lambda (2j+1) \chi^j \left(\mathds{1} \right)= \sum_j^\Lambda (2j+1)^2 = 
\frac{1}{6} \left(1 + 2 \Lambda\right)\left(2 + 2 \Lambda\right) \left(3 + 4 \Lambda\right) \approx \Lambda^3 \ .
\end{equation}
Let's consider the spin foam amplitude \eqref{amplitudeSE3DBF} with face weight $\mu=1$ first. One integral per edge is redundant and can be eliminated by a trivial change of variables. We are left with three integrals over copies of $SU(2)$:
\begin{equation}
W^{3D}_{\text{bubble}} = \int \left(\prod_{l=1}^3\mathrm{d} g_l\right) E\left(g_1,g_2,g_3\right) \delta \left( g_{1} g_{3}^{-1}\right)\delta \left( g_{2} g_{1}^{-1}\right)\delta \left( g_{3} g_{2}^{-1}\right)
\end{equation}
where we indicated with $E\left(g_1,g_2,g_3\right)$ the tensor product of the Wigner matrices of the external faces. If we denote with $i^{m_1 m_2 m_3}$ the tensor in the trivial three valent intertwiner space 
\begin{equation*}
E\left(g_1,g_2,g_3\right) = i^{m_1 m_2 m_3} D^{(k_1)}_{m_1 n_1} \left( g_1 \right)D^{(k_2)}_{m_2 n_2} \left( g_2 \right)D^{(k_3)}_{m_3 n_3} \left( g_3 \right) i^{n_1 n_2 n_3}
\end{equation*}
We can perform the integrals by using the definition of the Dirac delta over the group 
\begin{flalign*}
W^{3D}_{\text{bubble}} =& \int \left(\prod_{l=1}^2\mathrm{d} g_l\right) E\left(g_1,g_2,g_1\right) \delta \left( g_{2} g_{1}^{-1}\right)\delta \left( g_{1} g_{2}^{-1}\right)=\\
&\int \mathrm{d} g_1 E\left(g_1,g_1,g_1\right) \delta_\Lambda\left( \mathds{1} \right) \approx \Lambda^3 
\end{flalign*}

The computation of the spin foam amplitude \eqref{amplitudeSE4DBF} with face weight $\mu=1$ is very similar. In terms of $SU(2)$ integrals reads
\begin{equation}
W_{\text{bubble}^{4D}} = \int \left(\prod_{l=1}^4 \mathrm{d} g_l\right) E\left(g_1,g_2,g_3,g_{4}\right)
\delta \left( g_{1} g_{4}^{-1}\right)
\delta \left( g_{2} g_{1}^{-1}\right)
\delta \left( g_{3} g_{1}^{-1}\right)
\delta \left( g_{2} g_{3}^{-1}\right)
\delta \left( g_{4} g_{2}^{-1}\right)
\delta \left( g_{3} g_{4}^{-1}\right)
\end{equation}
where we indicated with $E\left(g_1,g_2,g_3,g_4\right)$ the tensor product of the Wigner matrices of the external faces. If we denote with $i_{(t_1)}^{m_1 m_2 m_3 m_4}$ the tensor in the four valent intertwiner space in the recoupling basis $(k_1,k_2)$ identified with the spin $t_1$:
\begin{equation*}
E\left(g_1,g_2,g_3,g_4\right) = i_{(t_1)}^{m_1 m_2 m_3 m_4} D^{(k_1)}_{m_1 n_1} \left( g_1 \right)D^{(k_2)}_{m_2 n_2} \left( g_2 \right)D^{(k_3)}_{m_3 n_3} \left( g_3 \right) D^{(k_3)}_{m_3 n_3} \left( g_4 \right) i_{(t_2)}^{n_1 n_2 n_3 n_4}
\end{equation*}
Performing the integrations over the group using the definition of the Dirac delta over the group we obtain
\begin{equation*}
W_{\text{bubble}^{4D}} = \int \mathrm{d} g_1 E\left(g_1,g_1,g_1,g_{1}\right) \delta_\Lambda\left( \mathds{1} \right)^3 \approx \delta_{t_1 t_2} \Lambda^9
\end{equation*}
%Details of the simplification if anyone is interested
%\begin{align}
%W =& \int \prod_{l=1}^4 \mathrm{d} g_l \delta \left( g_{1} g_{4}^{-1}\right)
%\delta \left( g_{2} g_{1}^{-1}\right)
%\delta \left( g_{3} g_{1}^{-1}\right)
%\delta \left( g_{2} g_{3}^{-1}\right)
%\delta \left( g_{4} g_{2}^{-1}\right)
%\delta \left( g_{3} g_{4}^{-1}\right)
%= \\
%& \int \prod_{l=2}^4 \mathrm{d} g_l 
%\delta \left( g_{2} g_{4}^{-1}\right)
%\delta \left( g_{3} g_{4}^{-1}\right)
%\delta \left( g_{2} g_{3}^{-1}\right)
%\delta \left( g_{4} g_{2}^{-1}\right)
%\delta \left( g_{3} g_{4}^{-1}\right)
%= \\
%& \int \prod_{l=3}^4 \mathrm{d} g_l 
%\delta \left( g_{3} g_{4}^{-1}\right)
%\delta \left( g_{4} g_{3}^{-1}\right)
%\delta \left( \mathds{1}\right)
%\delta \left( g_{3} g_{4}^{-1}\right)
%= \delta \left( \mathds{1}\right)^3 \approx  \Lambda^9
%\end{align}
For completeness, we also consider the two ball divergences of BF spin foam diagrams we studied in Section \ref{sec:bubble3D} and \ref{sec:bubble4D}. 
\begin{align}
W_{\text{ball}}^{3D} =& \int \left(\prod_{l=1}^6 \mathrm{d} g_l\right) E\left(g_1,g_2,g_3,g_4,g_5,g_6\right)
\delta \left( g_{1} g_{3}^{-1}g_{2}^{-1}\right)
\delta \left( g_{3} g_{5}^{-1} g_6 \right)
\delta \left( g_{4} g_{6} g_{2}^{-1}\right)
\delta \left( g_{1} g_{5}^{-1} g_4^{-1}\right) =\\
&\int \mathrm{d} g_1\mathrm{d} g_2 \mathrm{d} g_4 \  E\left(g_1,g_2,g_1 g_2^{-1},g_4,g_4 g_2^{-1},g_4 g_1^{-1}\right) \delta_\lambda \left( \mathds{1}\right) \approx \sixj{k_1 & k_4 & k_2}{k_5 & k_3 & k_6} \Lambda^3 \nonumber
\end{align}
Where 
\begin{flalign*}
E\left(g_1,g_2,g_3,g_4,g_5,g_6\right) =& i_1^{m_1 m_2 m_4}i_2^{n_1 n_3 n_6}i_3^{o_2 o_3 o_5}i_4^{p_4 p_5 p_6}\\
&D^{(k_1)}_{m_1 n_1}(g_1^{-1})  D^{(k_2)}_{m_2 o_2}(g_2^{-1})  D^{(k_3)}_{o_3 n_3}(g_3)  D^{(k_4)}_{p_4 m_4}(g_4)  D^{(k_5)}_{p_5 o_5}(g_5)  D^{(k_6)}_{p_6 n_6}(g_6)
%&D^{(k_1)}_{}(g_1^{-1})  D^{(k_2)}_{}(g_2^{-1})  D^{(k_3)}_{}(g_2 g_1^{-1})  D^{(k_4)}_{}(g_4)  D^{(k_5)}_{}(g_4 g_2^{-1})  D^{(k_6)}_{}(g_4 g_1^{-1})
\end{flalign*}
And finally for the four dimensional ball, omitting for simplicity the boundary representation matrices 
\begin{align*}
W_{\text{ball}}^{4D} =& \int \prod_{l=1}^{10} \mathrm{d} g_l \   
 E(g_e)\ \delta \left(g_{1} g_{5} g_{2}^{-1} \right)
 \delta \left(g_{1} g_{6} g_{3}^{-1} \right)
 \delta \left(g_{1} g_{7} g_{4}^{-1} \right)
 \delta \left(g_{2} g_{8} g_{3}^{-1} \right)
 \delta \left(g_{2} g_{9} g_{4}^{-1} \right)\\
 &\hspace{2.5cm}\delta\left(g_{3} g_{10} g_{4}^{-1} \right)
 \delta \left(g_{5} g_{8} g_{6}^{-1} \right)
 \delta \left(g_{5} g_{9} g_{7}^{-1} \right)
 \delta \left(g_{6} g_{10} g_{7}^{-1} \right)
 \delta \left(g_{8} g_{10} g_{9}^{-1}\right) =\\
 &\int \mathrm{d} g_4  \mathrm{d} g_5  \mathrm{d} g_6  \mathrm{d} g_7 \ 
  E(g_4, g_5, g_6, g_7) \ \delta_\Lambda \left(\mathds{1} \right)^4  \approx \Lambda^{12}\\ 
\end{align*}

%Case balls 4D
%\begin{align}
%W =& \int \prod_{l=1}^{10} \mathrm{d} g_l 
% &\delta \left(g_{1} g_{5} g_{2}^{-1} \right)
% \delta \left(g_{1} g_{6} g_{3}^{-1} \right)
% \delta \left(g_{1} g_{7} g_{4}^{-1} \right)
% \delta \left(g_{2} g_{8} g_{3}^{-1} \right)
% \delta \left(g_{2} g_{9} g_{4}^{-1} \right)\\
% &&\delta \left(g_{3} g_{10} g_{4}^{-1} \right)
% \delta \left(g_{5} g_{8} g_{6}^{-1} \right)
% \delta \left(g_{5} g_{9} g_{7}^{-1} \right)
% \delta \left(g_{6} g_{10} g_{7}^{-1} \right)
% \delta \left(g_{8} g_{10} g_{9}^{-1}\right) =\\ 
%  & \int \prod_{l=2}^9 \mathrm{d} g_l 
% &\delta \left(g_{2} g_{5}^{-1} g_{6} g_{3}^{-1} \right)
% \delta \left(g_{2} g_{5}^{-1} g_{7} g_{4}^{-1} \right)
% \delta \left(g_{2} g_{8} g_{3}^{-1} \right)
% \delta \left(g_{2} g_{9} g_{4}^{-1} \right)\\
% &&\delta \left(g_{3} g_{8}^{-1} g_{9} g_{4}^{-1} \right)
% \delta \left(g_{5} g_{8} g_{6}^{-1} \right)
% \delta \left(g_{5} g_{9} g_{7}^{-1} \right)
% \delta \left(g_{6} g_{8}^{-1} g_{9} g_{7}^{-1} \right)=\\ 
% & \int \prod_{l=3}^8 \mathrm{d} g_l 
% &\delta \left(g_{3} g_{8}^{-1} g_{5}^{-1}  g_{7} g_{4}^{-1} \right)
% \delta \left(g_{3} g_{8}^{-1} g_{5}^{-1}  g_{7} g_{4}^{-1} \right)
% \delta \left(g_{3} g_{8}^{-1} g_{5}^{-1}  g_{7} g_{4}^{-1} \right)\\
% && \delta \left(g_{8}^{-1} g_{5}^{-1} g_{6}\right)
% \delta \left(g_{5} g_{8} g_{6}^{-1} \right)
% \delta \left(g_{6} g_{8}^{-1} g_{5}^{-1} \right)=\\ 
%   & \int \prod_{l=4}^7 \mathrm{d} g_l
% &\delta \left(\mathds{1} \right)^2  \delta \left(\mathds{1} \right)^2=\delta \left(\mathds{1} \right)^4 \approx \Lambda^{12}\\ 
%\end{align}
\section{Vertex amplitudes of the 4D Ball diagrams}
\subsection{EPRLs model}
\label{AppC1}
\paragraph{4D Ball}
Here we write the five vertex amplitudes as a complement to equation \eqref{vertexamplitudeBallEPRLS}
\begin{flalign*}
A_1 =\sum_{i^{(1)}_{ev}} \left(\prod_{ev} (2 i_{ev}^{(1)} + 1) \right) \left\lbrace 15 j\right\rbrace_1 \ &  B_4(k_1, j_{1} ,j_{3},j_{2}; i_{1}, i^{(1)}_{1})B_4(k_2, j_{1} ,j_{5},j_{4}; i_{2}, i^{(1)}_{2})\\
&B_4(k_3, j_{6} ,j_{2},j_{4}; i_{3}, i^{(1)}_{3})B_4(k_4, j_{6} ,j_{3},j_{5}; i_{4}, i^{(1)}_{4})\ ,\\
A_2 =\sum_{i^{(2)}_{ev}} \left(\prod_{ev} (2 i_{ev}^{(2)} + 1) \right) \left\lbrace 15 j\right\rbrace_2 \ &  B_4(k_1, j_{1} ,j_{3},j_{2}; i_{1}, i^{(2)}_{1})B_4(k_5, j_{1} ,j_{8},j_{7}; i_{5}, i^{(2)}_{5})\\
&B_4(k_6, j_{7} ,j_{2},j_{6}; i_{6}, i^{(2)}_{6})B_4(k_7, j_{8} ,j_{3},j_{9}; i_{7}, i^{(2)}_{7})\ ,\\
A_3 =\sum_{i^{(3)}_{ev}} \left(\prod_{ev} (2 i_{ev}^{(3)} + 1) \right) \left\lbrace 15 j\right\rbrace_3 \ & B_4(k_2, j_{1} ,j_{5},j_{4}; i_{2}, i^{(3)}_{2})B_4(k_5, j_{1} ,j_{8},j_{7}; i_{5}, i^{(3)}_{5})\\
&B_4(k_8, j_{7} ,j_{4},j_{10}; i_{8}, i^{(3)}_{8})B_4(k_9, j_{8} ,j_{5},j_{10}; i_{9}, i^{(3)}_{9})\ ,\\
A_4 =\sum_{i^{(4)}_{ev}} \left(\prod_{ev} (2 i_{ev}^{(4)} + 1) \right) \left\lbrace 15 j\right\rbrace_4 \ &  B_4(k_3, j_{6} ,j_{2},j_{4}; i_{3}, i^{(4)}_{3})B_4(k_6, j_{7} ,j_{2},j_{6}; i_{6}, i^{(4)}_{6})\\
& B_4(k_8, j_{7} ,j_{4},j_{10}; i_{8}, i^{(4)}_{8}) B_4(k_{10}, j_{6} ,j_{9},j_{10}; i_{10}, i^{(4)}_{10})\ ,\\
A_5 =\sum_{i^{(5)}_{ev}} \left(\prod_{ev} (2 i_{ev}^{(5)} + 1) \right) \left\lbrace 15 j\right\rbrace_5 \ & 
B_4(k_1, j_{6} ,j_{3},j_{5}; i_{4}, i^{(5)}_{4}) B_4(k_7, j_{8} ,j_{3},j_{9}; i_{7}, i^{(5)}_{7}) \\
& B_4(k_9, j_{8} ,j_{5},j_{10}; i_{9}, i^{(5)}_{9}) B_4(k_{10}, j_{6} ,j_{9},j_{10}; i_{10}, i^{(5)}_{10})\ .\\
\end{flalign*}
The full change of variables on all the auxiliary intertwiners is the following:
\begin{equation*}
\bgroup
\setlength{\arraycolsep}{2em} 
\begin{array}{ccccc}
A_1:& \iota^{(1)}_1 = i^{(1)}_1 -j_1 & \iota^{(1)}_2 = i^{(1)}_2 -j_1 & \iota^{(1)}_3 = i^{(1)}_3 -j_6 & \iota^{(1)}_4 = i^{(1)}_4 -j_6 \\
A_2:&\iota^{(2)}_1 = i^{(2)}_1 -j_1 & \iota^{(2)}_5 = i^{(2)}_5 -j_1 & \iota^{(2)}_6 = i^{(2)}_6 -j_7 & \iota^{(2)}_7 = i^{(2)}_7 -j_8 \\
A_3:&\iota^{(3)}_2 = i^{(3)}_2 -j_1 & \iota^{(3)}_5 = i^{(3)}_5 -j_1 & \iota^{(3)}_8 = i^{(3)}_8 -j_7 & \iota^{(3)}_9 = i^{(3)}_9 -j_8 \\
A_4:&\iota^{(4)}_3 = i^{(4)}_3 -j_6 & \iota^{(4)}_6 = i^{(4)}_6 -j_7 & \iota^{(4)}_8 = i^{(4)}_8 -j_7 & \iota^{(4)}_{10} = i^{(4)}_{10} -j_6\\ 
A_5:&\iota^{(5)}_4 = i^{(5)}_4 -j_6 & \iota^{(5)}_7 = i^{(5)}_7 -j_8 & \iota^{(5)}_9 = i^{(5)}_9 -j_8 & \iota^{(5)}_{10} = i^{(5)}_{10} -j_6 
\end{array} \ .
\egroup
\end{equation*}
In terms of which expanding at the first order in $\lambda_f$ the amplitudes read:
\begin{flalign*}
A_1 \approx \lambda_1^2\lambda_6^2 \left\lbrace 15 j\right\rbrace_1 \ &
B_4(k_1, \lambda_{1} ,\lambda_{3},\lambda_{2}; \lambda_{1}, \lambda_{1}) B_4(k_2, \lambda_{1} ,\lambda_{5},\lambda_{4}; \lambda_{1}, \lambda_{1})\\
& B_4(k_3, \lambda_{6} ,\lambda_{2},\lambda_{4}; \lambda_6, \lambda_6)B_4(k_4, \lambda_{6} ,\lambda_{3},\lambda_{5}; \lambda_6, \lambda_6)\ ,\\
A_2  \approx \lambda_1^2\lambda_6 \lambda_8 \left\lbrace 15 j\right\rbrace_2 \ & 
 B_4(k_1, \lambda_{1} ,\lambda_{3},\lambda_{2}; \lambda_{1}, \lambda_{1})B_4(k_5, \lambda_{1} ,\lambda_{8},\lambda_{7}; \lambda_{1}, \lambda_{1})\\
&B_4(k_6, \lambda_{7} ,\lambda_{2},\lambda_{6}; \lambda_{7}, \lambda_{7})B_4(k_7, \lambda_{8} ,\lambda_{3},\lambda_{9}; \lambda_{8}, \lambda_{8})\ ,\\
A_3  \approx \lambda_1^2\lambda_7\lambda_8 \left\lbrace 15 j\right\rbrace_3 \ & B_4(k_2, \lambda_{1} ,\lambda_{5},\lambda_{4}; \lambda_{1},\lambda_{1})B_4(k_5, \lambda_{1} ,\lambda_{8},\lambda_{7};\lambda_{1},\lambda_{1})\\
&B_4(k_8, \lambda_{7} ,\lambda_{4},\lambda_{10}; \lambda_{7}, \lambda_{7})B_4(k_9, \lambda_{8} ,\lambda_{5},\lambda_{10}; \lambda_{8},\lambda_{8})\ ,\\
A_4 \approx \lambda_6^2 \lambda_7^2 \left\lbrace 15 j\right\rbrace_4 \ &
B_4(k_3, \lambda_{6} ,\lambda_{2},\lambda_{4}; \lambda_{6}, \lambda_{6})B_4(k_6, \lambda_{7} ,\lambda_{2},\lambda_{6}; \lambda_{7},\lambda_{7})\\
& B_4(k_8, \lambda_{7} ,\lambda_{4},\lambda_{10}; \lambda_{7},\lambda_{7}) B_4(k_{10}, \lambda_{6} ,\lambda_{9},\lambda_{10}; \lambda_{6},\lambda_{6})\ ,\\
A_5  \approx \lambda_6^2\lambda_8^2 \left\lbrace 15 j\right\rbrace_5 \ & 
B_4(k_1, \lambda_{6} ,\lambda_{3},\lambda_{5}; \lambda_{6},\lambda_{6}) B_4(k_7, \lambda_{8} ,\lambda_{3},\lambda_{9}; \lambda_{8},\lambda_{8}) \\
& B_4(k_9, \lambda_{8} ,\lambda_{5},\lambda_{10}; \lambda_{8},\lambda_{8}) B_4(k_{10}, \lambda_{6} ,\lambda_{9},\lambda_{10}; \lambda_{6},\lambda_{6})\ .\\
\end{flalign*}
\subsection{EPRL model}
\label{AppC2}
\paragraph{3D Ball}
Here we write the four vertex amplitudes as a complement to equation \eqref{vertexamplitude3DBallEPRL}
\begin{flalign*}
A_1= \sum_{\Delta l_1^{(1)}, \Delta l_3^{(1)}, \Delta l_4^{(1)}}&\sixj{k_{1} & k_{2} & k_{3}}{j_4+\Delta l_{4}^{(1)} & j_1+\Delta l_{1}^{(1)} & j_3+\Delta l_{3}^{(1)}} B_3(k_1,j_1 + \Delta l_1^{(1)},j_3+ \Delta l_3^{(1)})\\
&\quad B_3(j_4+ \Delta l_4^{(1)},k_2,j_3+ \Delta l_3^{(1)})B_3(j_4+ \Delta l_4^{(1)},j_1+ \Delta l_1^{(1)},k_3) \ ,\\
A_2=\sum_{\Delta l_1^{(2)}, \Delta l_2^{(2)}, \Delta l_4^{(2)}}&\sixj{k_{3} & k_{4} & k_{5}}{j_{2} +\Delta l_2^{(2)}& j_{4}+\Delta l_4^{(2)} & j_{1} +\Delta l_1^{(2)}} B_3(k_3,j_4+\Delta l_4^{(2)},j_1+\Delta l_1^{(2)})\\
&\quad B_3(j_2+\Delta l_2^{(2)},k_4,j_1+\Delta l_1^{(2)})B_3(j_2+\Delta l_2^{(2)},j_4+\Delta l_4^{(2)},k_5) \ , \\
A_3 =\sum_{\Delta l_2^{(3)}, \Delta l_3^{(3)}, \Delta l_4^{(3)}}&\sixj{k_{2} & k_{5} & k_{6}}{j_{2} +\Delta l_2^{(3)}& j_{3} +\Delta l_3^{(3)}& j_{4}+\Delta l_4^{(3)}} B_3(k_2,j_3+\Delta l_3^{(3)},j_4+\Delta l_4^{(3)}) \\
&\quad B_3(j_2+\Delta l_2^{(3)},k_5,j_4+\Delta l_4^{(3)})B_3(j_2+\Delta l_2^{(3)},j_3+\Delta l_3^{(3)},k_6) \ ,\\
A_4 =\sum_{\Delta l_1^{(4)}, \Delta l_2^{(4)}, \Delta l_3^{(4)}}&\sixj{k_{1} & k_{4} & k_{6}}{j_{2}+\Delta l_2^{(4)} & j_{3}+\Delta l_3^{(4)}& j_{1}+\Delta l_1^{(4)}} B_3(k_1,j_3+\Delta l_3^{(4)},j_1+\Delta l_1^{(4)}) \\
&\quad B_3(j_2 +\Delta l_2^{(4)},k_4,j_1+\Delta l_1^{(4)})B_3(j_2+\Delta l_2^{(4)},j_3+\Delta l_3^{(4)},k_6) \ . 
\end{flalign*}
The full change of variables on all the auxiliary spins is the following:
\begin{equation*}
\bgroup
\setlength{\arraycolsep}{2em} 
\begin{array}{llll}
A_1: & \delta_{1}^{(1)}=\Delta l_{1}^{(1)} & \delta_{3}^{(1)}=\Delta l_{3}^{(1)}-\Delta l_{1}^{(1)} & \delta_{4}^{(1)}=\Delta l_{4}^{(1)}-\Delta l_{1}^{(1)}\\ 
A_2: &\delta_{1}^{(2)}=\Delta l_{1}^{(2)} & \delta_{2}^{(2)}=\Delta l_{2}^{(2)}-\Delta l_{1}^{(2)} & \delta_{4}^{(2)}=\Delta l_{4}^{(2)}-\Delta l_{1}^{(2)}\\ 
A_3: &\delta_{2}^{(3)}=\Delta l_{2}^{(3)} & \delta_{3}^{(3)}=\Delta l_{3}^{(3)}-\Delta l_{2}^{(3)} & \delta_{4}^{(3)}=\Delta l_{4}^{(3)}-\Delta l_{2}^{(3)}\\ 
A_4: &\delta_{1}^{(4)}=\Delta l_{1}^{(4)} & \delta_{2}^{(4)}=\Delta l_{2}^{(4)}-\Delta l_{1}^{(4)} & \delta_{3}^{(4)}=\Delta l_{3}^{(4)}-\Delta l_{1}^{(4)}
\end{array} 
\egroup
\end{equation*}
In terms of which expanding at the first order in $\lambda_1$ and the umbounded variable $\delta^{(v)}_f$ the amplitudes read:
\begin{flalign*}
A_1\approx & \sum_{\delta_1^{(1)}} \sixj{k_{1} & k_{2} & k_{3}}{\lambda_1+\delta_1^{(1)} & \lambda_1+\delta_1^{(1)} & \lambda_1+\delta_1^{(1)}} B_3(k_1,\lambda_1+\delta_1^{(1)},\lambda_1+\delta_1^{(1)}) B_3(\lambda_1+\delta_1^{(1)}, k_2, \lambda_1+\delta_1^{(1)})B_3(\lambda_1+\delta_1^{(1)}, \lambda_1+\delta_1^{(1)}, k_3) \ ,\\
A_2\approx &\sum_{\delta_1^{(2)}}\sixj{k_{3} & k_{4} & k_{5}}{\lambda_1+\delta_1^{(2)}& \lambda_1+\delta_1^{(2)} & \lambda_1+\delta_1^{(2)}} B_3(k_3,\lambda_1+\delta_1^{(2)},\lambda_1+\delta_1^{(2)})B_3(\lambda_1+\delta_1^{(2)},k_4,\lambda_1+\delta_1^{(2)})B_3(\lambda_1+\delta_1^{(2)},\lambda_1+\delta_1^{(2)},k_5) \ ,\\
A_3\approx &\sum_{\delta_2^{(3)}}\sixj{k_{2} & k_{5} & k_{6}}{\lambda_1 +\delta_2^{(3)}& \lambda_1 +\delta_2^{(3)}& \lambda_1 +\delta_2^{(3)}} B_3(k_2,\lambda_1 +\delta_2^{(3)},\lambda_1 +\delta_2^{(3)})B_3(\lambda_1 +\delta_2^{(3)},k_5,\lambda_1 +\delta_2^{(3)})B_3(\lambda_1 +\delta_2^{(3)},\lambda_1 +\delta_2^{(3)},k_6) \ ,\\
A_4\approx &\sum_{\delta_1^{(4)}}\sixj{k_{1} & k_{4} & k_{6}}{\lambda_1 +\delta_1^{(4)}&\lambda_1 +\delta_1^{(4)}&\lambda_1 +\delta_1^{(4)}} B_3(k_1,\lambda_1 +\delta_1^{(4)},\lambda_1 +\delta_1^{(4)})B_3(\lambda_1 +\delta_1^{(4)},k_4,\lambda_1 +\delta_1^{(4)})B_3(\lambda_1 +\delta_1^{(4)},\lambda_1 +\delta_1^{(4)},k_6) \ . 
\end{flalign*}
\paragraph{4D Ball}
Here we write the five vertex amplitudes as a complement to equation \eqref{vertexamplitudeBallEPRL}
\begin{flalign*}
A_1 =\sum_{\Delta l^{(1)}_{fv}, i^{(1)}_{ev}} \left(\prod_{ev} (2 i_{ev}^{(1)} + 1) \right) \left\lbrace 15 j\right\rbrace_1 \ &  B_4(k_1, j_{1} +\Delta l^{(1)}_{1} ,j_{3} +\Delta l^{(1)}_{3},j_{2} +\Delta l^{(1)}_{2}; i_{1}, i^{(1)}_{1})\\[-1.5em]
&B_4(k_2, j_{1} +\Delta l^{(1)}_{1},j_{5}+\Delta l^{(1)}_{5},j_{4}+\Delta l^{(1)}_{4}; i_{2}, i^{(1)}_{2})\\
&B_4(k_3, j_{6} +\Delta l^{(1)}_{6},j_{2}+\Delta l^{(1)}_{2},j_{4}+\Delta l^{(1)}_{4}; i_{3}, i^{(1)}_{3})\\
&B_4(k_4, j_{6} +\Delta l^{(1)}_{6} ,j_{3} +\Delta l^{(1)}_{3},j_{5}+\Delta l^{(1)}_{5}; i_{4}, i^{(1)}_{4})\ ,\\
A_2 =\sum_{\Delta l^{(2)}_{fv}, i^{(2)}_{ev}} \left(\prod_{ev} (2 i_{ev}^{(2)} + 1) \right) \left\lbrace 15 j\right\rbrace_2 \ &  B_4(k_1, j_{1} +\Delta l^{(2)}_{1},j_{3}+\Delta l^{(2)}_{3},j_{2}+\Delta l^{(2)}_{2}; i_{1}, i^{(2)}_{1})\\[-1.5em]
&B_4(k_5, j_{1} +\Delta l^{(2)}_{1},j_{8}+\Delta l^{(2)}_{8},j_{7}+\Delta l^{(2)}_{7}; i_{5}, i^{(2)}_{5})\\
&B_4(k_6, j_{7} +\Delta l^{(2)}_{7},j_{2}+\Delta l^{(2)}_{2},j_{6}+\Delta l^{(2)}_{6}; i_{6}, i^{(2)}_{6})\\
&B_4(k_7, j_{8} +\Delta l^{(2)}_{8},j_{3}+\Delta l^{(2)}_{3},j_{9}+\Delta l^{(2)}_{9}; i_{7}, i^{(2)}_{7})\ ,\\
A_3 =\sum_{\Delta l^{(3)}_{fv}, i^{(3)}_{ev}} \left(\prod_{ev} (2 i_{ev}^{(3)} + 1) \right) \left\lbrace 15 j\right\rbrace_3 \  
& B_4(k_2, j_{1} +\Delta l^{(3)}_{1}, j_{5} +\Delta l^{(3)}_{5}, j_{4} +\Delta l^{(3)}_{4}; i_{2}, i^{(3)}_{2})\\[-1.5em]
& B_4(k_5, j_{1} +\Delta l^{(3)}_{1}, j_{8} +\Delta l^{(3)}_{8}, j_{7} +\Delta l^{(3)}_{7}; i_{5}, i^{(3)}_{5})\\
& B_4(k_8, j_{7} +\Delta l^{(3)}_{7}, j_{4} +\Delta l^{(3)}_{4}, j_{10} +\Delta l^{(3)}_{10}; i_{8}, i^{(3)}_{8})\\
& B_4(k_9, j_{8} +\Delta l^{(3)}_{8}, j_{5} +\Delta l^{(3)}_{5}, j_{10} +\Delta l^{(3)}_{10}; i_{9}, i^{(3)}_{9})\ ,\\
A_4 =\sum_{\Delta l^{(4)}_{fv}, i^{(4)}_{ev}} \left(\prod_{ev} (2 i_{ev}^{(4)} + 1) \right) \left\lbrace 15 j\right\rbrace_4 \ 
& B_4(k_3, j_{6} +\Delta l^{(4)}_{6}, j_{2} +\Delta l^{(4)}_{2}, j_{4} +\Delta l^{(4)}_{4}; i_{3}, i^{(4)}_{3})\\[-1.5em]
& B_4(k_6, j_{7} +\Delta l^{(4)}_{7}, j_{2} +\Delta l^{(4)}_{2}, j_{6} +\Delta l^{(4)}_{6}; i_{6}, i^{(4)}_{6})\\
& B_4(k_8, j_{7} +\Delta l^{(4)}_{7}, j_{4} +\Delta l^{(4)}_{4}, j_{10} +\Delta l^{(4)}_{10}; i_{8}, i^{(4)}_{8}) \\
& B_4(k_{10}, j_{6} +\Delta l^{(4)}_{6}, j_{9} +\Delta l^{(4)}_{9}, j_{10} +\Delta l^{(4)}_{10}; i_{10}, i^{(4)}_{10})\ ,\\
A_5 =\sum_{\Delta l^{(5)}_{fv}, i^{(5)}_{ev}} \left(\prod_{ev} (2 i_{ev}^{(5)} + 1) \right) \left\lbrace 15 j\right\rbrace_5 \ 
& B_4(k_1, j_{6} +\Delta l^{(5)}_{6},j_{3} +\Delta l^{(5)}_{3},j_{5} +\Delta l^{(5)}_{5}; i_{4}, i^{(5)}_{4}) \\[-1.5em]
& B_4(k_7, j_{8} +\Delta l^{(5)}_{8},j_{3} +\Delta l^{(5)}_{3},j_{9} +\Delta l^{(5)}_{9}; i_{7}, i^{(5)}_{7}) \\
& B_4(k_9, j_{8} +\Delta l^{(5)}_{8},j_{5} +\Delta l^{(5)}_{5},j_{10} +\Delta l^{(5)}_{10}; i_{9}, i^{(5)}_{9}) \\
& B_4(k_{10}, j_{6} +\Delta l^{(5)}_{6},j_{9} +\Delta l^{(5)}_{9},j_{10} +\Delta l^{(5)}_{10}; i_{10}, i^{(5)}_{10})\ .\\
\end{flalign*}
The full change of variables on all the auxiliary intertwiners and auxiliary face spins is the following:
\begin{equation*}
\bgroup
\setlength{\arraycolsep}{0.5em} 
\begin{array}{ccccc}
A_1:& \iota^{(1)}_1 = i^{(1)}_1 -j_1 - \Delta l^{(1)}_{1} & \iota^{(1)}_2 = i^{(1)}_2 -j_1 - \Delta l^{(1)}_{1}& \iota^{(1)}_3 = i^{(1)}_3 -j_6 - \Delta l^{(1)}_{6}& \iota^{(1)}_4 = i^{(1)}_4 -j_6 - \Delta l^{(1)}_{6}\\
A_2:&\iota^{(2)}_1 = i^{(2)}_1 -j_1  - \Delta l^{(2)}_{1}& \iota^{(2)}_5 = i^{(2)}_5 -j_1  - \Delta l^{(2)}_{1}& \iota^{(2)}_6 = i^{(2)}_6 -j_7  - \Delta l^{(2)}_{7}& \iota^{(2)}_7 = i^{(2)}_7 -j_8  - \Delta l^{(2)}_{8}\\
A_3:&\iota^{(3)}_2 = i^{(3)}_2 -j_1  - \Delta l^{(3)}_{1}& \iota^{(3)}_5 = i^{(3)}_5 -j_1  - \Delta l^{(3)}_{1}& \iota^{(3)}_8 = i^{(3)}_8 -j_7  - \Delta l^{(3)}_{7}& \iota^{(3)}_9 = i^{(3)}_9 -j_8  - \Delta l^{(3)}_{8}\\
A_4:&\iota^{(4)}_3 = i^{(4)}_3 -j_6  - \Delta l^{(4)}_{6}& \iota^{(4)}_6 = i^{(4)}_6 -j_7 - \Delta l^{(4)}_{7}& \iota^{(4)}_8 = i^{(4)}_8 -j_7 - \Delta l^{(4)}_{7}& \iota^{(4)}_{10} = i^{(4)}_{10} -j_6- \Delta l^{(4)}_{6}\\ 
A_5:&\iota^{(5)}_4 = i^{(5)}_4 -j_6 - \Delta l^{(5)}_{6}& \iota^{(5)}_7 = i^{(5)}_7 -j_8 - \Delta l^{(5)}_{8}& \iota^{(5)}_9 = i^{(5)}_9 -j_8 - \Delta l^{(5)}_{8}& \iota^{(5)}_{10} = i^{(5)}_{10} -j_6 - \Delta l^{(5)}_{6} 
\end{array} 
\egroup
\end{equation*}
In terms of which expanding at the first order in $\lambda_f$ and $\delta^{(v)}_f$ the amplitudes read:
\begin{flalign*}
A_1 \approx \sum_{\delta^{(1)}_{fv}}\left(\lambda_1 + \delta_1^{(1)}\right)^2\left(\lambda_6 + \delta^{(1)}_6\right)^2 \left\lbrace 15 j\right\rbrace_1 \ 
& B_4(k_1, \lambda_{1}  + \delta^{(1)}_1,\lambda_{3} + \delta^{(1)}_3,\lambda_{2} + \delta^{(1)}_2; \lambda_1, \lambda_{1} + \delta^{(1)}_1)\\[-1.5em]
& B_4(k_2, \lambda_{1} + \delta^{(1)}_1,\lambda_{5}+ \delta^{(1)}_5,\lambda_{4}+ \delta^{(1)}_4; \lambda_1, \lambda_{1}+ \delta^{(1)}_1)\\
& B_4(k_3, \lambda_{6} + \delta^{(1)}_6,\lambda_{2}+ \delta^{(1)}_2,\lambda_{4}+ \delta^{(1)}_4; \lambda_6, \lambda_6+ \delta^{(1)}_6)\\
& B_4(k_4, \lambda_{6} + \delta^{(1)}_6,\lambda_{3}+ \delta^{(1)}_3,\lambda_{5}+ \delta^{(1)}_5; \lambda_6, \lambda_6+ \delta^{(1)}_6)\ ,\\[0.5em]
A_2  \approx \sum_{\delta^{(2)}_{fv}}\left(\lambda_1 + \delta_1^{(2)}\right)^2\left(\lambda_6 + \delta^{(2)}_6\right)\left(\lambda_8 + \delta^{(1)}_8\right) \left\lbrace 15 j\right\rbrace_2 \ 
&B_4(k_1, \lambda_{1} + \delta^{(2)}_1,\lambda_{3} + \delta^{(2)}_3,\lambda_{2} + \delta^{(2)}_2; \lambda_{1}, \lambda_{1} + \delta^{(2)}_1)\\[-1.5em]
&B_4(k_5, \lambda_{1} + \delta^{(2)}_1,\lambda_{8}+ \delta^{(2)}_8,\lambda_{7}+ \delta^{(2)}_7; \lambda_{1}, \lambda_{1}+ \delta^{(2)}_1)\\
&B_4(k_6, \lambda_{7} + \delta^{(2)}_7,\lambda_{2}+ \delta^{(2)}_2,\lambda_{6}+ \delta^{(2)}_6; \lambda_{7}, \lambda_{7}+ \delta^{(2)}_7)\\
&B_4(k_7, \lambda_{8}+ \delta^{(2)}_8 ,\lambda_{3}+ \delta^{(2)}_3,\lambda_{9}+ \delta^{(2)}_9; \lambda_{8}, \lambda_{8}+ \delta^{(2)}_8)\ ,\\[0.5em]
A_3  \approx \sum_{\delta^{(3)}_{fv}}\left(\lambda_1 + \delta^{(3)}_1\right)^2\left(\lambda_7 + \delta^{(3)}_7\right)\left(\lambda_8 + \delta^{(3)}_8\right) \left\lbrace 15 j\right\rbrace_3 \ 
& B_4(k_2, \lambda_{1} + \delta^{(3)}_1,\lambda_{5}+ \delta^{(3)}_5,\lambda_{4}+ \delta^{(3)}_4; \lambda_{1},\lambda_{1}+ \delta^{(3)}_1) \\[-1.5em]
& B_4(k_5, \lambda_{1} + \delta^{(3)}_1,\lambda_{8}+ \delta^{(3)}_8,\lambda_{7}+ \delta^{(3)}_7;\lambda_{1},\lambda_{1}+ \delta^{(3)}_1)  \\
& B_4(k_8, \lambda_{7}+ \delta^{(3)}_7 ,\lambda_{4}+ \delta^{(3)}_4,\lambda_{10}+ \delta^{(3)}_{10}; \lambda_{7}, \lambda_{7}+ \delta^{(3)}_7) \\
& B_4(k_9, \lambda_{8} + \delta^{(3)}_8,\lambda_{5}+ \delta^{(3)}_5,\lambda_{10}+ \delta^{(3)}_{10}; \lambda_{8},\lambda_{8}+ \delta^{(3)}_8)\ ,\\[0.5em]
A_4 \approx \sum_{\delta^{(4)}_{fv}}\left(\lambda_6 + \delta^{(4)}_{6}\right)^2\left(\lambda_7 + \delta^{(4)}_{7}\right)^2 \left\lbrace 15 j\right\rbrace_4 \ 
& B_4(k_3, \lambda_{6}+ \delta^{(4)}_{6} ,\lambda_{2}+ \delta^{(4)}_{2},\lambda_{4}+ \delta^{(4)}_{4}; \lambda_{6}, \lambda_{6}+ \delta^{(4)}_{6})\\[-1.5em]
& B_4(k_6, \lambda_{7}+ \delta^{(4)}_{7} ,\lambda_{2}+ \delta^{(4)}_{2},\lambda_{6}+ \delta^{(4)}_{6}; \lambda_{7},\lambda_{7}+ \delta^{(4)}_{7})\\
& B_4(k_8, \lambda_{7}+ \delta^{(4)}_{7} ,\lambda_{4}+ \delta^{(4)}_{4},\lambda_{10}+ \delta^{(4)}_{10}; \lambda_{7},\lambda_{7}+ \delta^{(4)}_{7}) \\
& B_4(k_{10}, \lambda_{6}+ \delta^{(4)}_{6} ,\lambda_{9}+ \delta^{(4)}_{9},\lambda_{10}+ \delta^{(4)}_{10}; \lambda_{6},\lambda_{6}+ \delta^{(4)}_{6})\ ,\\[0.5em]
A_5  \approx \sum_{\delta^{(5)}_{fv}}\left(\lambda_6 + \delta^{(5)}_{6}\right)^2\left(\lambda_8 + \delta^{(5)}_{8}\right)^2 \left\lbrace 15 j\right\rbrace_5 \ 
& B_4(k_1, \lambda_{6}  + \delta^{(5)}_{6},\lambda_{3} + \delta^{(5)}_{3},\lambda_{5} + \delta^{(5)}_{5}; \lambda_{6},\lambda_{6} + \delta^{(5)}_{6}) \\[-1.5em]
& B_4(k_7, \lambda_{8}  + \delta^{(5)}_{8},\lambda_{3} + \delta^{(5)}_{3},\lambda_{9} + \delta^{(5)}_{9}; \lambda_{8},\lambda_{8} + \delta^{(5)}_{8}) \\
& B_4(k_9, \lambda_{8}  + \delta^{(5)}_{8},\lambda_{5} + \delta^{(5)}_{5},\lambda_{10} + \delta^{(5)}_{10}; \lambda_{8},\lambda_{8} + \delta^{(5)}_{8}) \\
& B_4(k_{10}, \lambda_{6}  + \delta^{(5)}_{6},\lambda_{9} + \delta^{(5)}_{9},\lambda_{10} + \delta^{(5)}_{10}; \lambda_{6},\lambda_{6} + \delta^{(5)}_{6})\ .
\end{flalign*}

%----------------------------------------------------------------------------
\section{Details of the numeric analysis}\label{AppB}
%----------------------------------------------------------------------------
All the computation are done with Wolfram Mathematica and a C++ code. The computation of the $B_3$ booster functions use the formula in terms of $SL(2,\mathbb{C})$ Clebsch-Gordan coefficients reported in \cite{speziale_boosting_2016}. The computation of the $B_4$ booster functions use the formula \eqref{boosters} and the integral over the rapidity is done numerically using arbitrary precision artimetic libraries GMP \cite{granlund_gnu_2012}, MPFR \cite{fousse_mpfr:_2007} and MPC \cite{enge_mpc:_2015}. The details on how the code works and what kind of techniques are used will be illustrated in a future work \cite{citaFrancois}. 

\begin{figure}[h!t]
\centering
    \includegraphics[width=0.5 \textwidth]{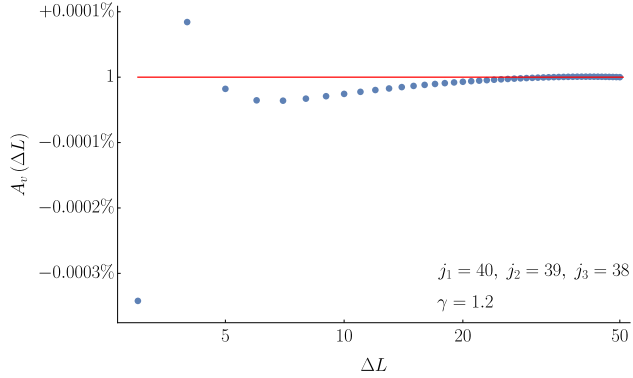}
\caption{\label{fig:convergence} {\small 
Convergence of the sum over the virtual spins of the vertex amplitude $A_1 = A_2 \equiv A_v$ for face spins $j_1=40$, $j_2=39$, $j_3=38$. We plot $A_v$ as a function of the cutoff $\Delta L$ imposed uniformly on the three virtual $\Delta l_f \leq \Delta L$ for $f=1,2,3$. For readability we also rescale the plot such that $A_v(50)=1$. 
}}
\end{figure}

To be sure that the summation over the virtual spins in \eqref{amplitudeA1A2} we picked a ``large'' number $\Delta L=50$ and trucate the sum over the $\Delta l_f$ at that value. We then go back and check that for each configuration of face spins $j_f$ the sum $A_v$ ``converged''. Numerically we decided to be satisfied with the truncation if the sum chaged only by a 0.0001\% (we choose this number arbitrarly). The convergence of those sum is quite fast, to be concrete we plot in Figure \ref{fig:convergence} the values of of the vertex amplitude $A_v$ as a function of the truncation $\Delta L$ the configuration with the slowest convergence.

\newpage

\providecommand{\href}[2]{#2}

\begingroup\raggedright
\endgroup

\end{document}